\documentclass[useAMS,usenatbib]{mn2e}

\usepackage{natbib}
\usepackage{cite}
\usepackage{amssymb}
\usepackage{graphicx}
\usepackage{graphics}
\usepackage{lscape}
\usepackage{color}
\newcommand{\mnras}{MNRAS}
\newcommand{\apj}{ApJ}
\newcommand{\apjl}{ApJL}
\newcommand{\apjs}{ApJS}
\newcommand{\aj}{AJ}
\newcommand{\aap}{A\&A}

\newcommand{\pasp}{PASP}
\newcommand{\aaps}{AAPS}
\newcommand{\nat}{Nature}

\newcommand{\ha}{H$\alpha$}

\newcommand{\msol}{M$_{\small{\sun}}$}

\title[On the Nature of Star Formation at Large Galactic Radii]{On the Nature of Star Formation at Large Galactic Radii}
\author[Q. E. Goddard, R. C. Kennicutt, E. V. Ryan-Weber]{Q. E. Goddard$^{1}$\thanks{E-mail:
goddard@ast.cam.ac.uk; qeg20@cam.ac.uk}, R. C. Kennicutt$^{1}$ \& E. V. Ryan-Weber$^{2}$\\
$^{1}$Institute of Astronomy, University of Cambridge, Madingly Road, Cambridge. CB3 0HA\\
$^{2}$Centre for Astrophysics \& Supercomputing, Swinburne University of Technology, \\Mail H39, PO Box 218, Hawthorn, 3122 VIC, Australia}

\begin{document}

\date{}

\pagerange{\pageref{firstpage}--\pageref{lastpage}} \pubyear{2010}

\maketitle

\label{firstpage}

\begin{abstract}

  We have compared far-ultraviolet (FUV), near-ultraviolet (NUV), and \ha\ 
  measurements for star forming regions in 21 galaxies, in order to 
  characterise the properties of their discs at radii beyond the main
  optical radius ($R_{25}$).  \
  
  In our representative sample of extended 
  and non-extended UV discs we find that half of the extended UV discs 
  also exhibit extended \ha\ emission. We find that extended UV discs fall 
  into two categories, those with a sharp truncation in the \ha\ disc
  close to the optical edge (R$_{25}$), and those with extended emission
  in \ha\ as well as in the ultraviolet.  Although most galaxies with
  strong \ha\ truncations near R$_{25}$ show a significant corresponding
  falloff in UV emission (factor 10--100), the transition tends to be
  much smoother than in \ha, and significant UV emission often extends
  well beyond this radius, confirming earlier results by \citet{thilker07a} and
  others.  \
  
  After correcting for dust attenuation the median fraction of total FUV emission 
  from regions outside of $R_{25}$ is 1.7\%, but it can be as high as 35\% in the most
  extreme cases.  The corresponding fractions of \ha\ emission are 
  approximately half as large on average.  This difference reflects
  both a slightly lower ratio of \ha\ to UV emission in the HII
  regions in the outer discs, as well as a lower fraction of star
  clusters showing HII regions.  Most HII
  regions in the extended disc have fluxes consistent with small numbers of
  ionising O-type stars, and this poor sampling of the upper initial
  mass function in small clusters can probably account for the differences
  in the emission properties, consistent with earlier conclusions by
  \citet{zaritsky07}, without needing to invoke a significant
  change in the stellar IMF itself. Consistent \ha/FUV ratios and brightest HII region 
  to total \ha\ fluxes in the inner and extended discs across our whole galaxy sample
  demonstrate no evidence for a change in the cluster luminosity function or the IMF in the low gas density outer disc.

\end{abstract}


\section{Introduction}

One of the most interesting discoveries of the Galaxy Evolution Explorer
(GALEX) satellite has been the presence of very extended ultraviolet (UV)
emitting discs around many nearby galaxies.
\citet{thilker05} first reported
an extended UV disc for the spiral galaxy M83. Since then extended
discs have be found in many galaxies, most notably NGC 4625
\citep{gildepaz05}. Extended discs typically exhibit emission well
beyond the classical optical edge of the galaxy, which is usually
defined by R$_{25}$, the radius at which the surface brightness in the
B band drops below 25 magnitudes arcsec$^{-2}$.  In M83
UV knots are found extending to 4 R$_{25}$, and are 
associated with large scale filamentary HI structures.  These structures 
are now commonly termed extended ultraviolet discs
(XUV-discs).

Subsequent studies have begun to characterise the frequency of XUV-discs
and the properties of the star forming regions in these
discs. \citet{zaritsky07} examined a sample of 11 galaxies and found
an excess of blue (FUV -NUV $<$ 1, NUV $<$ 25) sources out to 2
R$_{25}$ for $\sim 25\%$ of their sample. Based on and analysis of
the GALEX Nearby Galaxies Survey (NGS; \citet{gildepaz07a}),
\citet{thilker07a} concluded that extended discs are common, and can
be divided into two distinct types. Type 1 discs ($\ga$ 20\%
occurence) show structured UV bright regions beyond the typical star
forming threshold; Type 2 discs ($\approx$10\% occurrence) display
diffuse regions of UV emission, but not reaching extreme
galactocentric radii. The vast majority of galaxies do not exhibit any
significant emission beyond R$_{25}$, and are not classified as
extended, though it should be noted that there were no quantitative
definitions to identify extended discs.

HII regions are often but not always associated with the extended UV
emission. HII regions located well beyond
the main star forming disc were identified in one of the first \ha\
surveys of galaxies (e.g. \citet{hodge69, hodge74, hodge83}), and have been
studied in detail by \citet{ferguson98}, \citet{vanzee98},
and \citet{lelievre00}.  Isolated intergalactic
HII regions have also been identified by \citet{gerhard02}, Sakai et al.
(2002) and \citet{cortese04} in the Virgo and Abell 1367 clusters,
and by \citet{ryanweber04a} for nearby galaxies in groups. 
The lifetimes of massive ionising
stars responsible for \ha\ emission are much lower than those of 
the stars responsible for most of the near-UV emission
(of order 10 Myr vs 100 Myr), and this is reflected in
the relative numbers of UV and \ha\ knots observed by \citet{zaritsky07}.
 
The formation of massive stars beyond the classical 'edge' of galaxies
raises questions about their formation and properties, as well as their
possible impacts on their low gas density environment. Spectroscopy of
HII regions at extreme radii in M83 and NGC4625 by \citet{gildepaz07b}
showed them to be consistent with regions dominated by a single ionising
source star with masses in the range 20-40 \msol.  Ages for most of 
the UV knots are in the range 0--200 Myr, based on their integrated
fluxes and colors (e.g., \citet{thilker05}, \citet{gildepaz05},
\citet{zaritsky07}, \citet{dong08}).  
Masses for the regions are more difficult to
constrain; \citet{gildepaz05} report a mass range of $10^{3}-10^{4}$
\msol\ for NGC4625, and \citet{werk08} gives an upper mass limits in
the range 600\msol\ to 7000\msol\ for HII regions associated with
NGC1533. \citet{dong08} used the UV to IR spectral energy
distributions to estimate the masses of UV knots and measured a range
of $10^{3}-3\times 10^{6}$ \msol with a peak $\approx 10^{4.7}$ \msol
\ for M83. In a study of the HII regions of NGC628 \citet{lelievre00}
postulated that the cluster mass function of these HII regions may be
significantly different to that of the disc.  Relatively little is
known yet about the chemical abundance properties of the outer discs, but
preliminary results show a range of behaviours.  Abundance studies of
a few systems such as M101 and NGC 628 show a continuous exponential
abundance gradient extending into the outermost discs \citep{kennicutt03,ferguson98}.  However a recent study of
HII regions in the XUV disc of M83 show chemical properties that are decoupled from
those of the brighter inner disc, with a nearly constant oxygen abundance
of $\sim$0.3 (O/H)$_\odot$ \citep{bresolin09}.


The XUV-discs are also important for understanding star formation
thresholds in galaxies. Radial profiles in the \ha\ often show sharp
turnovers, usually located near the optical R{$_{25}$
radius. \citet{kennicutt89,martin01} showed that the distribution of
gas density appears to be roughly continuous across this truncation
in contrast to the massive star formation threshold visible in \ha\
profiles. These observation have been interpreted as arising from a
threshold surface gas density, which could be attributed to
gravitation instability \citep{kennicutt89,martin01}, gas phase
instabilities \citep{schaye04} or gas cloud fragmentation
instability \citep{krumholz08}. However UV emission, which also
traces star formation does not display truncations as pronounced as
those seen in \ha\ \citep{thilker05}. This raises questions about the
nature and interpretation of the apparent \ha\ thresholds.

A closely related question is the physical origin of the
XUV-discs. \citet{dong08,bush08} have argued that the XUV -discs can
be understood in the context of normal gravitational threshold
picture. Although the average gas densities beyond the \ha\ thresholds are too low in theory to allow star formation, in
localised regions the density may be high enough to prompt star
formation in a small volume. Others have suggested that interactions
play a role. Interactions may disturb the gas creating regions
that collapse and thus form pockets of
stars. \citet{gildepaz05} proposed such an interaction between NGC
4625 and its neighbor NGC 4618 and possibly NGC 4625A as a likely
trigger for star formation at large radii. Similar processes may be
able to account for the presence of HII regions at very large radii in some
interacting and merging systems
\citep{ryanweber04a,oosterloo04,werk08}, but other factors may
contribute such as galactic outflows and spiral density waves. \citet{elmegreen06} modelled star formation in the outer discs, citing the effects of compression and turbulence as well as the propagation of gaseous arms as continual drivers for low level star formation.



A common question underlying these studies is whether star formation
at large radii represents a simple continuation of star formation
with in the inner disc, or is it a separate mode of star formation
altogether?  Perhaps the most radical interpretation of the XUV discs
is that they represent low surface density environments in which the
stellar initial mass function is truncated at masses well below the
$\sim$100 \msol\ \citep{kroupa08, meurer09}.  A strong
preferential suppression of massive star formation in the discs could
account for the putative excesses of UV emission relative to \ha\ 
emission, and possibly account for the differences in the outer
disc profiles at the respective wavelengths.

Before these questions posed above can be answered we need uniform
measurements of the UV and \ha\ properties of the extended discs  
for a representative sample
of nearby galaxies.  Although the published studies to date have
highlighted the remarkable properties of a handful of galaxies,
most have focussed on case studies of extreme examples such as
M83 and NGC 4625.  The goal of this study is to address the phenomenon
in a broader context, by carrying out detailed photometric measurements
in the UV and \ha\ for the discs and their individual star-forming
regions, for 20 galaxies selected from the Spitzer Infrared Nearby
Galaxies Survey (SINGS; \citet{kennicutt03}) as well as the prototype
galaxy M83.  Some of our work parallels an important study of UV radial
profiles of nearby galaxies by \citet{boissier07}, but our study
differs in emphasis in focussing especially on the relation of the
UV and \ha\ discs, and on measurements of the individual star-forming
knots.


The remainder of this paper is organized as follows.  In Section 2
we describe the GALEX and groundbased \ha\ observations that were 
used in this study, and in Section 3 we describe the processing
and analysis of these data, including the treatment of dust
attenuation.  In Section 4 we present the observed properties of
the extended discs in terms of their UV emission, \ha\ emission,
and star formation properties, and in Section 5 we describe the
properties of the populations of star-forming regions in both
the inner and outer discs of the galaxies.  In Section 6 we 
discuss the results in the context of the questions raised above,
and in Section 7 we present a brief summary of results and conclusions.

\section[]{The Data}

\subsection{Sample Selection}


All the galaxies in our sample were taken from the Spitzer infrared Nearby Galaxies Survey (SINGS) and the GALEX Nearby Galaxy Survey (NGS). SINGS is a multi-wavelength imaging and 
spectroscopic survey of 75 nearby (D$<$30Mpc) galaxies, including visible,
near-infrared, ultraviolet and radio observations.  The galaxies were
selected to span a range of galaxy types, luminosities, and 
infrared/optical properties among normal galaxies in this volume.
We selected a subset of 20 spiral galaxies, with emphasis on nearby,
face-on or moderately inclined objects with major axis diameters in the range
$2\arcmin<a<13\arcmin$; this was large enough to allow surface photometry 
of the inner disc whilst allowing for a reasonable background estimation
outside the disc.  We included all galaxies with known extended discs,
but also included a representative sampling of galaxies selected independently
from the outer disc struture.  The prototype XUV disc galaxy M83 is
not in the SINGS sample, so we added it to our sample for purposes of comparison.

The full list of galaxies along with their
associated properties is presented in Table \ref{tab:gals}. Although this sample of
galaxies does not cover the full range of possible galaxy types, it
provides a representative sample of galaxies both with and without
extended UV emission.

\begin{table*}
\begin{minipage}{160mm}
 \caption{Properties of selected galaxies}
 \label{tab:gals}
 \begin{tabular}{@{}lccccccccc}
  \hline
  \hline

 NGC no. & Other & Type & B-Band & FUV & NUV & D25 & Distance & Ref & E(B-V) \\
   & Names(s) & & Magnitude & Mag & Mag& (arcmin) & (Mpc) & & \\
   (1) & (2) & (3) & (4) & (5) & (6) & (7) & (8) & (9) & (10)\\

  \hline
628 	& M74 & 	SAc & 	$9.95\pm0.01$ & 	$11.81\pm0.01$ & 	$11.50\pm0.01$ & 	$10.5 \times 9.5$ & 11.4	&$_{1}$ & 0.07 \\
925 	& & 		SABd & 	$10.96\pm0.11$ & 	$12.19\pm0.01$ & 	$11.96\pm0.01$ &	$10.5 \times 5.9$ & 10.1	&$_{2}$ & 0.08 \\
1097 & & 		SBd & 	$10.23\pm0.07$ & 	$12.60\pm0.01$ & 	$12.21\pm0.01$ &	$9.3 \times 6.3$ & 16.9	&$_{3}$ & 0.03 \\
1291 & & 		SBa & 	$9.39\pm0.04$ & 	$14.83\pm0.01$ & 	$13.81\pm0.01$ &	$9.8 \times 8.1$ & 9.7	&$_{4}$ & 0.01\\
1512 & & 		SBab & 	$11.13\pm0.10$ & 	$13.89\pm0.01$ & 	$13.58\pm0.01$ &	$8.9 \times 5.6$ & 10.4	&$_{4}$ & 0.01\\
1566 & & 		SABbc & 	$10.33\pm0.03$ & 	$12.18\pm0.01$ & 	$11.96\pm0.01$ & 	$8.3 \times 6.6$ & 18.0	&$_{3}$ & 0.01\\
2841 & & 		SAb & 	$10.09\pm0.10$ & 	$13.73\pm0.01$ & 	$13.20\pm0.01$ & 	$8.1 \times 3.5$ & 14.0	&$_{5}$ & 0.02 \\
3198 & & 		SBc & 	$10.87\pm0.10$& 	$13.13\pm0.01$ & 	$12.91\pm0.01$ & 	$8.5 \times 3.3$ & 9.8	&$_{6}$ & 0.01 \\
3351 & M95 & 	SBb & 	$10.53\pm0.10$& 	$13.36\pm0.01$ & 	$12.82\pm0.01$ & 	$7.4 \times 5.0$ &  9.3	&$_{3}$ & 0.03 \\
3521 & & 		SABbc & 	$9.83\pm0.10$& 	$13.09\pm0.01$ & 	$12.33\pm0.01$ & 	$11.0 \times 5.1$ & 9.0	&$_{4}$ & 0.06 \\
3621 & & 		Sad & 	$10.05\pm0.08$& 	$11.97\pm0.01$ & 	$11.51\pm0.01$ & 	$12.3 \times 7.1$ & 6.2	&$_{3}$ & 0.08 \\
4321 & M100, V & SABbc & $10.05\pm0.08$ & $...$ & 	$12.10\pm0.01$ & 	$7.4 \times 6.3$ & 16.5	&$_{12}$ & 0.03 \\
4536 & V& 	SABbc & 	$11.16\pm0.08$ & 	$13.39\pm0.01$ & 	$13.10\pm0.01$ & 	$7.6 \times 3.2$ & 16.5	&$_{12}$ & 0.02 \\
4579 & M58, V & SABb & $10.48\pm0.08$& 	$14.49\pm0.01$ & 	$13.71\pm0.01$ & 	$5.9 \times 4.7$ & 16.5	&$_{12}$ & 0.04 \\
4625 & & 		SABmp & $12.92\pm0.04$ & 	$14.93\pm0.01$ & 	$14.57\pm0.01$ & 	$2.2 \times 1.9$ & 9.5	&$_{4}$ & 0.02 \\
5194 & M51a & SABbc & $8.96\pm0.06$ & 	$10.95\pm0.01$ & 	$10.42\pm0.01$ & 	$11.2 \times 6.9$ & 8.2	&$_{8}$ & 0.03 \\
5236 & M83 & 	SABc & 	$8.20\pm0.03$ &  	$10.13\pm0.01$ & 	$9.55\pm0.01$ & 	$12.9 \times 11.5$ & 4.5	&$_{9}$ & 0.07 \\
5398 & & 		SABdm & $12.78\pm0.17$& 	$14.25\pm0.01$ & 	$13.83\pm0.01$ & 	$2.8 \times 1.7$ & 15.0	&$_{3}$ & 0.07 \\
5474 & & 		SAcd & 	$14.36\pm0.14$ & 	$13.16\pm0.01$ & 	$13.01\pm0.01$ & 	$4.8 \times 4.3$ & 6.9	&$_{10}$ & 0.01 \\
7552 & & 		SAc & 	$11.25\pm0.13$ & 	$14.26\pm0.01$ & 	$13.53\pm0.01$ & 	$3.4 \times 2.7$ & 22.3	&$_{4}$ & 0.01 \\
7793 & & 		SAd & 	$9.63\pm0.05$&  	$11.23\pm0.01$ & 	$11.05\pm0.01$ & 	$9.3 \times 6.3$ & 3.2	&$_{11}$ & 0.02 \\
 \hline
 \end{tabular}
 \medskip
 Column (1) lists the NGC number.  Column (2) shows other
common names for the galaxy, and the symbol 'V' notes galaxies
associated with the Virgo cluster. Column (3) indicates galaxy type, taken
from the NASA-IPAC Extragalactic Database (NED)\footnote{The NASA/IPAC 
Extragalactic Database (NED) is operated by the Jet Propulsion Laboratory, 
California Institute of Technology, under contract with the National 
Aeronautics and Space Administration.}. The B-band magnitude in Column (4) and D25 ellipse information in Column (5) are both
taken from the RC3 catalogue \citep{devaucouleurs91}. Columns (6) and (7) show the 
FUV and NUV AB magnitude taken from \citet{gildepaz07a}, errors are the same as 
previously published. 
Galaxy distances are shown in Column (8). Column (9)
indicates the reference from which the galaxy distance was taken; (1)
\citet{tully88}; (2) \citet{silbermann96}; (3) from the Virgo-infall
corrected radial velocity adopting H$_{0}$=70 km s$^{-1}$ Mpc$^{-1}$;
(4) \citet{kennicutt03}; (5) \citet{macri01}; (6) \citet{paturel02};
(7) \citet{gavazzi99}; (8) \citet{feldmeier97}; (9) \citet{thim03};
(10) \citet{drozdovsky00}; (11) assumed to be at the same distance as
NGC 300 tanken from \citet{freedman01}; (12) the distance to the Virgo cluster taken from \citet{mei07}.
Column (10) lists the Galactic colour excess taken from \citet{schlegel98}.
 \end{minipage}
 \end{table*}

\subsection{GALEX Data}

All ultraviolet images were taken from the GALEX data release 3
catalogue (Morrissey et al. 2007). 
For all but a few objects the galaxy is located at the centre
of the field of view. Exceptions to this are M83 where a
bright star to the north-east required repositioning of the field, and
NGC 4625, where the field was shifted to include both 4625 and its
companion 4618. The GALEX satellite boasts an impressive $1.2\degr$
circular field of view and a pixel scale of $1.5\arcsec$. The two
GALEX bands are centred at 1516\,\AA\ (FUV) and 2267\,\AA\ (NUV), with bandpass widths
of 269\,\AA\ and 616\,\AA\ FWHM.  The spatial resolutions are approximately
$4\arcsec.5$ and $5\arcsec$ in the FUV and NUV bands, respectively
(Morrissey et al. 2007).  Typical integration times for the NGS were
$\approx 1700$ s, one orbit time for GALEX. The observing time yielded limiting sensitivities of 26.6
(26.8) AB mag for the FUV (NUV), or 27.5 (27.6) AB mag arcsec$^{-2}$,
expressed in units of surface brightness evaluated at the scale of the
point-spread function. 
We used the background maps provided with the images as a first order
background subtraction on each field; the typical background levels in the
GALEX images are approximately 10$^{-4}$ counts per second for the FUV
and 10$^{-3}$ for the NUV, corresponding to surface brightnesses of $1.40\times10^{-19}$ ergs s$^{-1}$ cm$^{-2}$ \AA$^{-1}$ in the FUV and $2.06\times10^{-19}$ ergs s$^{-1}$ cm$^{-2}$ \AA$^{-1}$ in the NUV.
 
\subsection{\ha\ Data}
 
For most of the galaxies in our sample archival narrowband imaging in
the H$\alpha +$ [NII] emission lines was obtained from the SINGS survey.
These data were obtained using CCD imagers on the 2.1\,m telescope at
Kitt Peak National Observatory or the 1.5\,m telescope at Cerro Tololo
Interamerican Observatory.  Corresponding imaging in the $R$ band was
used for subtraction of underlying stellar continuum emission.  Details 
of the observing parameters and reduction procedures can be found in
the online documentation accompanying the SINGS data products.

For M83 we used new narrowband and $R$-band imaging of the galaxy obtained
with the 90Prime imager on the Steward Observatory Bok 2.3\,m telescope
(Williams et al. 2004).  The instrument images a field 1.16 $\times$ 1.16
degrees, using four 4096 $\times$ 4097 element CCD detectors with a pixel
scale of 0.45 arcsec.  A series of dithered exposures was obtained to
over a contiguous 1-degree field with an effective exposure time of 
75\,m in H$\alpha$, using a filter with bandpass 7 nm centered at 658 nm.
A shorter set of matching exposures with a Kron-Cousins $R$-band filter
was used to provide continuum subtraction.  Similarly NGC 4625 was observed
with the same imager, but since this galaxy easily fit into one of the
CCD fields ($\sim$30 arcmin square) we used only one of the detectors
for those observations.  Observations were calibrated using imaging of
spectrophotometric standard stars, and also checked using independently
calibrated H$\alpha$ imaging of the galaxies from Kennicutt et al. (2008).

All of our narrowband observations included emission contributions
from H$\alpha$ as well as the forbidden lines of [NII]$\lambda\lambda$6548,6583.
We have chosen to perform our analysis in terms of the combined
H$\alpha$ + [NII] fluxes, because our primary interest in this paper
is in the outer star-forming discs, where the [NII] emission is
expected to be weak (0.01 $\le$ [NII]/H$\alpha$ $\le$ 0.2), and 
Hence the line emission in these regions should be dominated by H$\alpha$. This assumption is confirmed in spectra of HII regions in the extended disc of M83 \citep{bresolin09}, in which [NII]/\ha\ typically ranges from 0.12 to 0.3 and NGC 4625 \citep{gildepaz07b} which has values between 0.05 - 0.15. \citet{ferguson98} also took spectra of HII in the extended discs of several galaxies, in particular NGC628 where the [NII]/\ha\ ratio drops from 0.4 in the inner disc to 0.06 in the most distant HII region. \citet{vanzee98} reports similar ratios in a sample of HII regions from the inner discs of mostly spiral face-on galaxies.
Our alternative choice would have been to apply a mean correction
for [NII] emission (e.g., Appendices 1 and 2 of Kennicutt et al. 2008),
but since the integrated [NII]/H$\alpha$ ratio of the galaxy is
dominated by the metal-rich inner discs, where the [NII] emission
is much stronger, this would cause the H$\alpha$ fluxes of the outer
disc regions to be systematically underestimated.  Rather than
introduce such systematic errors we have chosen to present the
observed H$\alpha$ + [NII] fluxes instead.  
 
\section{Data Processing \& Analysis}
 
Throughout this study we have used Source Extractor \citep{bertin96}
to estimate non-uniform background variations and to define and select
sources inside and outside the disc.

\subsection{Processing of GALEX Images}
\label{sec:galtreat}

SExtractor was first used to define a master catalogue of objects based on
detections from the NUV image, against which detected sources were subsequently
matched from the FUV and \ha\ images.  Both 
GALEX bands are broad, and thus we expect significant contamination
from background galaxies and foreground stars.  Fortunately for this 
study we are primarily interested in young stellar knots with relatively
blue UV colors ((FUV - NUV $<$ 1), and the low dust contents of the outer
discs mean that heavily reddened clusters should be rare.  Consequently 
we excluded as a likely foreground or background contaminant all objects
in the outer discs with FUV - NUV $>$ 1.5 mag.  This removed more than
98\% of the foreground stars.  
This process did not completely remove background sources, however,
as can be seen in Figure \ref{fig:628treat} which shows the south west
quadrant of NGC 628 in the NUV both before treatment (top panel) and
after (bottom).  The spatial distribution of such objects is 
expected to be homogeneous across each field, and we applied a statistical
correction for these in our radial profiles by fitting a constant residual
background at very large radii, beyond the UV and HI extents of the discs.
We discuss this further in Section \ref{sec:back}.

\begin{figure}
   \includegraphics[width= 84mm]{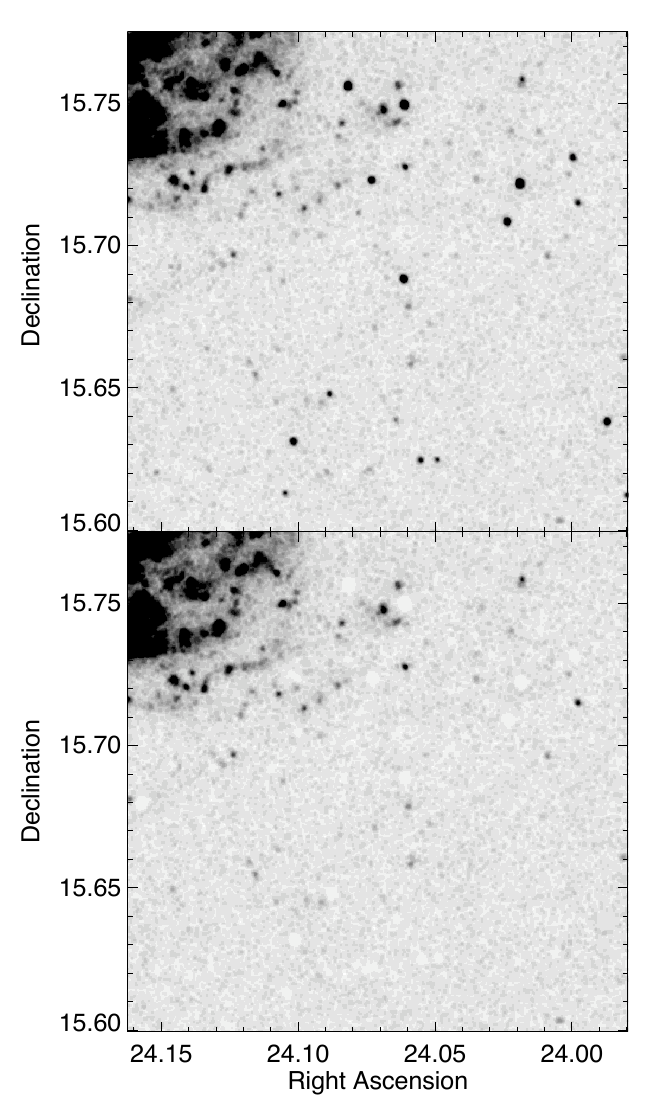}

   \caption{South west quadrant of NGC 628 in the GALEX NUV band, the
     top panel shows the image before we masked unwanted objects with
     UV colours greater than 1.5 and the bottom panel shows the same
     image after masking.}
\label{fig:628treat}
\end{figure}

\subsection{Processing of \ha\ images}

Contamination from background objects was less problematic
for the narrowband \ha\ observations, but foreground star contamination
is significant.  Most starlight is removed in the continuum subtraction
process, but residual flux is often present because of variations in
stellar colors and mismatches in the point spread function between the
narrowband and R-band exposures.  Stars were distinguished from HII
regions based on their H$\alpha$/R flux ratios, and were masked out 
with the local background value as determined by SExtractor. This
technique results in the removal of $\approx95\%$ of foreground stars,
any obvious remaining stars were removed by hand.

\subsection{Source Identification \& Photometry}

Several factors complicate the association of individual UV-emitting
regions with HII regions in the outer discs.  First, the UV images have lower
resolution ($4\arcsec.5 - 5\arcsec$) compared to the \ha\ images
($\approx 2 - 3\arcsec$) and have a larger pixel size ($1\arcsec.5$)
compared to the \ha\ images ($0\arcsec.31-0\arcsec.43$). Second, UV
light traces a larger proportion of the stellar mass function. In low mass star
clusters where we expect few ionising stars, the UV emission is
expected to be larger and more diffuse than the \ha\ emission; in
extreme cases one can find individual UV regions which 
are associated with several distinct HII objects (also see
\citet{gildepaz05}). Third, several earlier studies have shown that 
many UV knots in the extended disc do not have associated HII regions
\citep{gildepaz05,thilker05,zaritsky07}. With these
consideration in mind it is unwise to use \ha\ images for primary
object identification, we have used the catalogue of UV objects
produced by SExtractor.  This catalogue was created using the FUV 
and NUV images in the dual image mode of SExtractor.  
We measured isophotal fluxes, which uses an aperture calculated based on
pixels above the threshold detection limit.  Although this
method may not be the most accurate as it involves no aperture
correction, it does allow us to measure the total flux defined by the
same area for both UV and \ha\ images.

Using this catalogue of UV objects we search for corresponding objects
in the \ha\ images that occupy the same area. Fluxes were measured from
each area covered by all UV objects and statistically compared to
background levels in order to detect HII regions. Only objects with 
\ha\ fluxes measured to better than $3\sigma$ were recorded as 
detections. It should be noted that no attempt was made to determine
how many HII regions corresponded to each individual UV object, we
have only attempted to compare total fluxes of these star forming
regions.

We treated detections within the inner discs of the galaxies (R $<$ R$_{25}$)
in the same manner, i.e., by 
using the NUV image to define the spatial extent of the
star forming region, and measuring the \ha\ flux under the same
area.  In many of the galaxies in our sample the crowding of UV sources
and HII regions was such that several physical star clusters and/or HII
regions were blended at the resolution of the GALEX images.  Since
our primary interest is in the properties of the extended outer discs
this blending is unimportant for most applications.  However we 
will need to take this blending into account later when comparing
the fluxes of regions in the inner and outer discs.  


\subsection{Surface Photometry \& Radial Profiles}

Radial profiles in both far-ultraviolet and \ha\ have been computed for
all galaxies in our sample.  Most previous studies have been based
on areal surface photometry, i.e., by measuring the summed fluxes
of all pixels in radial annuli, and computing an average surface 
brightness (after background subtraction) as a function of radius.
However this method is susceptible to uncertainties when
applied in the low surface brightness outer discs, especially in \ha,
due to background noise, contaminants, and larger-scale background
variations, because only a tiny fraction of image
pixels contain HII regions \citep{martin01}.  To circumvent these problems
we also derived radial fluxes by summing the fluxes of
objects detected by SExtractor for each radial bin.  This latter
method has the disadvantage of being prone to to missing very faint 
and diffuse regions, so we measured UV and \ha\ radial profiles 
using both methods and evaluted the results.

Figure \ref{fig:radprofiles} shows FUV and \ha\ 
profiles for the galaxy NGC 628 calculated using both
methods.  Blue lines show the profiles measured from
areal surface photometry, while red profiles were derived from photometry
of individual sources.  Within a radius of 
R$_{25}$ (17kpc) we find there is no discernible difference in profiles
between the two methods in either \ha\ or FUV (our SeXtractor apertures
were defined in such a way as to include the diffuse component of the 
inner disc).  Similar results were seen for all galaxies in our sample.

Beyond the edge of the main star-forming disc differences between the two
methods become apparent.  In the FUV (top panel of Figure 2) the
profiles obtained using both methods show a small turnover at 1.4 R$_{25}$
with an extended UV disc extending out to $\sim$2 R$_{25}$.
However the surface brightness measured in the surface photometry
is approximately twice as high relative to the profile from resolved
UV knots.  Inspection of our image shows that much of the excess
emission can be attributed to diffuse FUV emission that is not
included in the source photometry measurements. 

 
The areal surface photometry is much more problematic for \ha, as
illustrated in the bottom panel of Figure 2.  Qualitatively the
profiles derived from the two methods are consistent, but the 
areal photometry misses a faint extension of the main star forming
disc at 1.4 R$_{25}$ and a faint outer HII region spiral arm near
1.9 R$_{25}$, as well as producing a spurious feature (albeit with
only 1-$\sigma$ significance) at larger radii.  All of these reflect
a limiting surface brightness (mainly caused by background variations
in the processes images) that precludes the accurate detection or
and measurement of star formation beyond the main, high surface brightness
disc.  For \ha\ the source photometry is clearly superior.
For the remainder of this paper we use areal FUV surface photometry  
to remove the possibility of missing any diffuse emission, whilst the \ha\ photometry
has been derived using HII region photometry, which is much more robust for 
those data.  We have measured individual FUV profiles from source
photometry in order to be able to make self-consistent comparisons
with the \ha\ profiles.

\subsection{Background Estimation}\label{sec:back}

Many of the extended disc objects without HII regions may well be background
sources not actually associated with the galaxy in question. It is important to measure the contribution of these background objects when calculating the radial profile of a galaxy or when counting the number of objects which might comprise the extended disc. The GALEX images have a large one degree field of view enabling us to sample an area well beyond even the most extended of discs. Although we mask old objects with UV colours greater than 1.5 there remained a small number of UV objects beyond the extent of the extended disc, these being suspected background contaminates. When calculating the number of UV objects beyond R$_{25}$ we also count the number of objects in an annulus well beyond the edge of the suspected extended disc. Both the cumulative flux of the objects and the total number of of these objects is scaled to the area of the XUV disc to calculate the true flux of the extended disc and give a better estimate of the number of UV objects which comprise the extended disc.

In the \ha\ images foreground stars appear as over-subtracted objects and are thus easy to identify and remove. Background galaxies have \ha\ emission red-shifted out of the narrow-band filter range and rarely appear in such images. As such the number of \emph{false} HII regions is tiny and so we do not correct the number of HII regions found in the XUV disc for unwanted objects.

In both the FUV and \ha\ images we subtract the background flux when making radial profiles. To accomplish this we estimate the background flux within an annulus well beyond the extent of the XUV disc. This background estimation is subtracted from all radii less than the background annulus.

 
\begin{figure}
   \includegraphics[width= 84mm,angle = 0]{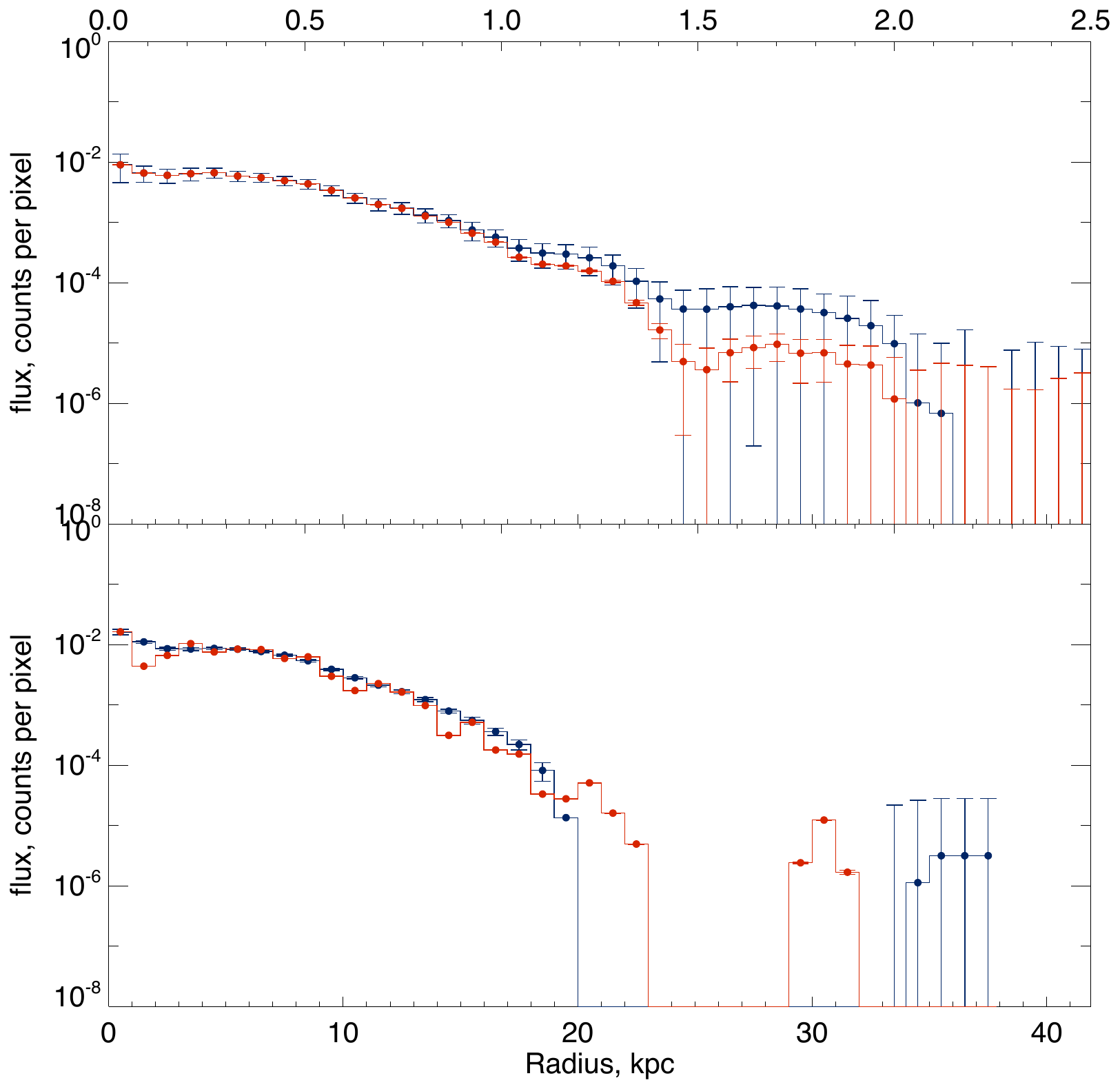}
   \caption{Surface photometry plots for the galaxy NGC628, the top
     panel shows the FUV emission in units of counts per second per
     pixel whilst the bottom panel shows the \ha\ emission. Blue lines
     show profiles derived using full annular area surface photometry, and 
     red lines show profiles measured from the addition of object fluxes for
     distinct radial bins. The top axis shows the radius
     in units of R$_{25}$.}
\label{fig:radprofiles}
\end{figure}

\subsection{Dust Extinction \& Reddening}

Interstellar dust attenuates the observed fluxes of star-forming regions.
This attenuation is lower in the extended discs (e.g., Boissier et al. 2007), 
but can be large in the inner star-forming discs, especially in the
ultraviolet.  Since our main interests in this paper are the outer
star-forming discs, where dust attenuation is low, we do not attempt
to solve the rather complex and difficult problem of dust obscuration,
but we do apply a first-order correction so the properties of the
inner and outer discs can be compared on a consistent basis.

Following many previous authors we have estimated the extinction
in our objects using the calibrated relation between UV color and 
extinction, with the latter measured from the ratio of total-infrared
to UV flux-- the IRX-$\beta$ relation (e.g. \citet{heckman95,meurer95,
meurer99,calzetti00,kong04,panuzzo07,cortese06,cortese08}).
Various authors have calibrated the relation in terms:

\begin{eqnarray}
	A_{FUV} = C (FUV-NUV) + Z
\end{eqnarray}

\noindent
where the slope of the extinction curve $C$ ranges from 
4.37 to 5.12 (\citet{calzetti01,seibert05,cortese06,hao09}.
The differences arise from differences in the bandpasses of
the original datasets and in the galaxy populations.  We
have chosen to adopt a simple average of these slopes,
giving a value $C = 4.82$; none of the conclusions of
this analysis are sensitive to this choice within the
range of values given above.  The zeropoint colour $Z$
depends on the average age of the regions of interest.
As discussed below we estimate our extinction corrections
using background-subtracted measurements of resolved UV
knots which dominate the integrated UV emission..  
For the range of ages of these regions (0--100 Myr)
evolutionary synthesis models predict corresponding colours
in the range $-0.2 \le FUV-NUV \le 0.2$ (see Section 5
and Figure 9).  For simplicity we adopt a mean colour for
unreddened young clusters of $FUV-NUV = 0.0$


It is unrealistic to apply individual attenuation corrections to
each individual UV knot or HII region, because the photometry is
not accurate enough, and the colors of the regions are influenced
by variations in age and dust geometry.  Instead we chose to derive
a mean correction as a function of galactocentric radius, as is 
illustred in Figure 3.  For most of the galaxies in our sample
the UV regions show a smooth gradient in colour with radius within
$R_{25}$, and we fitted this to derive a mean attenuation profile.
Colours for objects in the extended disc generally did not show
such a radial trend; objects associated with HII regions show 
colors consistent with those of unreddened young clusters, and
this implies that the other UV sources were either older (and hence
unreddened) clusters or background objects.  Consequently we
applied no dust corrections beyond the radius where the linear
fit above crossed the zeropoint colour $FUV-NUV = 0$.
The values for the extended disc were close to zero and for the majority of galaxy this
resulted in no dust correction being necessary for these
objects. Figure \ref{fig:dust} shows the UV colour as a function of
radius for the galaxy NGC 3198, there is a strong linear relation in
the inner disc. Although the scatter of objects in the extended disc
is large the average colour is close to zero.  In a few galaxies
no discernable colour gradient was apparent in the inner disc,
and in those cases we simply derived the mean colour of the young
clusters and applied a single average attenuation correction,
using equation (1) above.


To estimate the attenuation in the \ha\ waveband we applied the relation
given by \citet{calzetti01} between the extinction in the UV continuum
and the Balmer line emission as  shown in equation \ref{eqn:hyd}.

\begin{equation}	
	A_{FUV} = 1.78 A_{H\alpha}
	\label{eqn:hyd}
\end{equation}

\noindent
The resulting disc-averaged extinction corrections for our
sample range over 0--2.3 mag in the FUV and 0--1.3 mag in \ha.

\begin{figure}
   \includegraphics[width= 84mm]{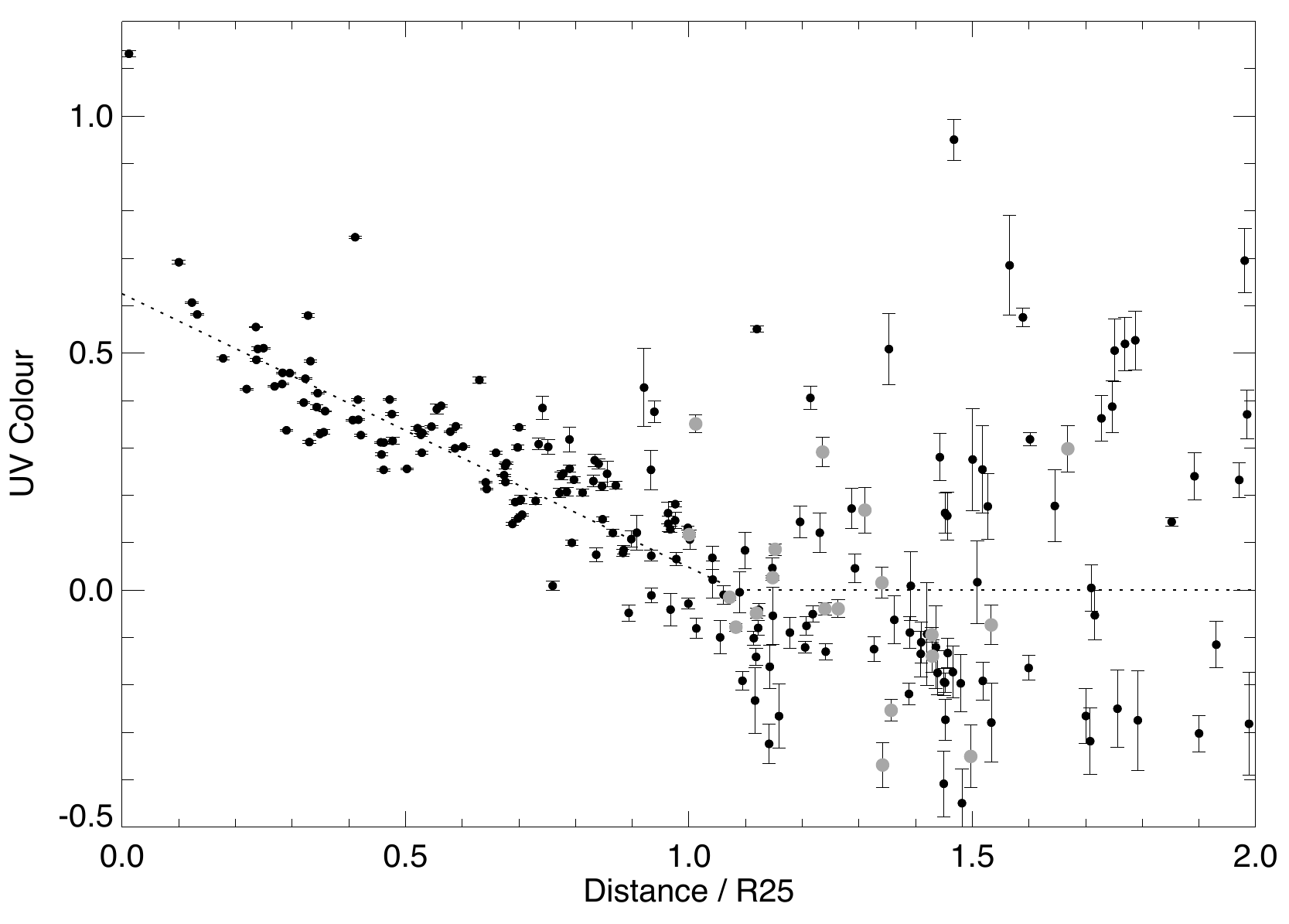}
   \caption{Plot showing the UV colour as a function of distance from
     the galactic centre for NGC 3198. Grey points indicate those
     objects in the extended disc found with HII regions, the dashed
     line shows a least squares fit to points in the inner disc.}
\label{fig:dust}
\end{figure}

We have used the UV colour to estimate the dust extinction but it would also have been possible to make this calculation using the IR to UV ratio. We have not attempted this in this study, because the SINGS MIPS images are not sensitive enough to detect most of the regions in the outer discs. However a recent study by \citet{munoz09} produced attenuation profiles for SINGS galaxies based on the IR/UV ratio. We used these attenuation profiles as a validation check for our method, re-calcualting the FUV and \ha\ flux of the extended disc as a fraction of the inner disc flux based on these new attenuation profiles.

Using the \citet{munoz09} attenuation curves we were able to make comparisons for 14 galaxies in our sample. Nine of these gave consistent answers between the two methods. For 4 galaxies (NGC 1566, 2841, 3198, 3351) we obtained higher fractions for both the FUV and Ha fluxes of the XUV disc using the IR/UV derived attenuation curves. This is primarily due to the high attenuation close to and beyond R$_{25}$ in the \citet{munoz09} sample. The UV colours of objects beyond R$_{25}$  in our sample tend to be close to zero indicating little dust extinction, an assumption confirmed by observations of the Balmer decrement of HII regions in the outer disc of M83 \citep{bresolin09}. We re-calculated the fractions for these 4 galaxies using the \citet{munoz09} attenuation curves but assumed an attenuation of zero beyond R25, this resulted in very similar numbers to those obtained using the UV colours.

There was one galaxy displayed different results, NGC 628. Using the UV colours we calculated a FUV flux fraction of 1.07\% compared to the inner-disc and 1.68\% for the \ha\ fraction. Using the IR/UV attenuation curves these fractions became 4.79\% in the FUV and 1.65\% in the \ha. The agreement was improved by assuming zero attenuation beyond R$_{25}$,  2.71\% in the FUV and Ha: 1.20\% in the \ha. We have not published these fractions in Table \ref{tab:galres} as most fractions are very similar, we report the fractions for NGC 628 here as an example of how different measures of the dust extinction may influence our results in the most extreme cases.

\section{Results}

\subsection{Classification of Extended Discs}

Before we analyze the properties of the discs and their star-forming
regions in detail it is useful to examine qualitatively the nature
and range in properties of the extended discs in our sample.
We began by classifying each of the UV discs as either \emph{normal} 
(with a prominent radial truncation in star formation and few
if any star-forming regions at larger radii) or \emph{extended}
(star formation extending well beyond the main inner disc, with
or without any turnover in emission near R$_{25}$).  The first galaxies
identified with extended UV regions (M83 \citet{thilker05} and NGC 4625
\citet{gildepaz07b}) are the prototypes of this class.  It was
quite straightforward to classify the discs as either normal or
extended based on visual inspection, and this subdivision is confirmed
by comparisons of quantitative measurements.  For example Figure
\ref{fig:super} plots the radial extents of the discs (the radius
of the most distant UV knot, in units of R$_{25}$) against the 
number of UV regions located at R $>$ R$_{25}$).  
Galaxies with more than 30 such regions show star formation extending 
out to 1.5 R$_{25}$ or further. Galaxies displaying few associated UV objects
have a very limited distribution and are often found close to the parent galaxy.  We have adopted these limits, indicated by the shaded
regions in Figure \ref{fig:super}, as distinguishing \emph{extended} and \emph{normal} UV discs.  Of course this
distinction is somewhat arbitrary, and a few galaxies are found
between these regions (we classified those based on visual inspection).  
However a similar bifurcation of the sample is seen when we
compare other properties of the discs, for example the radial
extents correlated with the fraction of total star formation
in the outer discs.  As can be seen in Figure \ref{fig:super}
our sample is roughly equally divided between galaxies with
\emph{normal} and \emph{extended} UV discs.
One galaxy in the sample, NGC 1291 has rather unique properties,
and is discussed separately in Section \ref{sec:1291}.

\begin{figure}
   \includegraphics[width= 84mm,angle = 0]{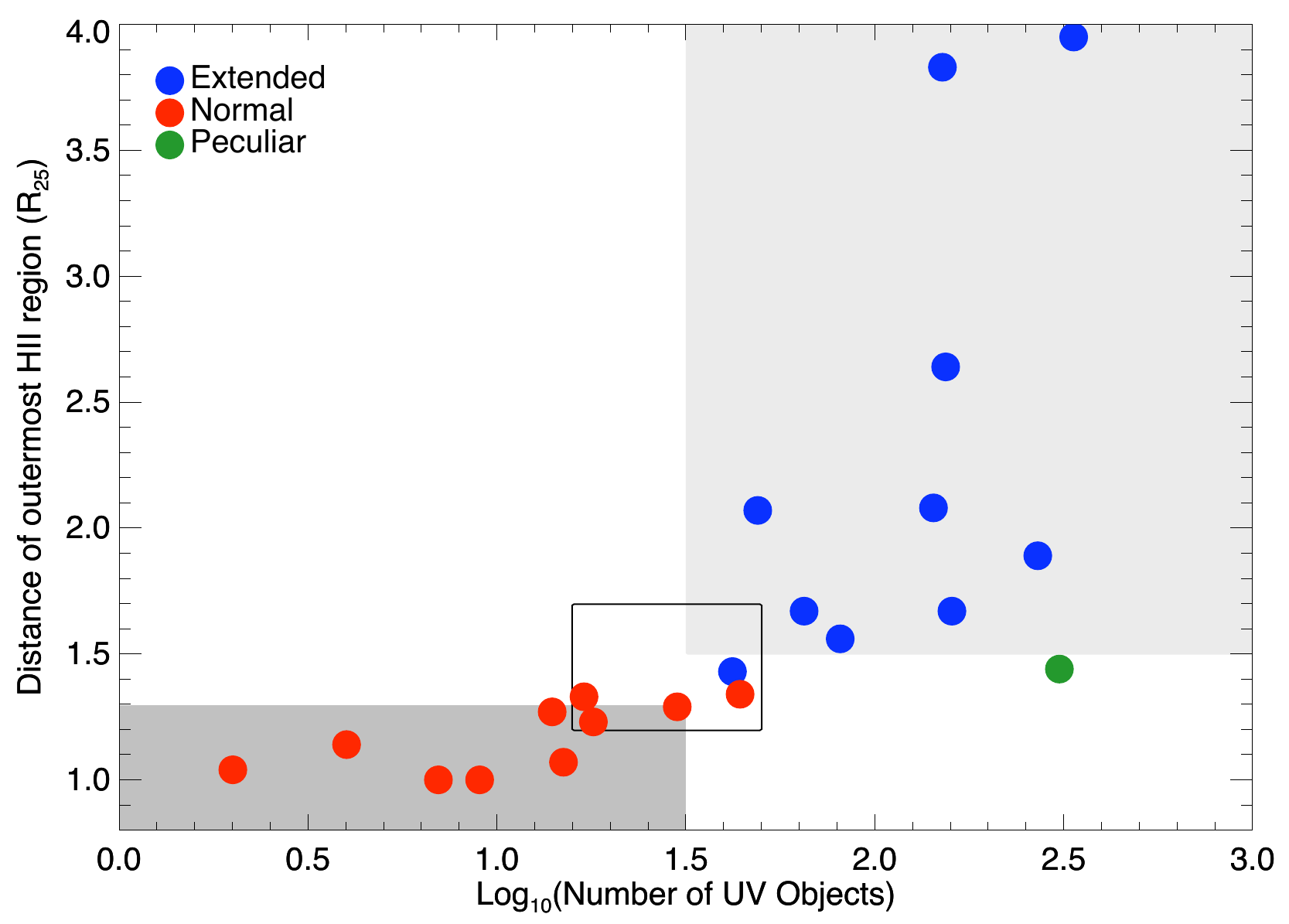}
   \caption{Plot showing the number of UV objects identified beyond
     R$_{25}$, after correcting for the expected number of background
     objects, against the radius of the outermost HII region given in
     terms of R$_{25}$. Galaxies classified as extended are shown as 
     blue dots and generally lie within the light grey area. Galaxies
     within the dark grey area were classed as \emph{normal} are shown
     in red. The black box shows the region in which visual inspection
     was required to classify galaxies.}
\label{fig:super}
\end{figure}

\subsection{Radial Profiles}

We now wish to examine the radial profiles of
our larger sample in the context of two key questions:  (1) How
extended is the star formation traced in the UV?; and (2)
How closely do the UV and \ha\ emission profiles track each other?
A large range of radial profiles is seen
in the UV discs and the \ha\ discs, and sometimes these behaviours
are coupled and sometimes they are not.

We have summarised the general trend of all the radial profiles
studied in coloumn (12) of table \ref{tab:galres}.  We have characterised
separately the behaviours of the FUV profiles and \ha\ profiles, separated by a
dash.  We subdivide the profiles into two general types, those with a
smoothly declining profile (Sm), and those with a truncation (Tr). We classify 
12 galaxies as having a smooth (Sm) FUV profile, three as truncated at or near to 
R$_{25}$ (Tr) and four as truncated at some point beyond R$_{25}$ (tr$^*$). Table \ref{tab:galres} 
also
denotes the degree of correspondence between the FUV and \ha\ profiles. We note that 9 galaxies have \ha\
profiles which match and trace the FUV profile, and these are denoted with
the prefix (Ma).  For two galaxies (NGC 1291, NGC 1512) the profiles are too 
erratic to be defined as smooth or having a single truncation, and these are left as
undefined (Un) and will be discussed separately.

Not surprisingly these disc profile types are closely related
in some (but not all) cases to our earlier classification of the discs
as \emph{normal} or \emph{extended}.  
All of the \emph{normal} galaxies display a clear turnover in the \ha\
radial profiles near the R$_{25}$ radius.  In these galaxies the FUV
radial profile also closesly traces the \ha\ profile, with a significant
drop in emission at the radius of the \ha\ truncation.
However the UV profiles tend to be less sharply truncated than their 
\ha\ counterparts. Figure \ref{fig:925} shows the profile of NGC 925, a galaxy we have classed as having a truncated profile in both the FUV and \ha\ (Tr-MaTr).  The top and middle panels show FUV and \ha\ images with
the radial scale indicated, while the bottom panel shows the respective
radial profiles.  A pronounced turnover in the \ha\ profile (red curve) is 
apparent near R$_{25}$ (note the logarithmic surface brightness
scale).  The FUV profile (blue points) shows a similar although less pronounced
feature at the same radius.  The same correspondence can be seen by
comparing the images in the upper two panels.  

The presence of strong radial turnovers in the the \emph{normal}
discs does not mean that star-forming regions are entirely absent.
As seen in Figure \ref{fig:super} and Table 2 regions are often
found extending well beyond R$_{25}$.  These can occur as isolated
regions or as extended regions of star formation.  For example
NGC 3521 display a faint spur of star formation to the north and NGC
3351 shows a few objects to the south-east.  However the total
amount of UV emission located beyond  R$_{25}$ in these galaxies
is always $<$2\%, this is discussed further in section \ref{sec:gr}.  

\begin{figure}
   \includegraphics[width= 84mm]{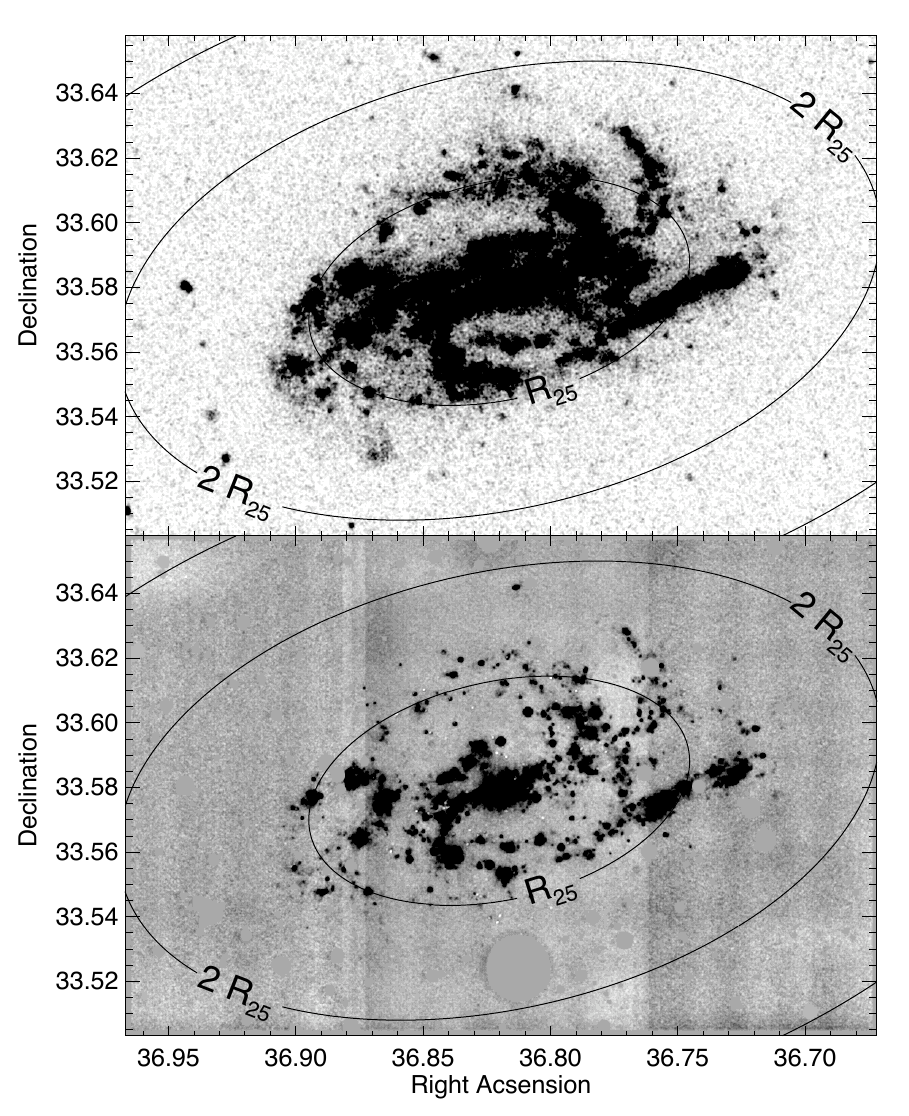}
      \includegraphics[width= 84mm]{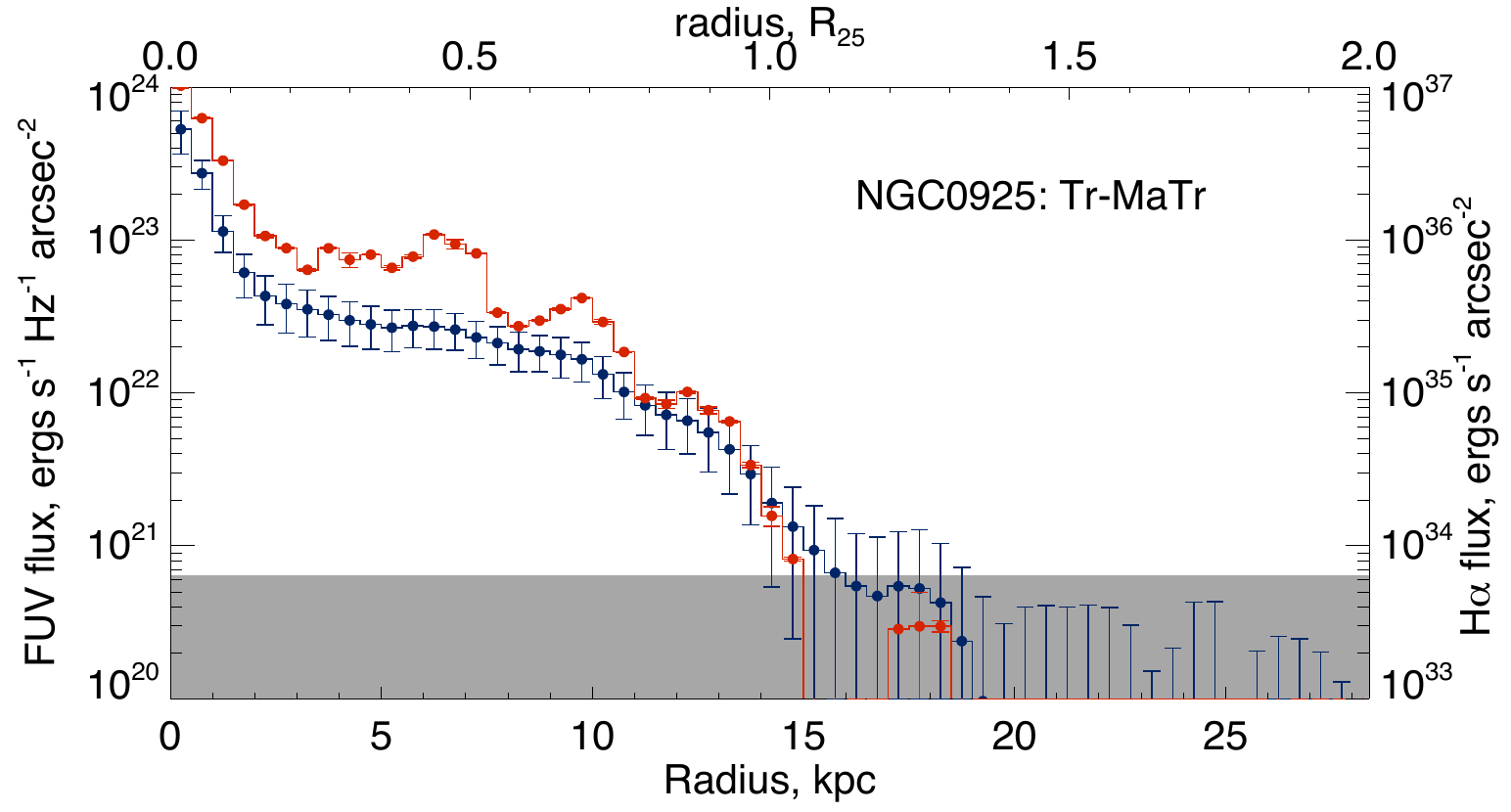}
      \caption{Top two panels show FUV and \ha\ images for NGC 925, the
        contours show multiples of the R$_{25}$ radius. The bottom
        panel shows the surface photometry for the galaxy, FUV in blue
        and \ha\ in red. The grey section of this plot indicates the
        level at which both FUV and \ha\ fluxes become untrustworthy. We have classed this galaxy as having a truncated profile in both the FUV and \ha (Tr-MaTr).}
\label{fig:925}
\end{figure}

We identified ten galaxies from our sample as having an extended
ultraviolet disc, including eight which were also identified by
\citet{thilker07a} as having a Type 1 extended disc (the others
are NGC 1097 and NGC 1566).  The radial profiles of these discs exhibit a much broader
range of behaviour than in the \emph{normal} discs. Six out of the ten \emph{extended} galaxies show emission in both the UV and \ha\ profiles well beyond R$_{25}$.  One of
of these galaxies, NGC 3621, is shown in figure \ref{fig:3621}.
The FUV image in the top left panel shows emission extending to
$\sim$1.7 R$_{25}$, and this behavior is traced closely in \ha,
as shown in the top right and bottom panels.  Both the UV and \ha\ profiles 
decline smoothly with no indication of a star forming
threshold close to R$_{25}$.  Overall the \ha\ profile is somewhat steeper than in 
the FUV, but most if not all of this can be attributed to higher
UV extinction in the inner disc. There are hints as well for a
systematically higher UV emission in the outermost region of the
disc, which is consistent with what is observed in many of the
\emph{normal} discs. In the case of NGC 3621 we do see a truncation 
in both the UV and \ha\ profiles but at a radius of $\sim$1.7 R$_{25}$, 
clearly in this case the optical \emph{'edge'} does not correspond with 
the extent of recent star formation. Other examples of galaxies in this sample
with extended but matched FUV and \ha\ profiles include 
NGC 628, NGC 1512, NGC 1566, NGC 3198, and NGC 5474.  The distinction between these galaxies
and normal discs is their radial extent of star formation and a tendency in some
cases for the ratio of UV/\ha\ emission to be somewhat higher in
the outer discs. In NGC 1512,
for example, the enormous radial extent to the star formation is
the result of a strong tidal interaction that has produced long
tidal arms.  The UV and \ha\ properties of these arms are indistinguishable
from those of the inner disc; the low surface brightnesses in azimuthally
averaged radial profiles are purely a product of the geometry of the
spiral structure, not of the physics of star formation.

\begin{figure}
   \includegraphics[width= 84mm]{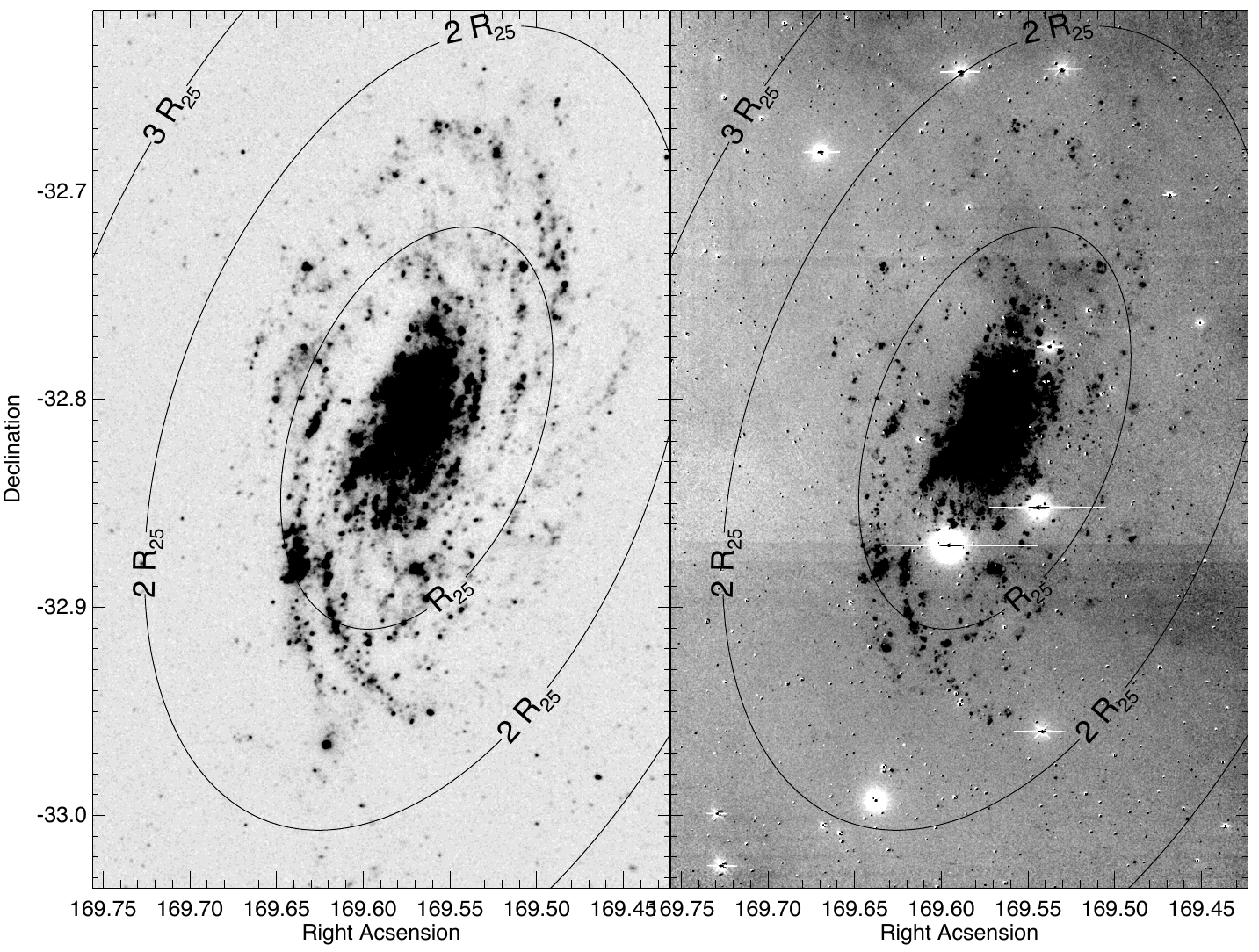}
      \includegraphics[width= 84mm]{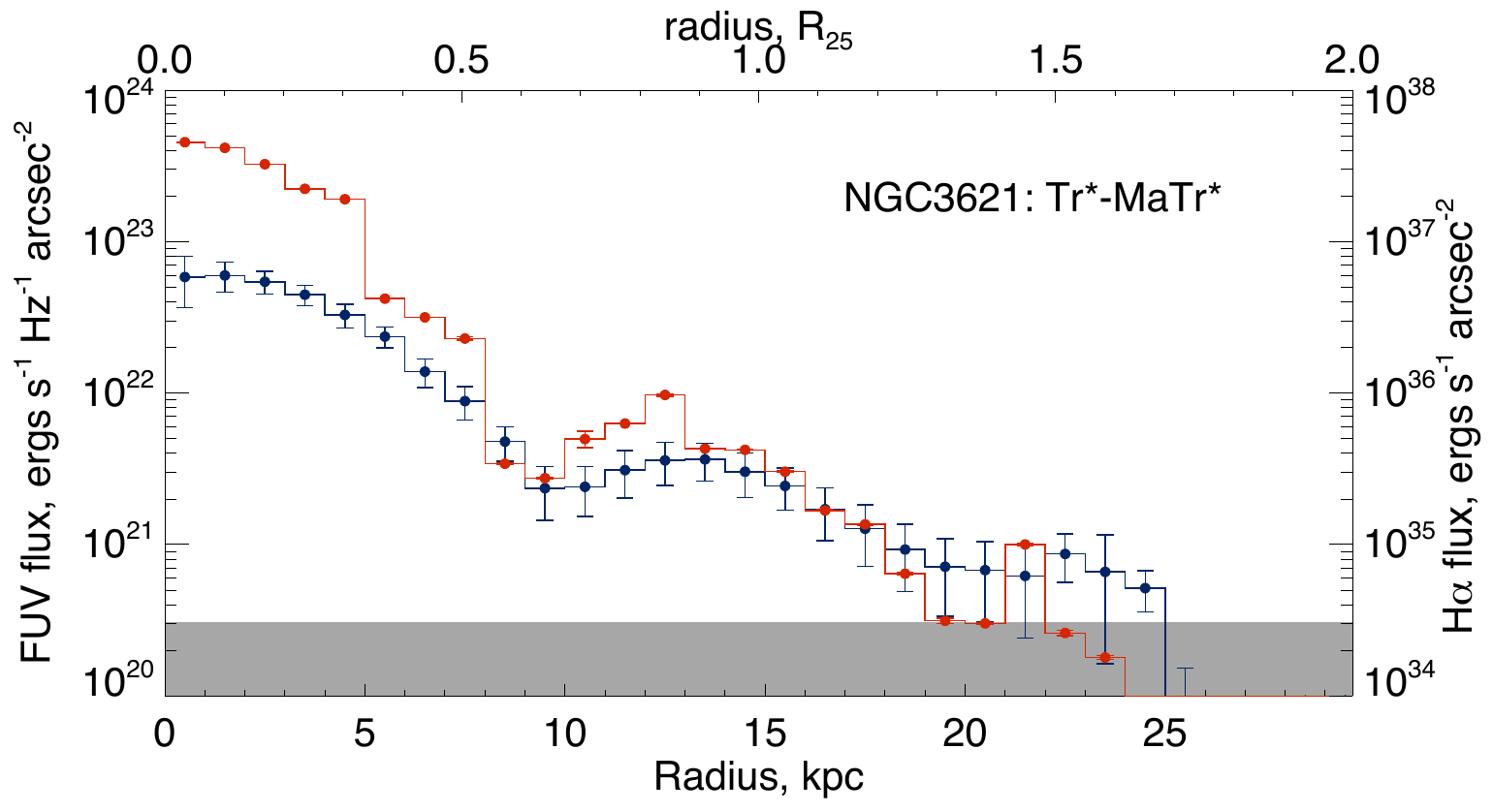}
      \caption{The top panel shows the FUV image on the left and \ha\
        image on the right, for the galaxy NGC 3621.  We have included
        contours indicating the optical edge R$_{25}$. Bottom panel
        shows the radial profile for the galaxy with FUV emission in
        blue and \ha\ in red, the distance is shown in kpc on the
        bottom axis and in terms of R$_{25}$ on the top axis. NGC 3621 was classed as having a truncation in both the \ha and FUV (Tr*-MaTr*), although this feature occurs beyond R$_{25}$.}
\label{fig:3621}
\end{figure}

Four of the \emph{extended} galaxies in our sample exhibit a smooth FUV profile
extending well beyond the truncated edge of the \ha\ disc, as reported
by \citet{thilker05} for M83; we have classed these galaxies as (Sm-Tr) in table \ref{tab:galres}.  Figure \ref{fig:2841} shows images and
radial profiles for the Sb galaxy NGC 2841.  The SFR throughout
this galaxy is quite low, and by the gravitational stability 
criteria of \citet{kennicutt89} and \citet{martin01} the entire
disc should be subcritical (below threshold).  Nevertheless the
\ha\ radial profile shows a turnover near R$_{25}$, with only a
few scattered HII regions present beyond that radius.  In contrast
the FUV profile shows no sign of truncation out to twice the
optical radius, though a falloff in surface brightness near the
edge of the \ha\ disc is visible.  Inspection of the FUV image
shows that the extended disk emission consists $>$40 discrete
knots, along with a very substantial contribution from diffuse
emission. XUV-discs have been shown to display a combination of discrete knots and diffuse emission. \citet{thilker07a} goes as far as specifying two types of XUV-disc; Type 1 disc with bright UV regions, and Type 2 with more diffuse areas of UV emission. Other galaxies in our sample showing this behaviour include
NGC 1097 and the two well-known prototypes, M83 and NGC 4625.

\begin{figure}
	   \includegraphics[width= 84mm]{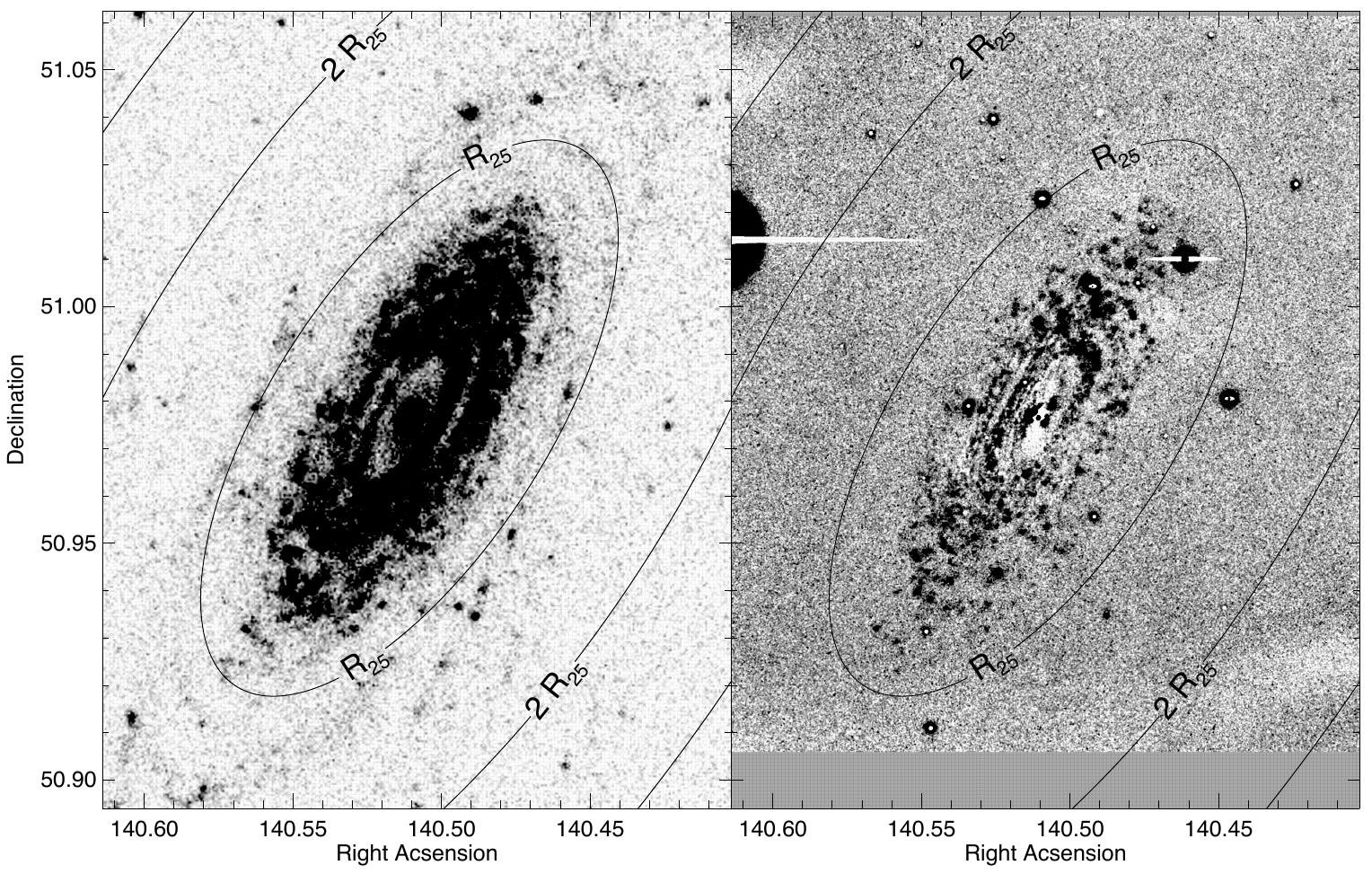}
      \includegraphics[width= 84mm]{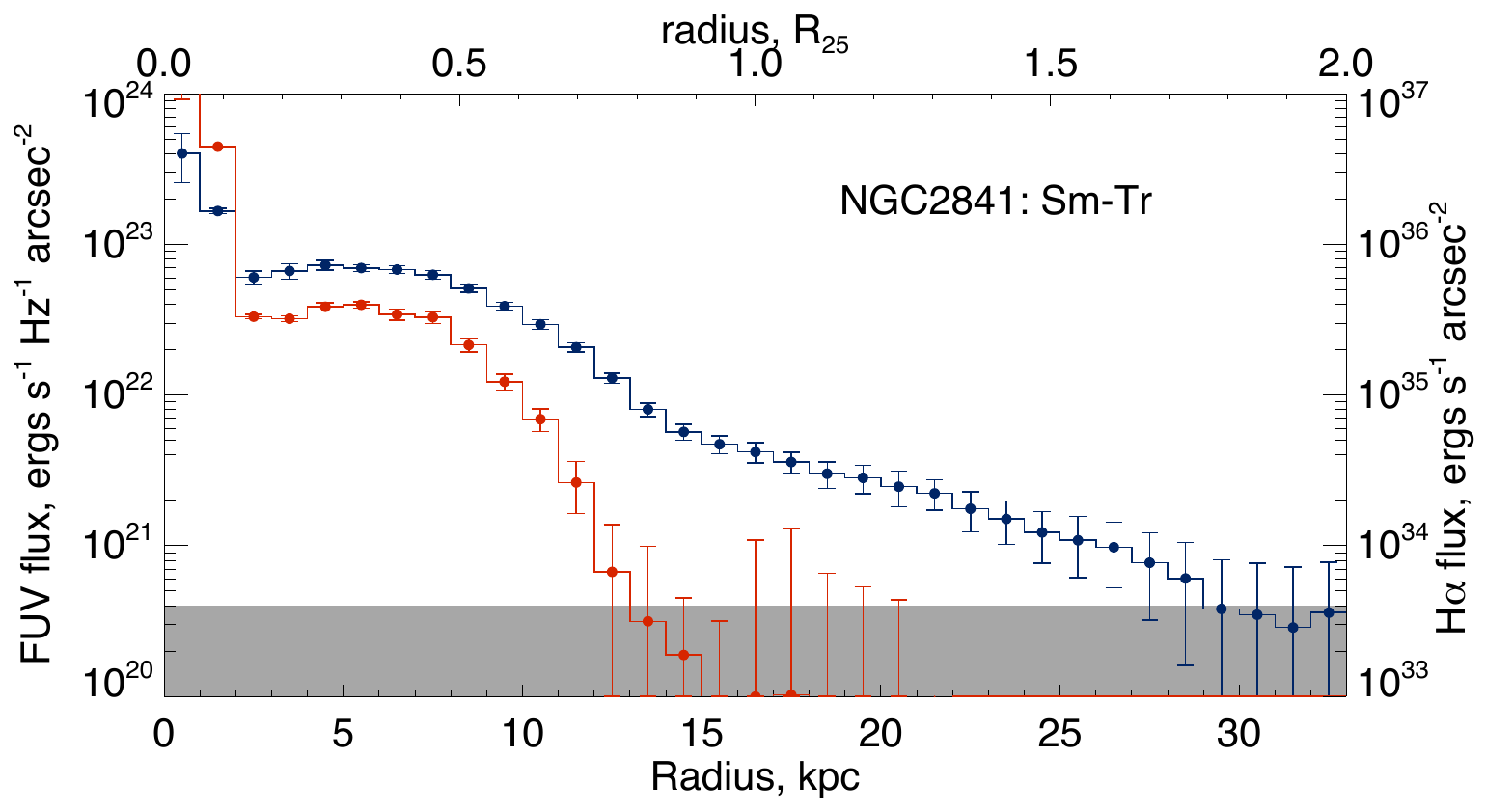}
      \caption{Top panel shows FUV (left) and \ha\ (right) images for
        NGC 2841 (profile classed Sm-Tr)}, contours indicate multiples of the optical radius
        R$_{25}$. Bottom panel shows the surface photometry for NGC
        2841, \ha\ emission is shown in red and FUV shown in blue.
\label{fig:2841}
\end{figure}

\subsection{The Effects of Azimuthal Averaging}
 
The interpretation of the radial profiles is often influenced
significantly by the azimuthal averaging.  As pointed out by
\citet{martin01}, the radial behaviour of the star formation
often varies strongly as a function of azimuth, with regions
showing strong radial truncations and others showing an extended
transition (or no transition at all).  Geometric distortions
introduced by spiral arms or lopsidedness can also cause 
truncations to be blurred and smoothed out in azimuthally
averaged data. For each galaxy in our sample surface photometry was
always accompanied by visual inspection of the images to ensure all
features presented by the photometry were visible in the images.

As an example of how galaxy asymmetries can affect the truncation radius
Figure \ref{fig:m83} shows the FUV and \ha\ images of the prototype
XUV galaxy M83 in the top panel; below this are 3 the average counts per 
pixel as a function of radius for 3 segments as labelled in the images. We have also shown 
a profile for the average of all the azimuthal segments shown as the top plot in the bottom panel of Figure  \ref{fig:m83}.
Along some axes, for example segment A in Figure \ref{fig:m83} shows a segment in which 
a spiral arm extends beyond R$_{25}$, this is reflected in both FUV and \ha\ profiles which 
trace emission out to 1.5 R$_{25}$ in both cases. In fact the profile of segment A is remarkably similar to that of the complete profile. However the same is not true for segments B and C, both of these
segments show a steep truncation in the \ha\ profile close to R$_{25}$. The extent of the FUV also varies in both these segments, from 1R$_{25}$ in segment B to around 1.3 in segment C. Apart from an isolated HII region in segment B there is no emission beyond the truncation in either the \ha\ or FUV for either of these segments, a feature which is lost in the combined profile. The loss of such features due to an asymmetric galactic distribution is not new and was first reported by \citet{vanderkruit88}.


\begin{figure}
   \includegraphics[width= 84mm]{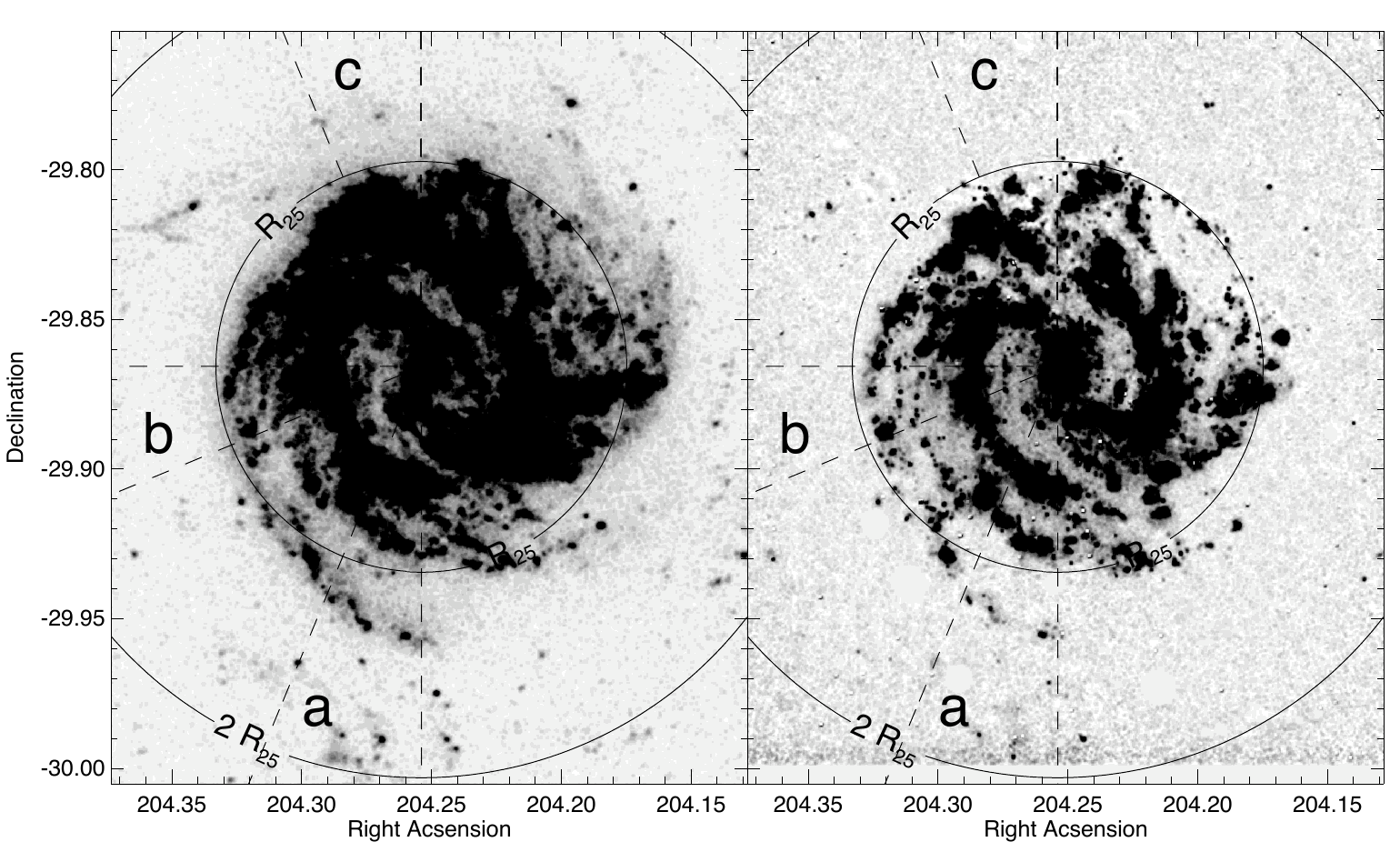}
      \includegraphics[width= 84mm]{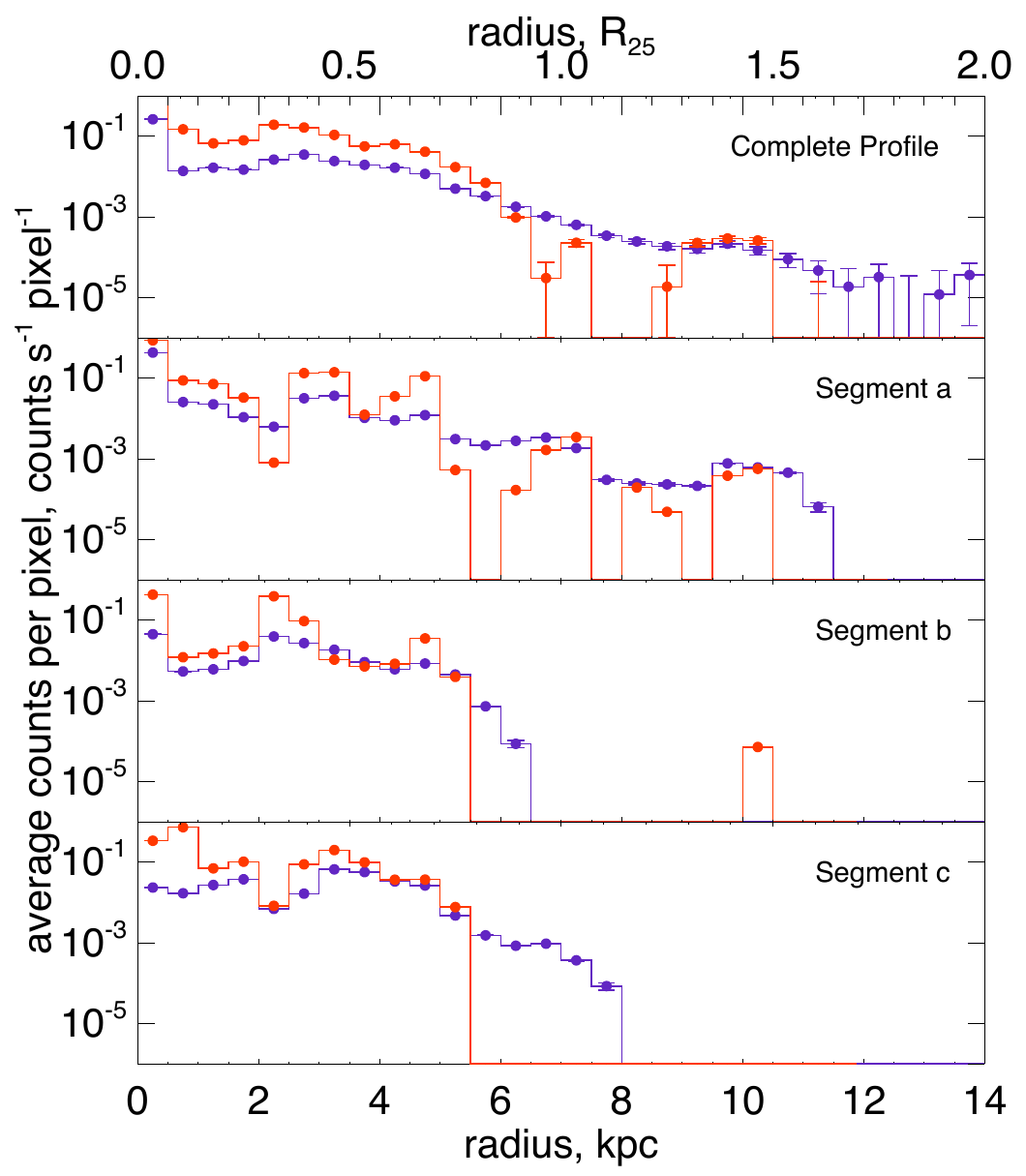}
\caption{Top panel shows FUV (left) and \ha\ Images (right) for the galaxy M83, contours show multiples of the R$_{25}$ isophote, dashed lines and letters correspond to segments used in the surface photometry in the panel below. Bottom panel shows surface photometry for the 3 segments labelled on the the images as well as the averaged profile for all segments; blue points represent the FUV emission whilst red shows the \ha.}
\label{fig:m83}
\end{figure}

 
When all of the profiles from individual segments were 
averaged to create the \emph{'Complete Profile'} shown in figure  \ref{fig:m83} the typical sharp truncation in the H$\alpha$ is 
largely preserved, while the FUV profile is smoothed into
such an extended transition radially that little turnover
is observed at the same radius.  This is produced by a 
combination of effects, the presence of UV arms and 
features outside of the main turnover in HII regions,
as well as azimuthal asymmetry in the galaxy, which
also tends to smooth the UV transitions together. 
 
\subsection{A Note on NGC 1291} \label{sec:1291}
\label{sec:1291}

We have classed NGC 1291 as \emph{Peculiar} in table \ref{tab:galres}
this is because of the unusual morphology of the galaxy. NGC 1291
contains a bright central disc and bulge with little or no star formation,
but a faint ring of star forming regions visible both in the UV and \ha.  The definition of the
optical radius R$_{25}$ for this galaxy is based predominantly on the bright central bulge. This results in the R$_{25}$ \emph{edge} bisecting the faint ring ring of star formation, so the
the distinction between ``inner" and ``outer" discs is superficial.
NGC 1291 is an example of a population of S0 and Sa galaxies (including
many barred galaxies) with extended low surface density HI discs \citep{vandriel88,vandriel89} and
low levels of star formation \citep{caldwell91,caldwell94}.
In this respect NGC 1291 can be regarded as a galaxy in which all
of the star formation is taking place in an extended UV disc.
Within this ring the radial profiles in the UV and \ha\ are very similar.
This galaxies serves as an example
of how the optical measure of the the extent of a galaxy (R$_{25}$)
can be unreliable when used to define edge of the star forming region.
Star forming regions
present in the ring with \ha\ emission appear to have a masses in the
range $10^{3.5-5.5}$ \msol\ and ages less than 10 Myr.

\subsection{Integrated Properties of XUV Discs} \label{sec:gr}

Table \ref{tab:galres} summarizes the most relevant properties
of the extended discs of our galaxies.  These include the disc
classifications as described in Sections 4.1 and 4.2, the number
of detected UV knots and HII regions along with the corresponding
detection limits, the radii of the most distant detected HII regions,
and the fractions of total UV and \ha\ luminosity contributed by
the extended discs ($R > R_{25}$), expressed as a percentage.
Two sets of the latter numbers are listed, those as observed
and after correcting for dust attenuation.

HII regions outside of R$_{25}$ are found in 19 of the 21 galaxies 
in our sample, and UV knots beyond R$_{25}$ were found in all of them,
although it is possible that a small fraction of the latter are
background contaminants (Section 3.1).  There is a pronounced difference in
the average number of objects detected in both UV and \ha\ in the extended disc between galaxies
classed as \emph{normal} and those classed as \emph{extended}, $16\pm14$
objects compared to $145\pm96$ objects, respectively.  These differences
are also reflected in the fractions of star formation located in the
outer discs.  These fractions vary over more than two orders of magnitude,
ranging from less than 1\% (after dust corrections) for NGC 925, NGC 4321,
NGC 5194 (M51), and (most notably) M83 to 10--35\% in NGC 1512, NGC 3621,
and NGC 4625 (as discussed earlier NGC 1291 is a special case).
As expected there is a strong tendency for galaxies with \emph{extended}
discs to have larger fractions of star formation located outside of R$_{25}$,
with an average fraction of 6\% compared to 1\% in the \emph{normal}
discs.  However there are pronounced exceptions; some of the most extended
discs in our sample, such as M83 and NGC 2841, contribute less than 1\%
of the total star formation in the respective galaxies.
The observed UV light fractions, before correcting for extinction,
are much higher, averaging 23\% for the \emph{extended} discs and 2.3\%
for the \emph{normal} discs, but these are influenced by the strong
dust attenuation in the inner discs.
 
Inspection of Table 2 also shows that the fraction of \ha\ emission contributed
by the outer discs is significantly lower on average than the
corresponding FUV fractions, even after correcting for dust
attenuation at both wavelengths. The median \ha\ fraction
is 2.5 times lower than the corresponding FUV fraction, a result
that is consistent with the original studies of \citet{thilker05} 
and \citet{gildepaz05} for M83 and NGC 4625.  Part of this difference
arises from the different methods used to measure the outer disc
profiles in the UV and \ha; since the latter were based on photometry
of individual HII regions faint sources and diffuse emission were
not included.  We can estimate the magnitude of this bias by 
comparing the FUV profiles measured using full-area surface photometry
(given in Table 2) and photometry of the resolved UV knots.
On average the latter are lower by a factor of 1.08, accounting
for part of the differences in FUV and \ha\ fractions.
The remaining difference (approximately a factor of 2.3), probably
reflects an intrinsic change in the FUV/\ha\ flux ratio between
the inner and outer discs.  We discuss this further after we
examine the UV and \ha\ properties of the individual HII regions
and UV knots in the galaxies.


\subsection{Properties of Star Forming Regions}

These same data provide valuable information on 
the properties of the individual star forming regions which
make up the extended discs. The \ha\ fluxes of the HII regions found in the
extended discs range from $10^{35.9}$ to $10^{39.2}$ ergs s$^{-1}$.
For the most distant galaxies, we are unable to detect 
the faintest regions, but for galaxies with distance
$\leq 12$ Mpc we are able to detect regions with \ha\ fluxes 
corresponding to the ionizing flux a single B0 type star,
the lowest mass star expected to produce significant ionizing radiation
\citep{vacca96}. In regions where both UV and \ha\ emission is
detected, the corresponding range in FUV fluxes is found to be $10^{22.9}$ 
to $10^{26}$ ergs s$^{-1}$ Hz$^{-1}$.  These are 10 to 1000 times larger
than expected for single ionising O-type stars, which confirms
the presence of large numbers of lower-mass B-type stars in the
associated star clusters.

The fraction of UV clusters which exhibit \ha\ emission varies 
somewhat among the galaxies, as shown in column (4) of Table
\ref{tab:galres}.  However the mean fraction is similar for
galaxies with \emph{extended} discs ($44\pm19$\%) and
\emph{normal} discs ($38\pm31$\%), and does not appear to
correlate systematically with any property of the galaxies,
apart perhaps from the depth of the photometry (with a higher
fraction of HII regions at higher luminosities, as might be expected).

 \begin{landscape}
\begin{table}
 \caption{Summary of results for all galaxies in our sample.}
 \label{tab:galres}
 \begin{tabular}{@{}l c c c c c c c c c c c c}
  \hline
  \hline
  NGC & No. of UV & No. with \ha & \% of objects& \multicolumn{2}{c}{Limiting Disc Flux (\emph{log})} &\multicolumn{4}{c}{Extended Disc Flux (\%)} & Dist of furthest & Class & Profile  \\
 & Objects & Emission & with \ha &FUV & \ha &\multicolumn{2}{c}{Observed} & \multicolumn{2}{c}{Dust Corrected} & HII region & & UV - \ha \\
  & & &  &  & & FUV & \ha & FUV & \ha & (R$_{25}$) & & \\
   & (1) & (2) & (3) & (4) & (5) & (6) & (7) & (8) & (9) & (10) & (11) & (12) \\
  \hline
628 	 & 	160  	& 	60	& 37.5	& $28.4$	& $41.7$ & $7.86 \pm 0.11$	& $2.2\pm 0.2$	& $1.07\pm 0.15$	&$ 0.71\pm 0.08$ & 1.67	&	Extended	&	Tr$^*$ - Tr$^*$	\\
925 	 & 	18  	& 	6	& 33.3	& $27.9$	& $41.3$ & $0.94 \pm 0.10$	& $0.41\pm 0.07$ & $0.51 \pm 0.13$& $0.2\pm 0.05$& 	1.23	&	Normal	&	Tr - MaTr	\\
1097 & 	81	& 	21	& 25.9	&$28.4$	& $41.7$ & $6.90\pm 0.12$	& $1.8\pm 0.2$	& $1.57\pm 0.47$ & $0.69\pm 0.14$ & 1.56	&	Extended	&	Sm - Tr	 \\
1291 & 	308 	& 	129	& 41.9	&$28.1$	& $41.9$ & $59.42\pm2.95$   & $6.23\pm 1.02$    & - & - & 	-	&	Peculiar	&	Un - Un	 \\
1512 & 	154 	& 	74	& 48.1	&$27.6$	& $41.0$ & $72.5 \pm 4.1$ & $36.9\pm1.6$	& $34.4\pm13.5$ & $21.6\pm4.75$ & 2.64	&	Extended	&	Un - MaUn	 \\
1566 & 	143 	& 	82	& 57.3	&$28.6$	& $41.8$ & $8.58\pm 0.84$	& $3.2\pm 0.3$	& $2.66\pm 0.71$ & $1.6\pm 0.3$  & 2.08	&	Extended	&	Tr$^*$ - MaTr$^*$	 \\
2841 & 	42 	& 	11	& 26.2	&$28.2$	& $41.0$ & $4.10\pm 0.20$	& $0.4\pm0.1$	& $0.50\pm0.06$ & $0.2\pm0.1$ & 1.43	&	Extended	&	Sm - Tr	 \\
3198 & 	65 	& 	19	& 29.2	&$28.4$	& $41.6$ & $12.87\pm 1.82$ & $2.7\pm0.2$	& $2.76\pm 0.50$ & $1.0\pm0.1$ & 1.67	&	Extended	&	Sm - MaSm	 \\
3351 & 	44 	& 	23 	& 52.3	&$28.2$	& $41.7$ & $5.60\pm 0.42$	& $1.1\pm0.1$	& $1.09\pm0.26$ & $0.43\pm0.09$ & 1.34	&	Normal	&	Sm - Tr	\\
3521 & 	14 	& 	1	& 7.1		&$28.3$	& $42.2$ & $1.78\pm0.11$   & $0.02\pm0.01$   &  $1.78\pm0.81$ & $0.02\pm0.02$ & 1.27	&	Normal	&	Sm - Tr	 \\
3621 & 	270 	& 	175	& 64.8	&$28.3$	& $42.8$ & $46.71\pm4.01$ & $2.41\pm0.30$& $19.84\pm4.76$ & $1.5\pm0.37$ & 1.89	&	Extended	&	Tr$^*$ - MaTr$^*$	 \\
4321 & 	30 	&  	3 	& 10.0	&$28.9$	& $41.9$ & $0.62\pm0.04$	& $0.04\pm0.03$	& $0.09\pm0.02$ & $0.01\pm0.02$ & 1.29	&	Normal	&	Sm - Tr	 \\
4536 & 	7	& 	0	& -		&$28.2$	& - & $1.76\pm0.13$	& 	-	& $0.47\pm0.13$ & - & -	&	Normal	&	Tr - MaTr		 \\
4579 & 	9 	& 	0	& - 		&$28.4$	& - & $0.16 \pm 0.03 $& 	-	& $0.13\pm0.05$ & - & -	&	Normal	&	Sm - Tr		 \\
4625 & 	151 	& 	49	& 32.5	&$26.9$	& $40.7$ & $49.65\pm6.07$ & $13.2\pm1.6$ & $20.51\pm8.72$ & $8.03\pm2.08$ & 3.83	&	Extended	&	Sm - Tr	 \\
5194 & 	17 	& 	9	& 52.9	&$28.9$	& $42.6$ & $3.61\pm0.24$	& $3.09\pm0.37$ & $1.67\pm0.77$ & $1.54\pm0.24$ & 1.33	&	Normal	&	Sm - Tr	 \\
5236 & 	336 	& 	122	& 36.3	& $29.0$	& $42.0$ & $6.02\pm0.56$	& $1.6\pm0.2$	& $0.72\pm0.20$ & $0.5\pm0.2$ & 3.95	&	Extended	&	Sm - Tr	 \\
5398 & 	4 	& 	1	& 25.0	&$27.8$	& $41.9$ & $1.30\pm0.09$	& $0.1\pm0.1$ & $1.30\pm0.39$ & $0.1\pm0.2$ & 1.14	&	Normal	&	Sm - Tr	 \\
5474 & 	49 	& 	24	& 49.0	&$27.6$	& $40.7$ & $13.78\pm1.08$ & $5.5\pm0.4$ & $3.69\pm0.87$ & $2.6\pm0.4$ & 2.07	&	Extended	&	Tr$^*$ - MaTr$^*$		 \\
7552 & 	2 	& 	2	& 100.0	&$28.6$	& $42.3$ & $1.77\pm0.20$ 	& $0.23\pm0.09$ & $1.77\pm0.82$ & $0.23\pm0.11$ & 1.04	&	Normal	&	Tr - MaTr	 \\
7793 & 	15 	& 	9	& 60.0	&$27.0$	&$40.3$ & $3.62\pm0.36$	& $3.6\pm0.2$	& $1.44\pm0.38$ & $2.1\pm0.3$ & 1.07	&	Normal	&	Sm -MaSm	\\ 
 \hline
 \end{tabular}
 \medskip (1) Number of UV objects associated with the galaxy in
 beyond R$_{25}$. (2) Number of UV found to have associated HII
 regions beyond R$_{25}$. (3) The percentage of UV objects beyond
 R$_{25}$ which have associated \ha\ emission. (4) \& (5) show the
 fluxes for the faintest object detected in the extended disc for FUV
 and \ha\ wavebands respectively in units of log(ergs s$^{-1}$
 Hz$^{-1}$) and log(ergs s$^{-1}$). (6) \& (7) flux of the extended
 disc as a percentage of the inner disc flux in the FUV and \ha\
 respectively before any correction for dust, (8) \& (9) show the same
 but after we corrected for dust. (10) Distance of the further most
 HII region in terms of the optical radius (R$_{25}$). (11) The
 classification we have applied to the galaxy to distinguish between
 those galaxies with an extended UV disc (\emph{extended}) and those
 without (\emph{normal}). (12) Summarises the surface photometry profiles in FUV and \ha respectively. Sm denotes a smoothly declining profile; Tr indicates a truncation at or near to R$_{25}$; Tr*, a truncation well beyond R$_{25}$; Ma, indicates the \ha\ profile closely resembles the FUV profile; Un, denotes an undefined profile.
 \end{table}
\end{landscape}

Following \citet{gildepaz05} we compare in Figure \ref{fig:objs}  
the \ha\ and FUV fluxes of the star forming regions for two galaxies
with especially extended discs, NGC 3621 (top left panel) and M83
(bottom left panel).  Red points represent HII regions identified
in the extended discs, while blue points denote comparison HII
regions in the inner star-forming discs (R $<$ R$_{25}$).
The fluxes have been corrected for dust attenuation following the
method described in Section 3.5.  
We also have generated a grid of synthesis models for young star
clusters using the Starburst99 code,  these models assume a 
Salpeter IMF with a stellar mass range of $0.1-100$\msol\ and solar 
metalicity, and for ages of 0 $-$ 15 Myr.  These models are superimposed
on the data in Figure \ref{fig:objs}, with lines of constant
age and constant cluster mass indicated.  Finally we show the
expected \ha\ and FUV luminosities of single main sequence stars
of spectral types B0 to O3 ($\sim 20-90$  \msol).  The offset
of the points from the single-star models illustrates the strong
contribution to the FUV flux from lower-mass stars.

As expected the FUV and \ha\ fluxes of the HII regions are
consistent with stellar populations with ages less than 10 Myr,
which is consistent with the main sequence lifetimes of the ionizing
stars. In general terms clusters detected beyond the edge
of the optical disc ($\geq$R$_{25}$) appear to be lower in mass
compared to objects identified in the inner disc. The mass range of
detected objects varies form galaxy to galaxy, with a lower limit
around 100\msol and an upper limit of $5\times10^4$\msol, while
the masses of the brightest star-forming complexes in the inner
discs are of order 10$^6$\,\msol.  These differences do not necessarily
imply a physical difference in the cluster populations, rather the
higher luminosities of the brightest clusters in the inner discs
could reflect the crowding of regions there (many of the objects
identified by SeXtractor are composites of many clusters and HII regions),
and the larger total populations of objects in those regions.
We examine this question in more detail later.  

The results shown so far only apply to the youngest UV knots that
coincide with HII regions, and it is instructive to examine the
properties of all of the UV sources in the extended discs.
For this we have constructed UV colour-magnitude diagrams, as shown 
in the right hand panels of Figure \ref{fig:objs}.  In this case  
the fluxes are not corrected for dust extinction.  Superimposed
on these diagrams are the same grid of synthesis models for
cluster masses of $10^3 - 10^6$ \msol, and ages (indicated by
black diamonds) of 1, 3, 10, 30, 100, and 300 Myr.  The extinction effects are
apparent from the colours of the inner disc objects (blue points).
All of these objects are associated with HII regions so we expect them 
to be relatively young and bright, the displacement of the objects
from the model curves can be attributed to reddening and extinction.
The effect of dust on the HII regions in the extended discs (black
crosses) is much lower, as is the case for the UV-only clusters in
the outer discs (red points).  In M83 there are also a large number
of UV clusters in the extended disc with relatively red colors.
Due to the low column densities of gas and dust in the extreme 
outer discs dust reddening is unlikely to be significant, so 
the red colors of these objects probably reflect older ages (30--300 Myr),
consistent with the conclusions of \citet{zaritsky07}.  
Precise age dating of these clusters is not possible however,
because of the degeneracy between dust extinction and age.


The galaxies NGC\,3621 and M83 were chosen for Figure \ref{fig:objs}
because they represent two extremes in the overlap of properties of
HII regions and UV clusters in the inner and outer discs.  In NGC\,3621
(top panels) the two populations appear to overlap in luminosities,
and are almost indistinguishable, apart from the slightly higher 
upper mass limit of the inner disc regions.  This similarity in
properties tends to be the norm for galaxies with \emph{normal}
discs, and also applies to some of the other galaxies with 
\emph{extended} discs including NGC\,1097, NGC\,1512, NGC\,1566,
and NGC\,3198.  In contrast to this behaviour, the regions detected
in the extended disc of M83, which extends out to as far as
4R$_{25}$, tend to be very faint as a population, with \ha\ luminosities
corresponding to the ionising outputs of single O-type stars, and 
cluster masses below $\sim$2000\,\msol.  In Figure \ref{fig:objs}
the HII regions found in the XUV dics of M83 appear to be segregated from 
the HII regions in the inner disc.
This separation arises in part from the extremely high density
of UV clusters and HII regions in the inner disc, so that the
fainter regions may be blended with brighter neighbours.  This confusion
is not a problem in the outer discs of either galaxy, due to the sparse nature of objects 
and the low level of background emission.  The 
outer disc population in M83 also shows a much lower fraction of
UV regions with \ha\ emission (36.3\% compared to 64.8\% in NGC\,3621),
and a much larger range in UV colours.  These observations are
all consistent with a larger proportion of relatively evolved
UV regions in M83 \citep{zaritsky07}.


\begin{figure*}
	\includegraphics[width=176mm]{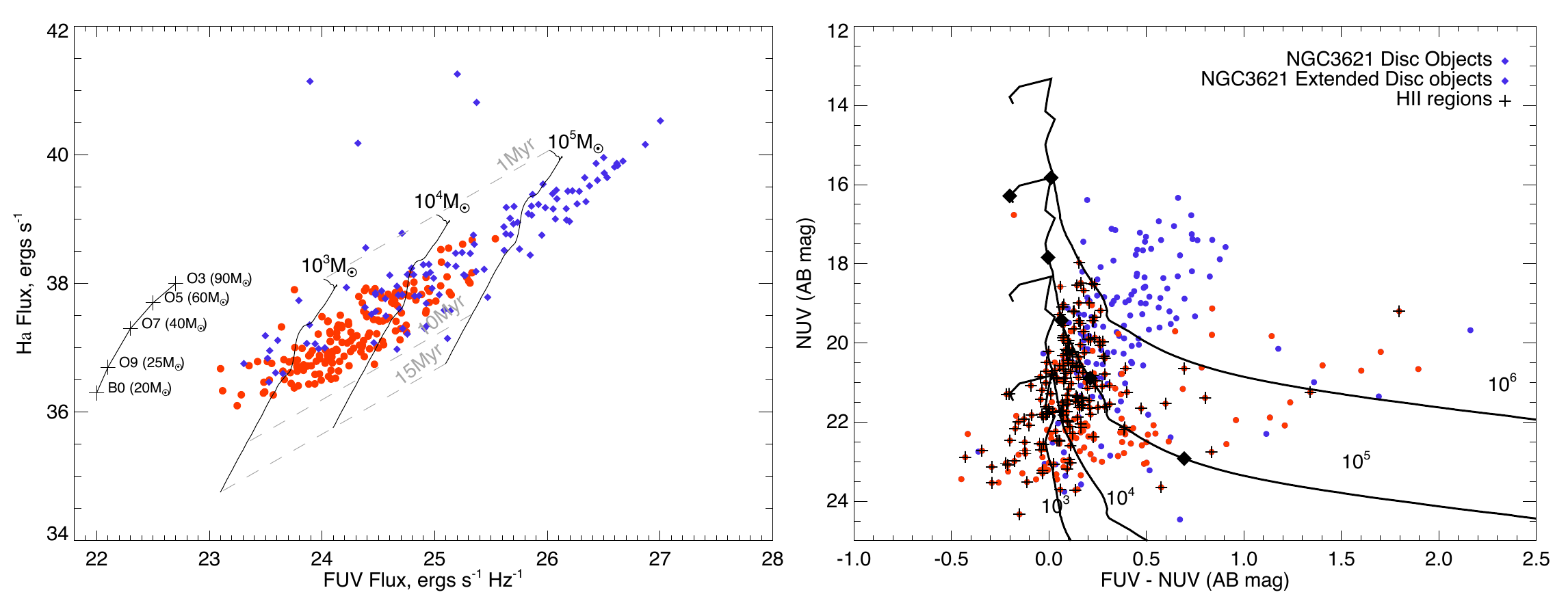}
		\includegraphics[width=176mm]{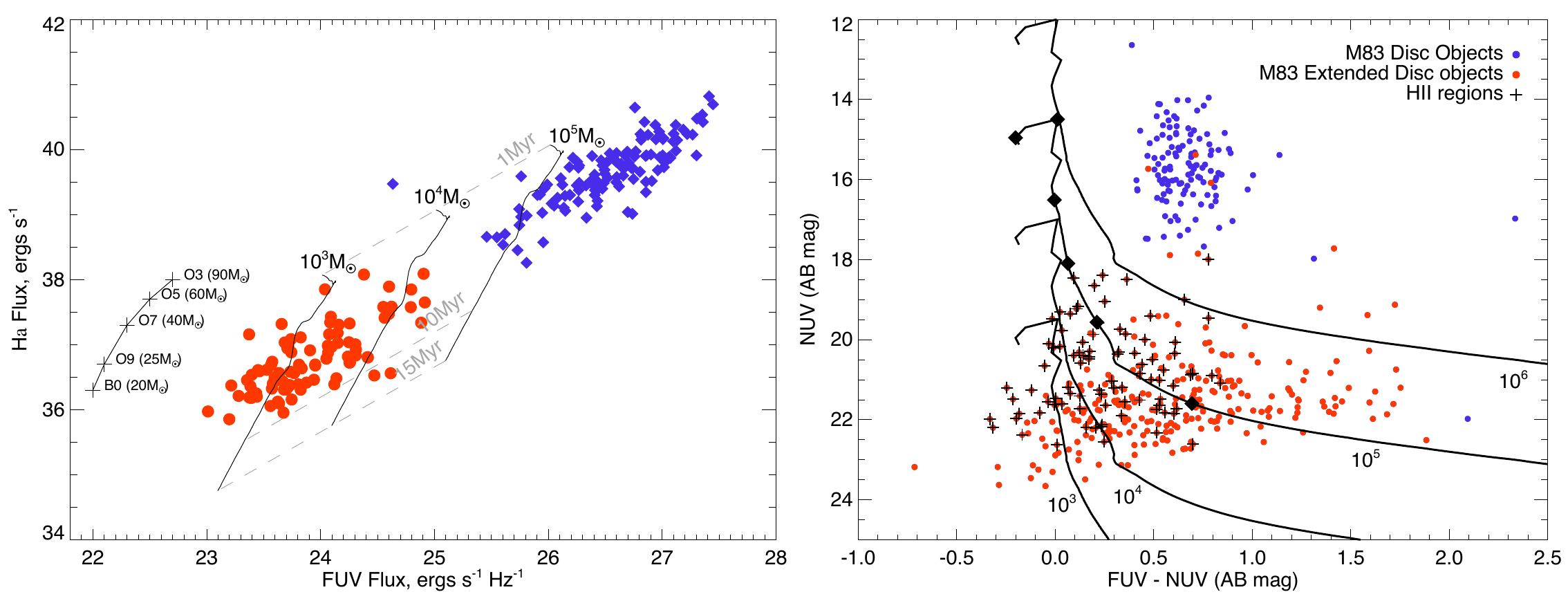}
                \caption{ Plots showing the observed properties of
                  star forming regions, NGC 3621 is shown in the top
                  two panels, and M83 is shown in the bottom
                  two. \emph{Left}: observed FUV luminosity against
                  the \ha\ luminosity after dust corrections. Objects
                  found in the inner disc (R$\le$R$_{25}$) are shown
                  in blue, whilst objects found beyond R$_{25}$ are in
                  red. We have marked on the FUV and \ha\ luminosity
                  for single O-type stars from
                  \citet{vacca96,sternberg03}. The solid black lines
                  are evolutionary models from the Starburst 99 code,
                  labelled for cluster masses of $10^4$, $10^5$ and
                  $10^6$ \msol, grey dotted lines link clusters at
                  common ages. \emph{Right}: UV colour magnitude
                  diagrams for all detected objects. The colours
                  represent the same distinction as the left hand
                  graphs. Black crosses mark the extended disc objects
                  found to have associated \ha\ emission. Solid black
                  lines show the same evolutionary models as before,
                  masses are labeled from top to bottom as $10^6$,
                  $10^5$ and $10^4$ and $10^3$\msol. In addition the
                  solid black diamonds represent cluster ages from
                  left to right of $10^{6}$, $10^{6.5}$, $10^{7}$,
                  $10^{7.5}$, $10^{8}$ and $10^{8.5}$years.}

\label{fig:objs}
\end{figure*}

Careful examination of the left hand panels in Figure \ref{fig:objs}
also shows a systematic shift in the ratio of \ha\ to FUV fluxes between
the HII regions in the inner and outer discs, with the regions in the
extended discs being fainter in \ha\ (for fixed FUV luminosity) by
factors of $\sim$1.9--2.6.  This is reminiscent of the offset in
the total fractions of \ha\ and FUV emission in the extended discs
from the previous section, and previous observations of the 
HII regions in NGC\,4625 and M83 \citep{gildepaz05,gildepaz07b}.  
In order to test whether this trend persists across our entire
sample, figure \ref{fig:pinion} shows the relation between dust-corrected
FUV and \ha\ fluxes for the approximately 2800 regions measured in 
all 21 galaxies.  Inner disc HII regions are plotted in blue 
whilst HII regions found beyond R$_{25}$ are shown in red.
The solid black line has a slope of unity, with the ratio of FUV to 
\ha\ flux expected for young star forming regions with a Salpeter
IMF, as described earlier.  The dot-dashed black line is a least 
squares best fit to the points.  There appears to be a continuous 
distribution of points in figure \ref{fig:pinion}, with no marked
change between the two populations of HII regions.  The difference
in slope between the least squares fit and a linear relation does
show a progressive decrease in the ratio of \ha\ flux to FUV flux
in the fainter HII regions, with a shift of $\sim$0.2 dex (factor 1.6)
over the 4 orders-of-magnitude range in luminosities probed in these
regions.  This is similar to trends seen in the integrated fluxes
of low-luminosity star-forming dwarf galaxies by \citet{lee09},
though not nearly as severe in magnitude as measured in the dwarf galaxies.
However it is interesting that this trend appears to be associated
with the ionising luminosities of the HII regions, and not with 
their presence in the outer discs.  We discuss the possible implications
of this result in \S5.


 \begin{figure}
	\includegraphics[width=84mm]{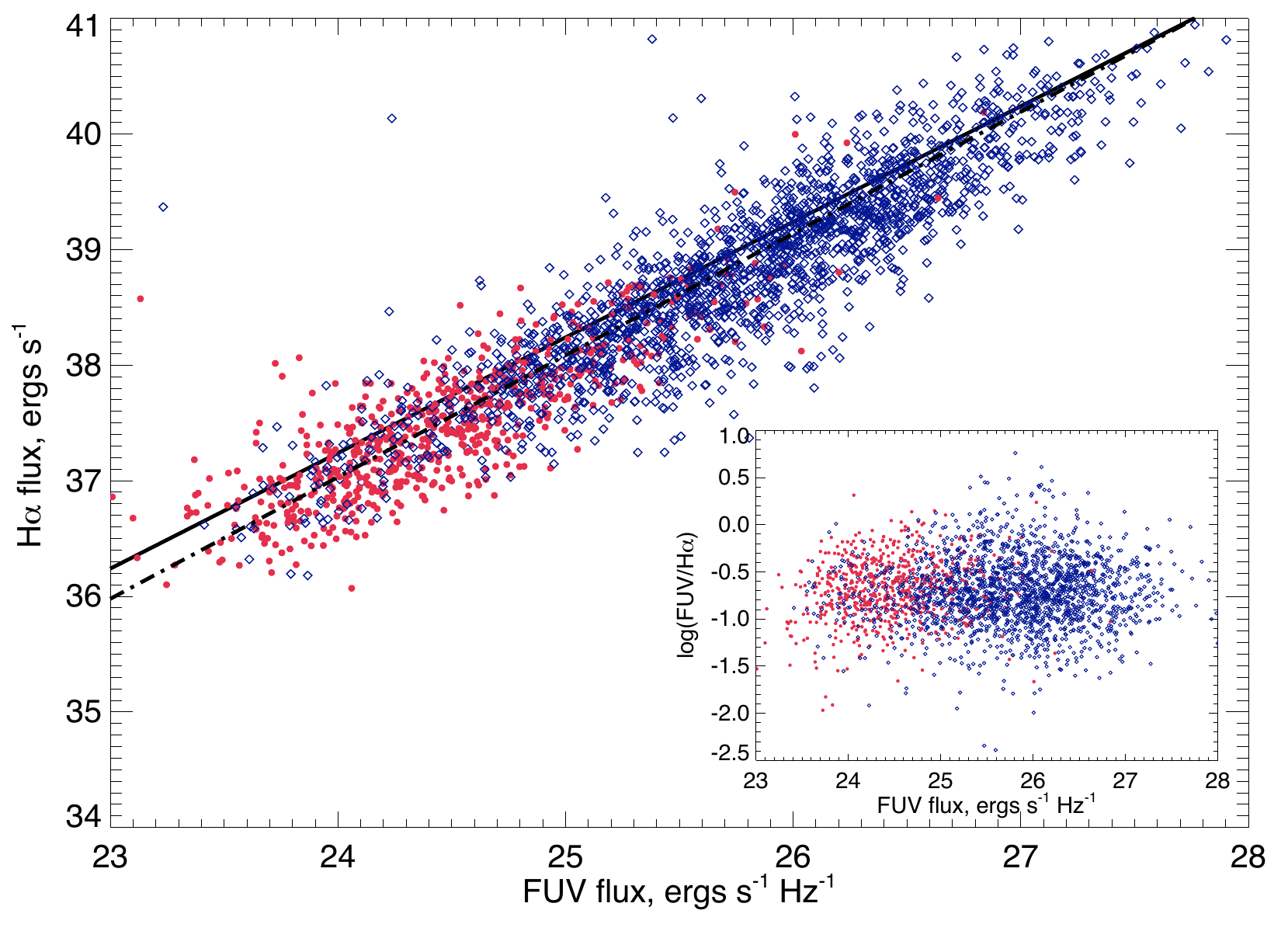}
\caption{FUV flux plotted against \ha\ flux for all HII regions found in all galaxies 
in our sample. Blue diamonds are objects found within R$_{25}$, red circles represent objects 
found beyond R$_{25}$. Inset panel shows the FUV to \ha\ ratio plotted against FUV flux.}
\label{fig:pinion}
\end{figure}

As remarked earlier the HII regions in the outer discs tend to be much
fainter than the relatively prominent HII regions in the inner discs.
Such a difference could arise from a physical difference in the populations
of star-forming regions (i.e., a steeper cluster luminosity function), but
the appearance of fainter regions could also arise from a simple size of
sample effect; if the total number of regions is much smaller the brightest
regions will tend to be fainter, simply because the upper limits of the
luminosity function will be less populated.  If the outer discs were 
sufficiently well populated in HII regions we could test for this latter
effect directly, by comparing the luminosity functions for the inner
and outer disc regions in each galaxy.  However the numbers of regions
in the outer discs are too low to derive reliable luminosity functions.
However we can test for this effect approximately, by using the statistics
of the brightest HII regions in the inner and outer discs.  \citet{kennicutt89b}
have shown that HII regions approximately follow a power law luminosity
function, with slope $dN(L)/dL = -2$.  For such a slope it is easy to
show that the expected luminosity of the brightest objects in the
population should scale linearly with the total size or luminosity of
the population \citep{kennicutt89b}.  

In Figure \ref{fig:rack} we compare the total \ha\ fluxes of all objects 
for each galaxies in the inner discs (blue points) and outer discs (red
points) with the fluxes of the brightest object in each disc region.
The inset panel shows the same comparison but using FUV fluxes.
The solid line shows a relation with unit slope, which is the expected
relation for a constant luminosity function with slope $dN(L)/dL \sim -2$. 
We have not plotted points for the two galaxies with no detected HII regions beyond
R$_{25}$. As expected both the total fluxes and the fluxes of the brightest objects
are systematically lower in the outer discs.  However there appears to
be no change in the ratio of these fluxes between the inner and outer
discs, which argues against a strong truncation or steepening of the
luminosity function in the outer discs.  Within the limited statistics
the two populations of regions are consistent with having been drawn
from the same distribution of luminosities as the inner discs.  
This does not rule out the possibility of a unique population in the
outer discs of specific galaxies (e.g., M83 as claimed by \citet{thilker05}),
but for the galaxy population as a whole there is appears to be no
need to invoke a peculiar luminosity function or IMF for the outer discs.

 \begin{figure}
	\includegraphics[width=84mm]{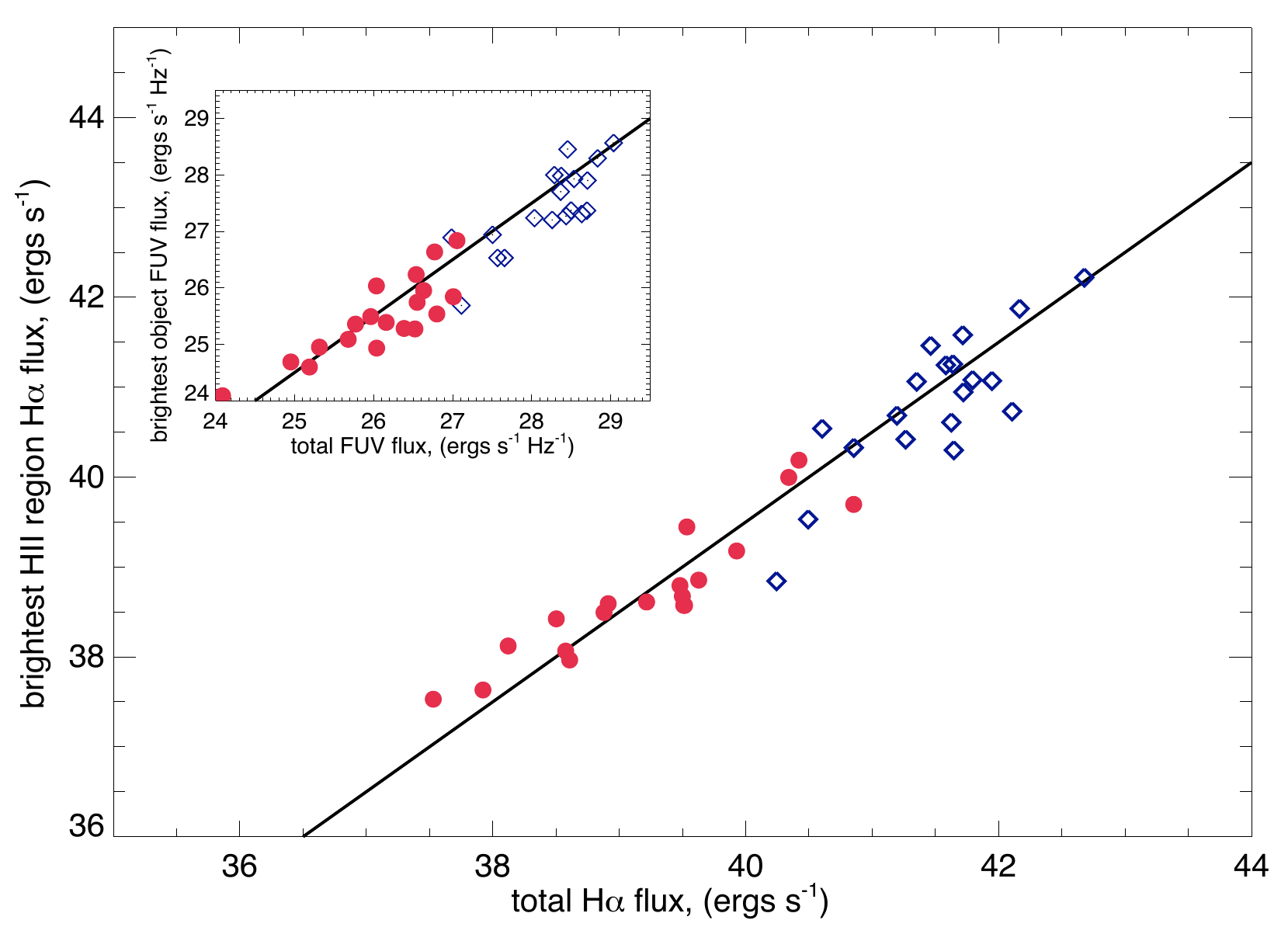}
\caption{Main figure shows a comparison of the total \ha\ flux for all objects in a given galaxy compared with the \ha\ flux of the brightest object in that galaxy. The blue diamonds represent the 
population of objects in the inner disc, whilst red dots represent the group of objects in the 
extended disc. The inset shows the same comparison but for FUV fluxes rather than H$\alpha$. 
In both cases the solid black line represents a slope of one. We have not included points for the two galaxies with no detected HII regions beyond R$_{25}$.}
\label{fig:rack}
\end{figure}


In Figure \ref{fig:objs} we are only able to plot objects for which
corresponding HII regions were identified and we might ask what are
the properties of those objects without detected HII regions. Figure
\ref{fig:fmags} shows the FUV fluxes for all objects detected in those
galaxies we have classified as extended. Objects found in the inner
disc at radii less than R$_{25}$ are shown in black, objects in the
extended disc found to have corresponding HII regions are in red and
those in the extended disc found without HII regions in green. We find
that objects found in the inner disc appear to be brighter than their
counterparts in the extended disc. For objects in the extended disc
with an FUV magnitude brighter than 18 mag we find equal numbers of objects with and without
HII regions, however for dimmer objects we see far more objects
without HII regions. This effect could be for many reasons, the lower
the FUV flux the lower the predicted cluster mass and so we expect a
lower \ha\ luminosity which might well be below the detection limit for
the farthest galaxies. Some of the fainter UV
objects might well be evolved objects with ages greater than 10Myr and
so without any \ha\ emission. Objects in the extended disc might be
statistically less likely to form HII regions than those in the inner
disc, a point we shall discuss further in the discussion. 

\begin{figure}
	\includegraphics[width=84mm]{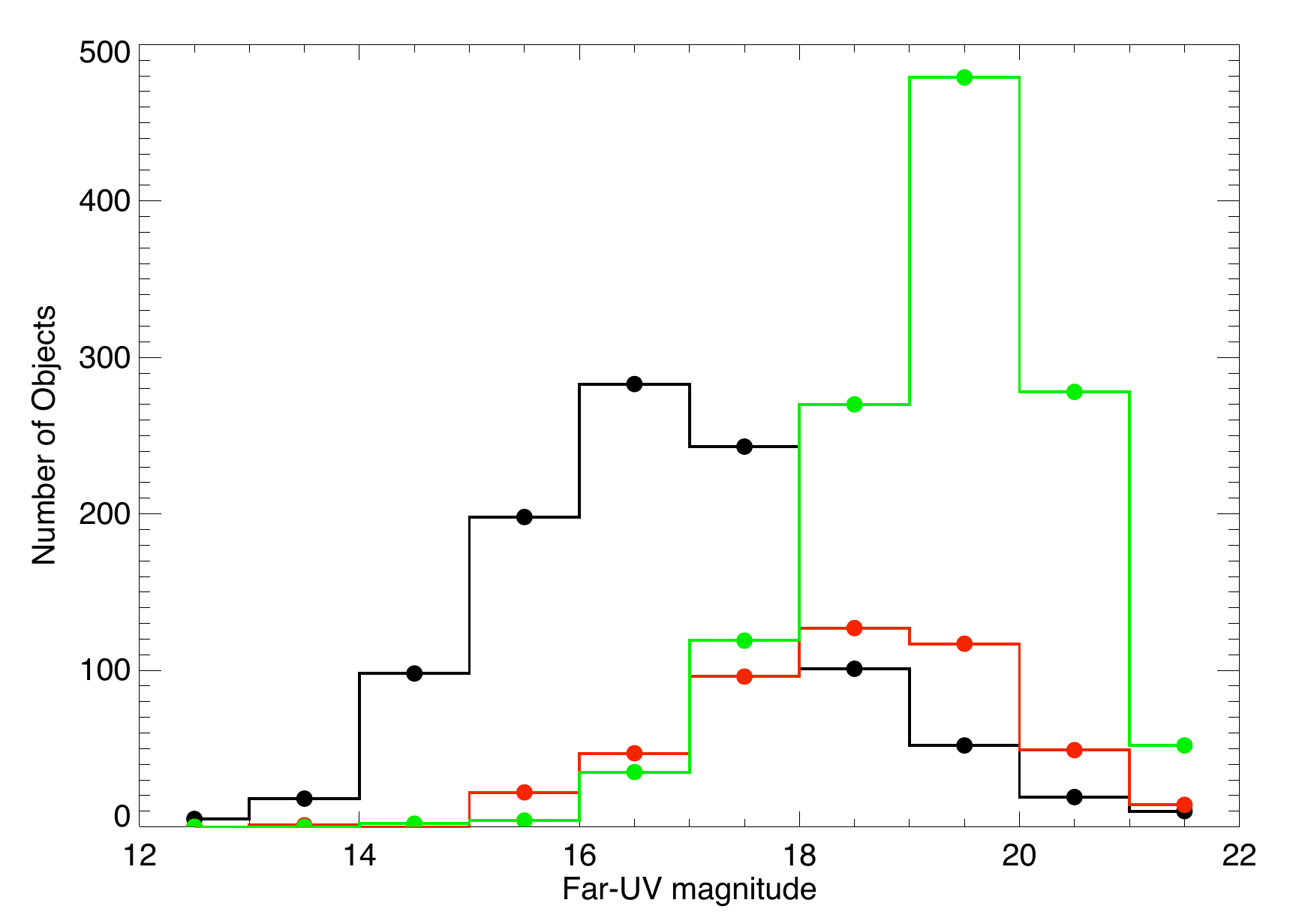}
        \caption{Histogram showing the FUV distribution of objects
          detected in all galaxies classified as extended. Objects in
          the inner disc shown as the black line, objects in the
          extended disc found to have corresponding HII regions is
          shown in red and those without HII regions shown in
          green. Magnitudes are AB magnitudes normalised to a distance
          of 5Mpc.}
\label{fig:fmags}
\end{figure}

\section{Discussion}

A primary objective of this study has been to use UV and \ha\ photometry 
of star forming regions in a large enough sample of nearby galaxies
to explore the diversity of star formation properties in their outer
discs.  We have found that extended star formation (defined as star
formation outside of  R$_{25}$) is nearly ubiquitous in nearby gas-rich
galaxies, and that there is an enormous diversity to the properties
of these extended discs.  Nearly all of the galaxies we studied
have UV clusters located beyond R$_{25}$, and most
show at least 1 HII region found beyond this radius.
In many cases these objects were found at large radii
even for the galaxies classed as normal, for example NGC 5194 displays
9 HII regions in the extended disc out to a 1.33 R$_{25}$.  
The star formation contained in these extended discs can be a 
significant fraction of the global star formation, but in most
instances represents a modest fraction ($0.5-10$\%), after correcting
for dust extinction.  Much more prominent extended discs are found
in specific examples including the well-studied
spiral NGC\,4625 and the interacting spiral NGC\,1512,  but they
tend to be the exception rather than the rule.

It has proven rather difficult to classify galaxies as \emph{extended} or not
based on a single parameter, the R$_{25}$ isophote may not encompass
all the HII regions and some may appear outside this superficial
boundary. The number of objects found in both UV and \ha\ provided a
better clue as to the nature of galaxies but this can be misleading,
we see some galaxies with as many as 44 possible UV objects associated
to the galaxy yet not classed as \emph{extended}. Identification and
classification is best done with many parameters as well as visual
inspection of the images. The trouble in defining a set of parameters
to determine the nature of extended discs lies in the rather diverse
range of properties they display.

As discussed in the introduction, observations of the extended
star-forming discs may have important implications for a range of
broader problems including star formation thresholds, the clustering
of star formation, the IMF, and the physical origins of the discs
themselves.  We conclude by briefly addressing each of these topics
in the light of our new results.

The apparent presence of strong radial truncations in the \ha\ discs
of spirals has been used by many authors to argue for the presence of
radial star formation thresholds \citep{kennicutt89}.
These in turn could be be explained by the presence of a large-scale
gravitational stability criterion \citep{safronov60,toomre64}, 
though other models have also been proposed to account for the 
thresolds \citep{elmegreen94,schaye04}.
However the absence of corresponding turnovers in the UV profiles
of many of these galaxies has called into question the reality
of the \ha\ truncations.  Are thresholds relevant any longer?

Our observations suggest that the detection of extended star formation
complicates what many treated previously as an overly simplified
picture of star formation thresholds, but the general observation
of a radial transition in star formation still holds up in most 
galaxies.  Approximately half of the galaxies in our sample show
strong radial turnovers in both \ha\ and UV emission.  Although these
turnovers are not always located near R$_{25}$, the presence of
consistent radial profiles in the UV and \ha\ is consistent with 
the general threshold picture.  Among the remaining 10 galaxies with
extended UV discs, half of those show similar profiles in the UV
and \ha.  Although there is no strong truncation radius, the
UV surface brightnesses often decline by up to two orders of magnitude
over a radius of a few kpc.  Examination of the images shows that
the extended star formation may be attributable to a variety of
causes, including azimuthal asymmetries, tidal arms, or
isolated regions of star formation \citep{martin01}.
For most of these galaxies the locations of the star-forming
regions can be understood in a general picture in which the
main inner \ha\ and UV discs are defined by the region where the 
interstellar gas is gravitationally (or thermally) unstable
in nearly all locations, whilst the star formation in the outer
discs is concentrated to locations where the gas disc is globally
stable, but locally unstable in regions where the gas is concentrated
or compressed \citep{kennicutt89, martin01,bush08}.
 

The most interesting and perplexing galaxies in our sample are those
which show a relatively strongly truncated \ha\ disc, but an extended disc of 
UV emission, with no sign of a corresponding radial truncation.
These include the prototype M83 as well as NGC\,1097, 2841, and 4625.
In some of these galaxies a sizeable fraction of the UV clusters have faint 
\ha\ counterparts, confirming their young ages, but in others only a
handful of clusters show \ha\ emission.  This may be the result of 
factors such as cluster age, a low mass
cluster forming without ionising stars or even a porous interstellar
medium.  It is difficult to distinguish between these alternatives with
our photometric data alone, and further exploration is beyond the scope
of this paper.  However we can apply a few simple tests to rule
out some of the simplest explanations.

A popular explanation for the properties of the extended UV discs
has been to invoke the presence of a truncated stellar IMF in regions
with low levels of star formation e.g., \citet{meurer09,kroupa08}
and references therein.  In this picture the formation of the most
massive stars is strongly suppressed in low-mass gas clouds, due
to the limited population of the IMF.  The observed \ha\ luminosities
of the outer disc HII regions (typically of order $10^{36} - 10^{38.5}$
ergs\,s$^{-1}$) are consistent with many falling into the regime
where the nebular ionisation is subject to small number 
statistics in O-type stars.  However neither the numbers of stars
nor the observed ratios of \ha\ to UV luminosities are consistent
with the dramatic IMF change that would be required to produce
radial truncations in the discs.  The azimuthally-averaged \ha\
surface brightness of M83 at the inner edge of its \ha\ turnover
(Figure 9) is equivalent to several tens of O6V stars per square
kiloparsec, with individual HII regions typically containing tens
to hundreds of such stars.  This is far to high to account for
a sharp radial turnover in HII regions at larger radii from 
statistical population effects alone.  Moreover the minimal changes
in the ratio of \ha\ to FUV luminosities of the HII regions 
are inconsistent with an IMF truncation that would sufficient
to mimic a star formation threshold.  Finally it is difficult
to understand why the IMF mechanism would only operate in the
handful of galaxies with deviant UV and \ha\ disc profiles, and
yet be absent in the majority of galaxies with consistent UV
and \ha\ profiles, despite their containing similar ranges of
luminosities in their star clusters and HII regions.
We suspect instead that the presence of extended UV emission
(in the absence of similarly extended \ha\ emission) in extreme
cases of XUV discs is due to a combination of the factors that
have been discussed by \citet{thilker05,lee09},
and others, including
age, ionisation bounding effects in some HII regions, and
small number statistics in the formation of
ionising stars in individual clusters, and possibly a modest
systematic deviation in the IMF.

Much has been made of a possible IMF truncation suggested by the UV to \ha\ ratio of dwarf galaxies \citep{meurer09} and we do not attempt a full discussion of the problem here. This is however a recurring theme and \citet{lee09} reports a similar trend in the local volume and notes no single explanation is capable of explaining this result. \citet{melena09} shows the same lack of \ha\ in the outer discs of 11 dwarf irregular galaxies which is attributed to the loss of ionising photons in these regions. We do note that our sample of HII regions which have formed in the low-gas environment of XUV discs displays only a small \ha\ deficit (Figure \ref{fig:pinion}). There could be several explanations for a small change; a loss of ionising photons in low density HII regions; long term low-rate star formation predicts that the the high mass end of the IMF may not be fully populated at all times; or even an effect of inaccurate dust measurements which \citet{boselli09} suggests might explain the trend seen by \citet{meurer09}.

Our study can not give definitive answers as to the possible mechanisms 
for the formation of these extended discs. We can say that the large variety 
in properties which we have highlighted indicate there may be more than one 
possible formation mechanism. In some cases like M83 the average gas density 
in the outer disc may be sub-critical for star formation, yet local variations in the gas 
density may produce pockets in which star formation may take place as suggested 
by \citet{bush08,dong08}. Mechanisms like this may produce HII regions 
with distinct properties from the inner disc, as seen in the flat and de-coupled 
abundance gradient of the outer disc in M83 \citep{bresolin09}. The outer discs of galaxies 
like NGC 3621 appear to be more of a continuation of the inner disc and it may simply 
be the case that the optical \emph{edge} does not correlate with the star formation 
threshold and so is not a \emph{true} extended disc. Investigation of the abundance 
gradient in galaxies like NGC 3621 might help to shed more light on this conundrum. 
Galaxies like NGC 4625 and NGC 1512 have obvious interactions and these may 
play a key part in the formation of some but not all extended discs.

\section{Summary \& Conclusions}

Over the course of this paper we have examined 21 galaxies by comparing the
 \ha\ and UV fluxes of objects in both the inner disc (R$<$R$_{25}$) and the outer 
 disc and by using detailed surface photometry in both these wavebands. We identified 
 ten of these galaxies as having an extended ultraviolet disc and investigated the properties 
 of this relatively new phenomenon. We have been able to quantify many of the characteristic of
\emph{extended} disc and compare the properties of these galaxies to
those with no \emph{extended} discs. Our conclusions are summarised below.
 
\begin{enumerate}
\item \emph{Properties of extended discs}: the galaxies in our sample show a
range of properties; extended disc fluxes range in value from a few percent up 
to almost 50\% compared to the flux of the host galaxy. We also note it is not
uncommon for galaxies to have a few HII regions lying beyond the
optical edge and at radii comparable to the extent of some
\emph{extended} discs. The number of both UV and \ha\ objects in beyond
the optical edge gives a clearer distinction between galaxies
classified as \emph{normal} and those thought to be \emph{extended} in
nature. The number of UV objects found with HII regions varies as well
from 26\% percent to 68\%, although this difference might be due to an
age discrepancy of different extended discs it might also be due to
the different environments associated with each galaxy. 
\item \emph{Star formation thresholds}: although we can confirm the findings of \citet{thilker05} for M83 for
which we see an \ha\ threshold at around R$_{25}$ and no UV truncation,
behaviour which is repeated for several galaxies in our sample it is
not a universal rule for \emph{extended} galaxies. Around half of the
galaxies we have classed as \emph{extended} show no threshold in
either \ha\ or the FUV close to the optical edge and have a gradually
declining surface brightness. This further emphasises the range of
behaviour seen in \emph{extended} galaxies.
\item \emph{Properties of HII regions}: in agreement with \citet{gildepaz05,thilker05} 
the HII regions found beyond the optical edge of galaxies appear to have \ha\ fluxes consistent
with single ionising stars; the UV fluxes indicate an underlying population of lower mass B-type stars.
These HII regions appear to
have lower than expected \ha\ luminosity compared to their inner disc
counterparts. However the same data show that the \ha\ to UV ratio is
similar to that for objects in the inner disc. This implies that any
excess UV emission from the extended disc mainly comes from those
clusters which exhibit no \ha\ emission.
\item \emph{Variations in the IMF}: we note that the low mass of many HII regions 
indicates that formation of massive stars is subject to stochastic variations. Comparison 
of the total flux and brightest object reveal no apparent change in the IMF between the inner 
and extended discs. The minimal change in the ratio of UV to \ha\ luminosities are inconsistent 
with an IMF truncation in the outer discs which might mimic a star formation threshold.

\end{enumerate}



As more and more extended disc are studied we are beginning to see
their full range of properties, such large variations might indicate
that more than one mechanism is capable of inducing star formation
beyond the classical edge of a spiral galaxy, be it a close interaction
disturbing a near critical density gas disc, the propagation of a
spiral density wave or ever more obscure mechanisms. To investigate the 
possible mechanisms for the formation of these extended discs further 
multi-wavelength analysis will be needed as well as further spectroscopic 
studies of HII regions to determine abundances.


\section*{Acknowledgments}

\appendix

\section{Radial Profiles}

Here we show radial profiles for all galaxies within our sample, profile classifications are shown as well. Once again FUV is shown in blue and \ha\ in red. The grey shaded area shows the flux level below which we determine the FUV flux to be too low to contribute to the disc, it should be noted that due to difference scales \ha\ emission in the grey region is still reliable. Below we also list some comments on individual galaxies to highlight some of their more unique features and interesting features in the radial profiles.

\begin{itemize}
\item NGC0628: Classified as Tr* - Tr*, the profile in both FUV and \ha\ are truncated at $\sim 2$ R$_{25}$ and $\sim 1.3$ R$_{25}$ respectively. However there are few HII regions found scattered beyond the \ha\ truncation.
\item NGC0925: Although classified as having a truncated profile in both FUV and \ha\ (Tr-MaTr), there is some evidence of an extremely faint FUV shelf extending to 1.3 R$_{25}$ with some associated \ha\ emission.
\item NGC1097: The bumps in the FUV and \ha\ profiles are caused by the asymmetric spiral arms. We classed the \ha\ profile as truncated due to the sharp decline seen at $\sim 1$R$_{25}$ and at $\sim1.4$R$_{25}$.
\item NGC1291: The original \ha\ image was under-subtracted toward the centre of the galaxy. To create the \ha\ profile we re-subtracted the continuum from the raw \ha\ image which makes the absolute \ha\ flux profile unreliable. It should also be noted that the FUV emission in the centre of the galaxy due primarily to an old stellar population rather than recent star formation. This is confirmed by the exceptionally red UV colour found toward the centre of the galaxy, this can be see in the GALEX composite images. 
\item NGC1512: This is the most extended galaxy in the sample and many of the HII regions fall close to the edge of the \ha\ image where edge effects can make the determination of the radial profile difficult. As a consequence we stop the \ha\ as soon before an annulus overlaps with the edge of the image, this is defined by the vertical dot-dashed line. Individual HII regions are detected beyond this radius and included in the analysis of this paper. 
\item NGC2841: There is a noticeable change in the slope of the FUV profile at the truncation of the \ha\ profile, showing a clear distinction between the behaviour of the inner and extended discs.
\item NGC3198: Despite some sharp truncations in the \ha\ we have classified this as following the smooth FUV profile. These sharp \ha\ truncations are due in part to the occasion sparse distribution of HII objects and the small annulus used to make the profile, with a larger annulus the \ha\ is smoother out to the extent of the FUV emission.
\item NGC3521: Although there is no single definitive edge to the \ha\ profile we find only 1 HII region beyond R$_{25}$ we class the profile as truncated.
\item NGC4321: Here the classification of a truncated \ha\ profile represents the clear change in the nature of \ha\ profile at the edge of the galaxy.
\item NGC4536: This galaxy shows an \ha\ profile extending beyond that of the FUV profile, a feature which is not evident from inspection of the images. This is in fact due to a difficult background estimation due to edge effects in the images.
\end{itemize}

\begin{figure}

	\includegraphics[width=84mm]{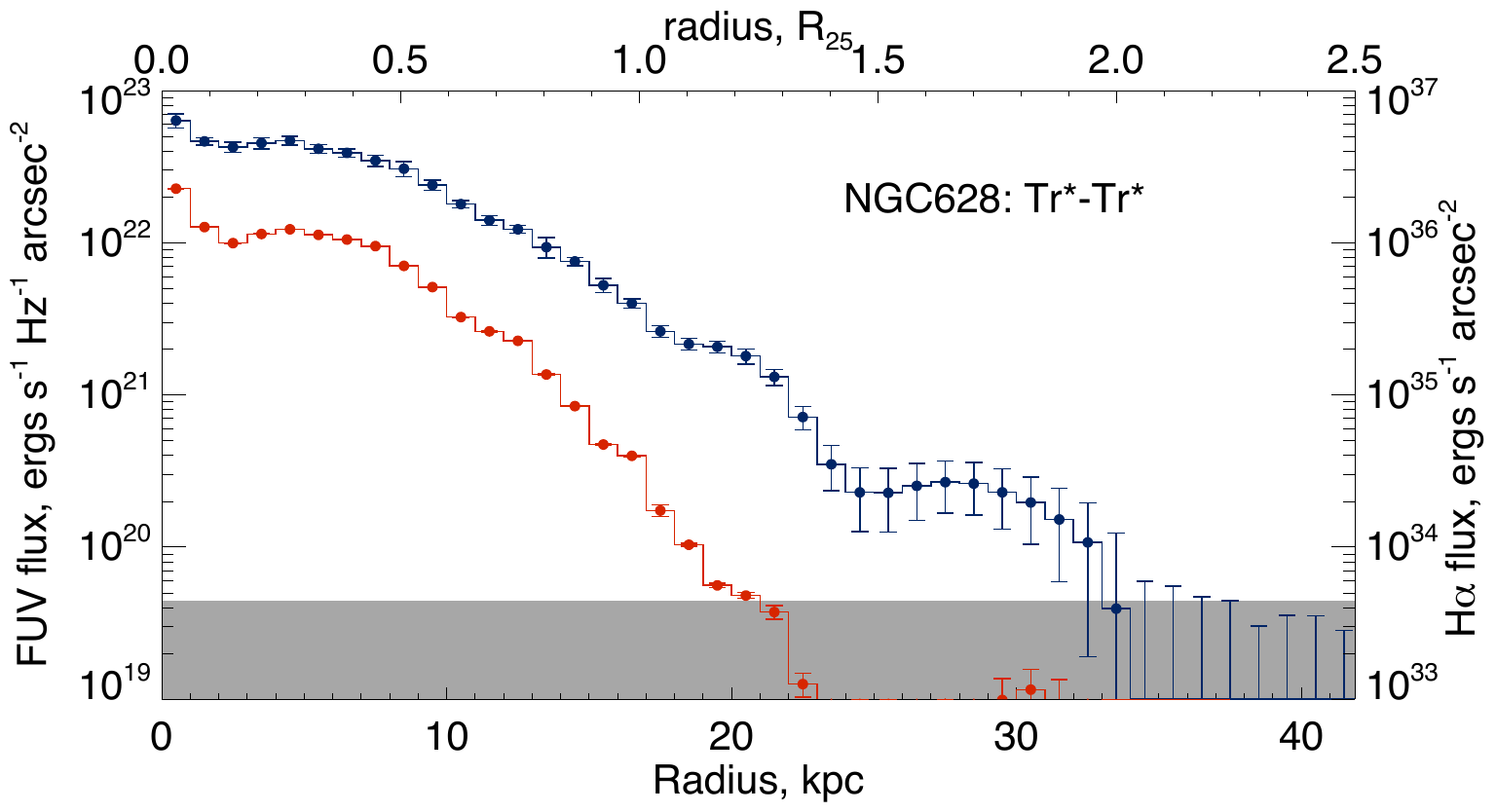}
	\includegraphics[width=84mm]{NGC0925_blurrnew_final.pdf}
	\includegraphics[width=84mm]{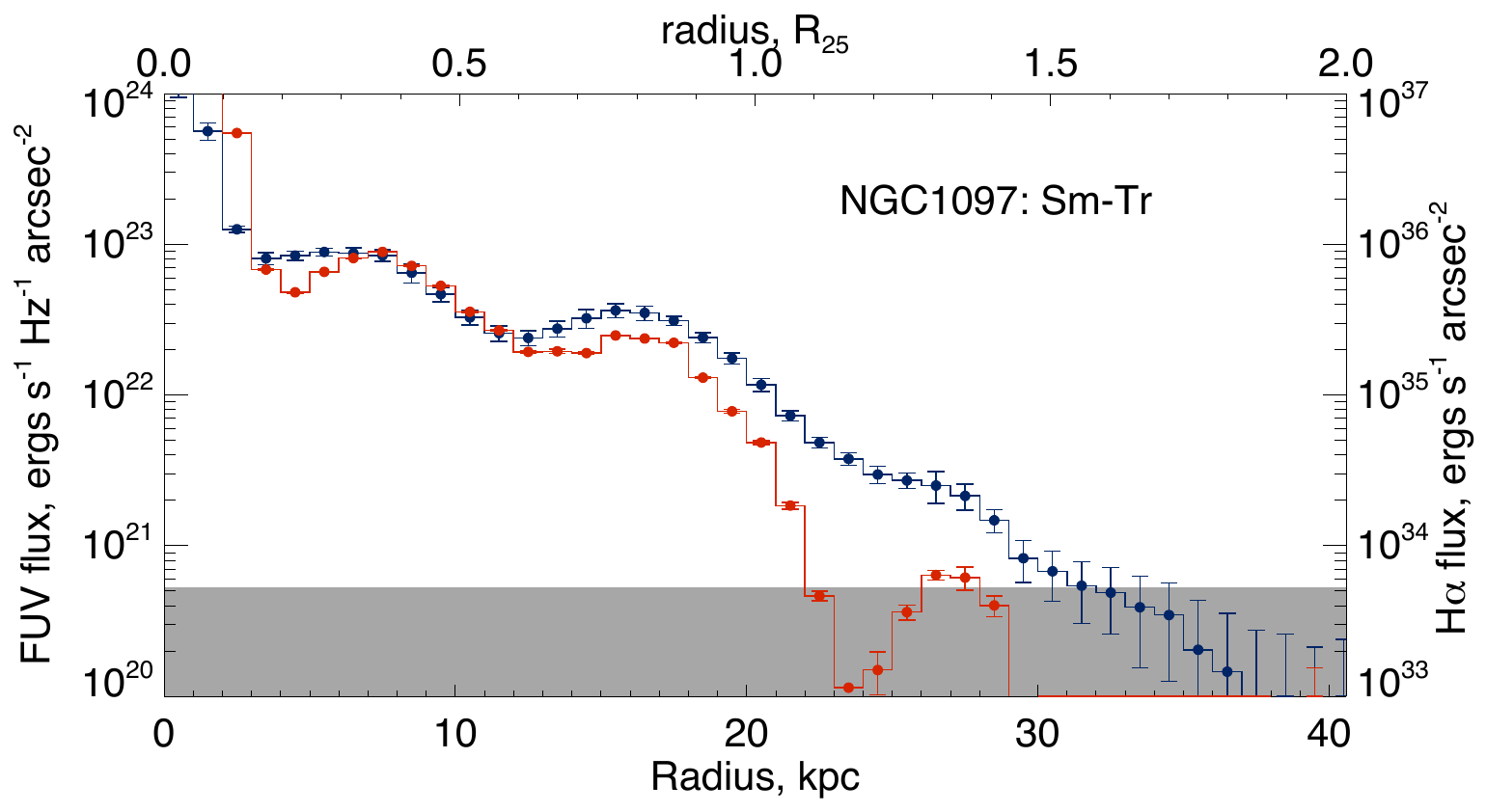}	
	\includegraphics[width=84mm]{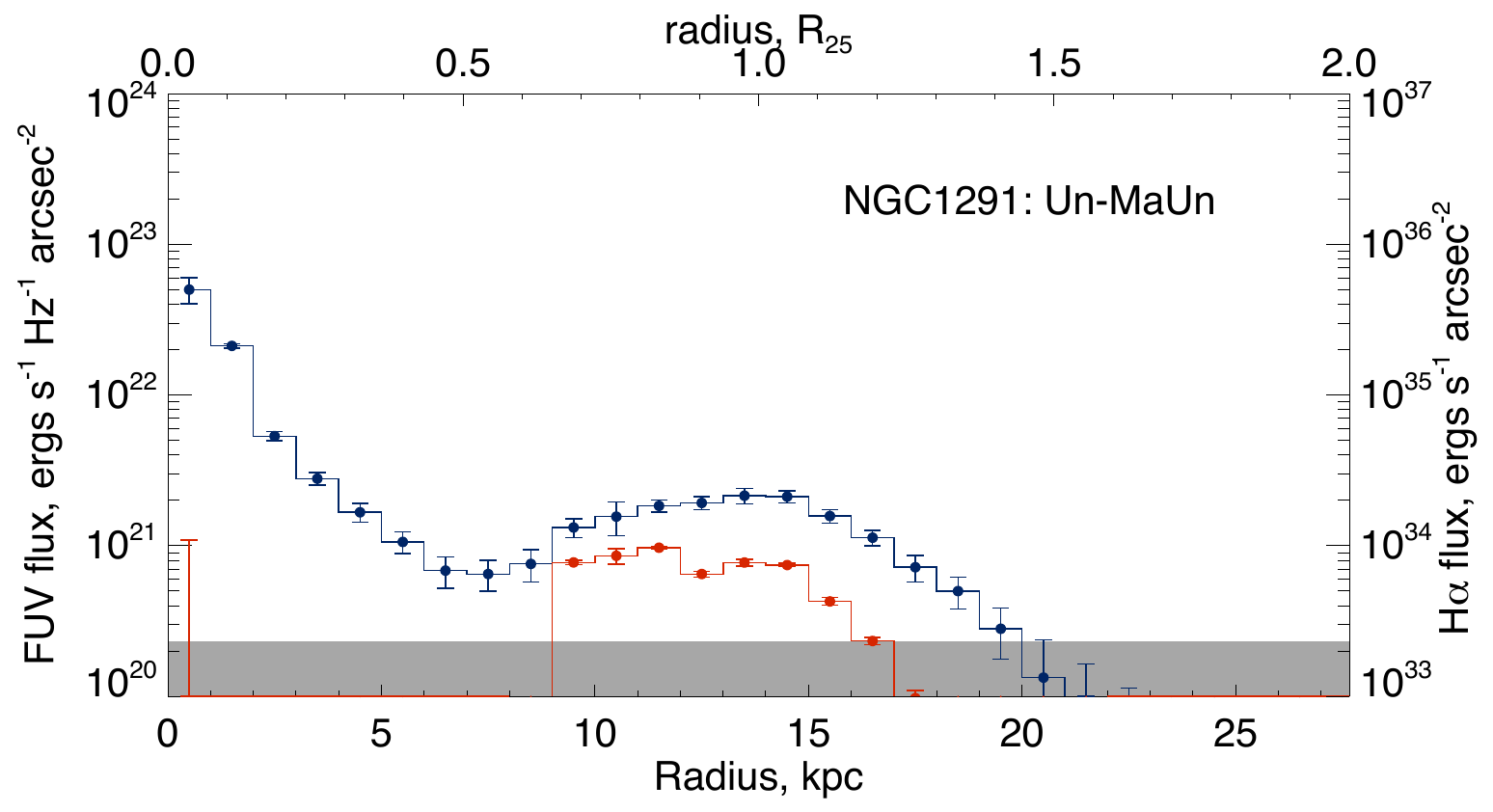}	
	\includegraphics[width=84mm]{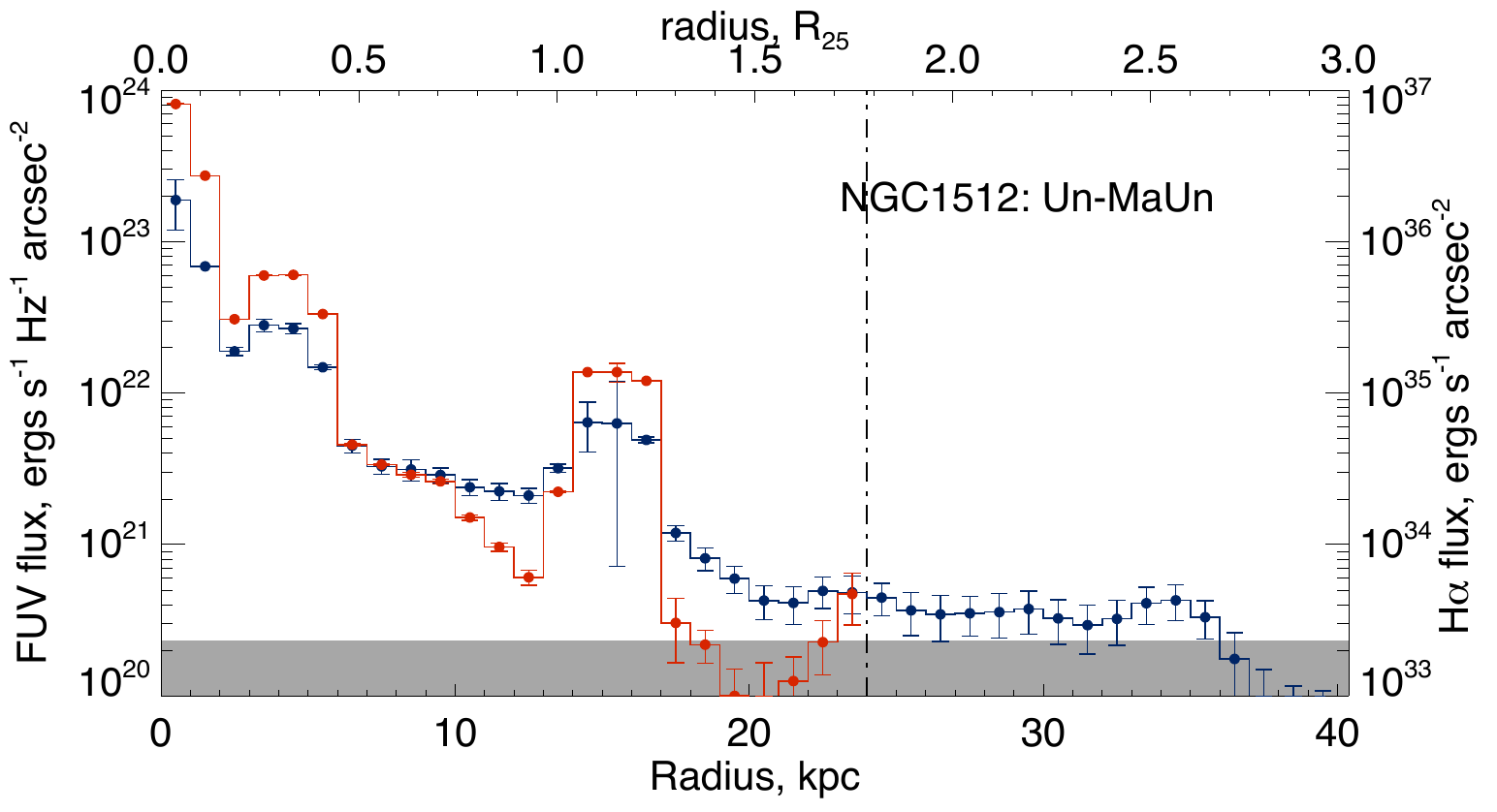}
\end{figure}

\begin{figure}	
	\includegraphics[width=84mm]{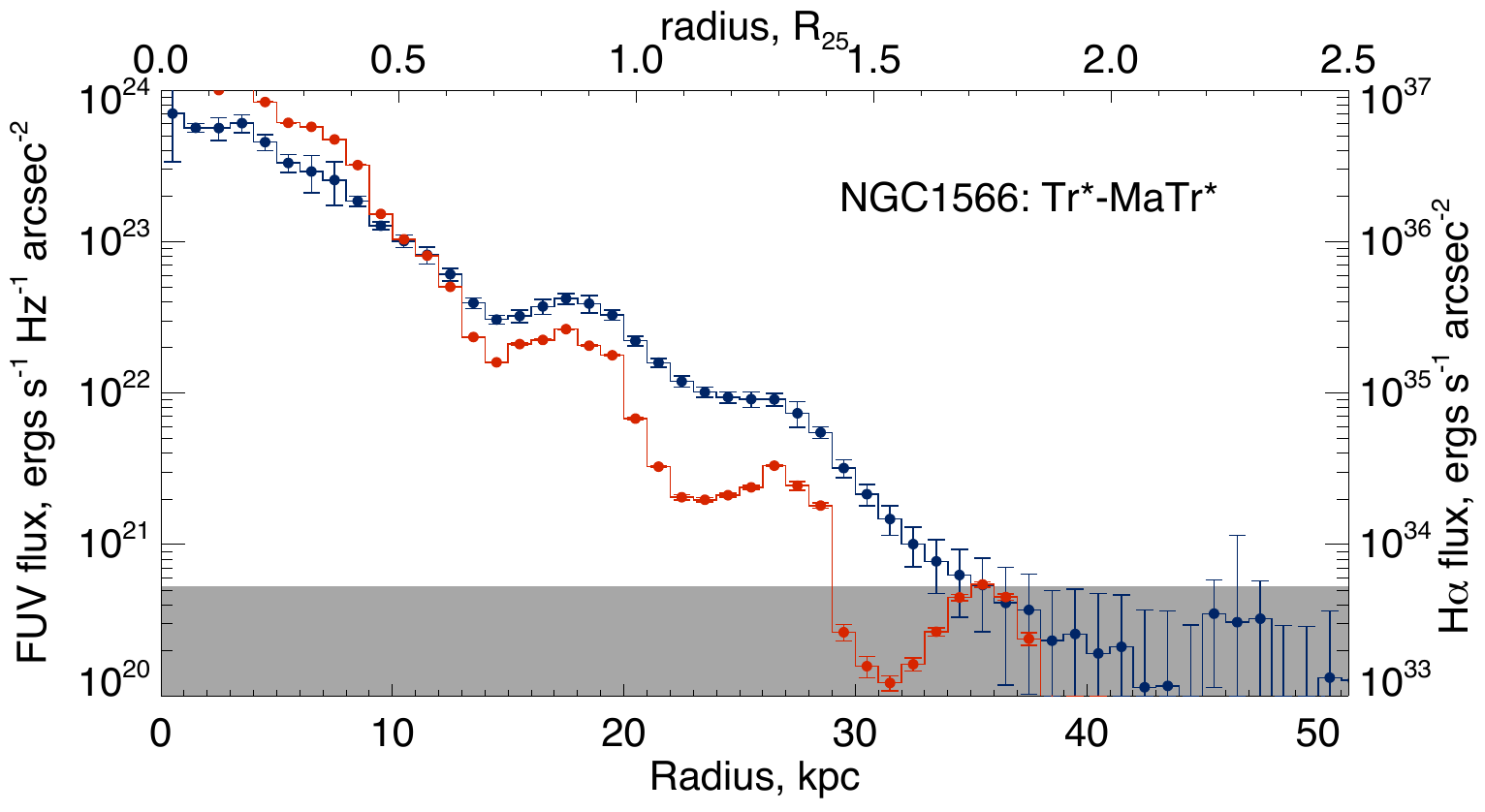}	
	\includegraphics[width=84mm]{NGC2841_blurrnew_final.pdf}
	\includegraphics[width=84mm]{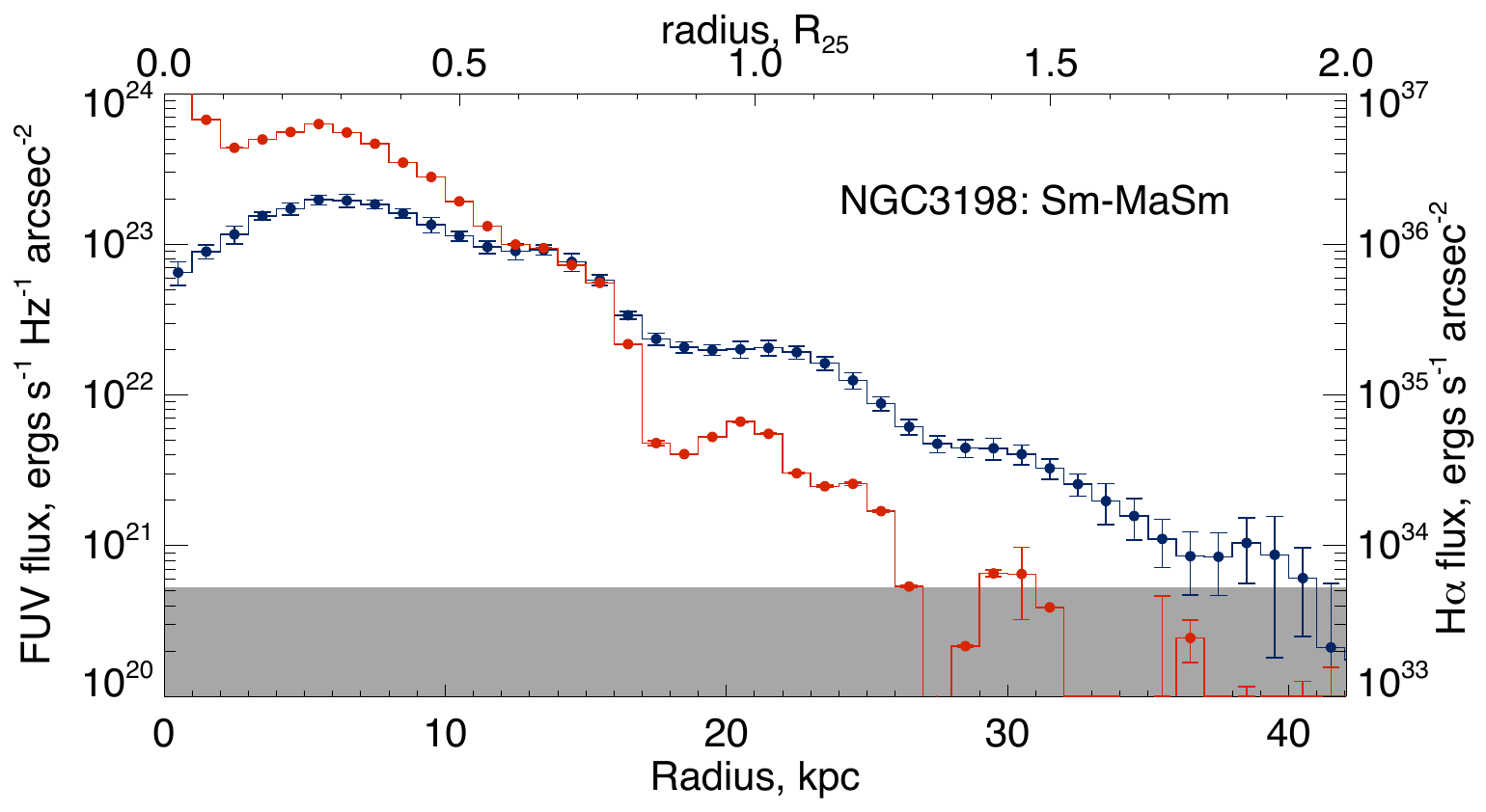}
	\includegraphics[width=84mm]{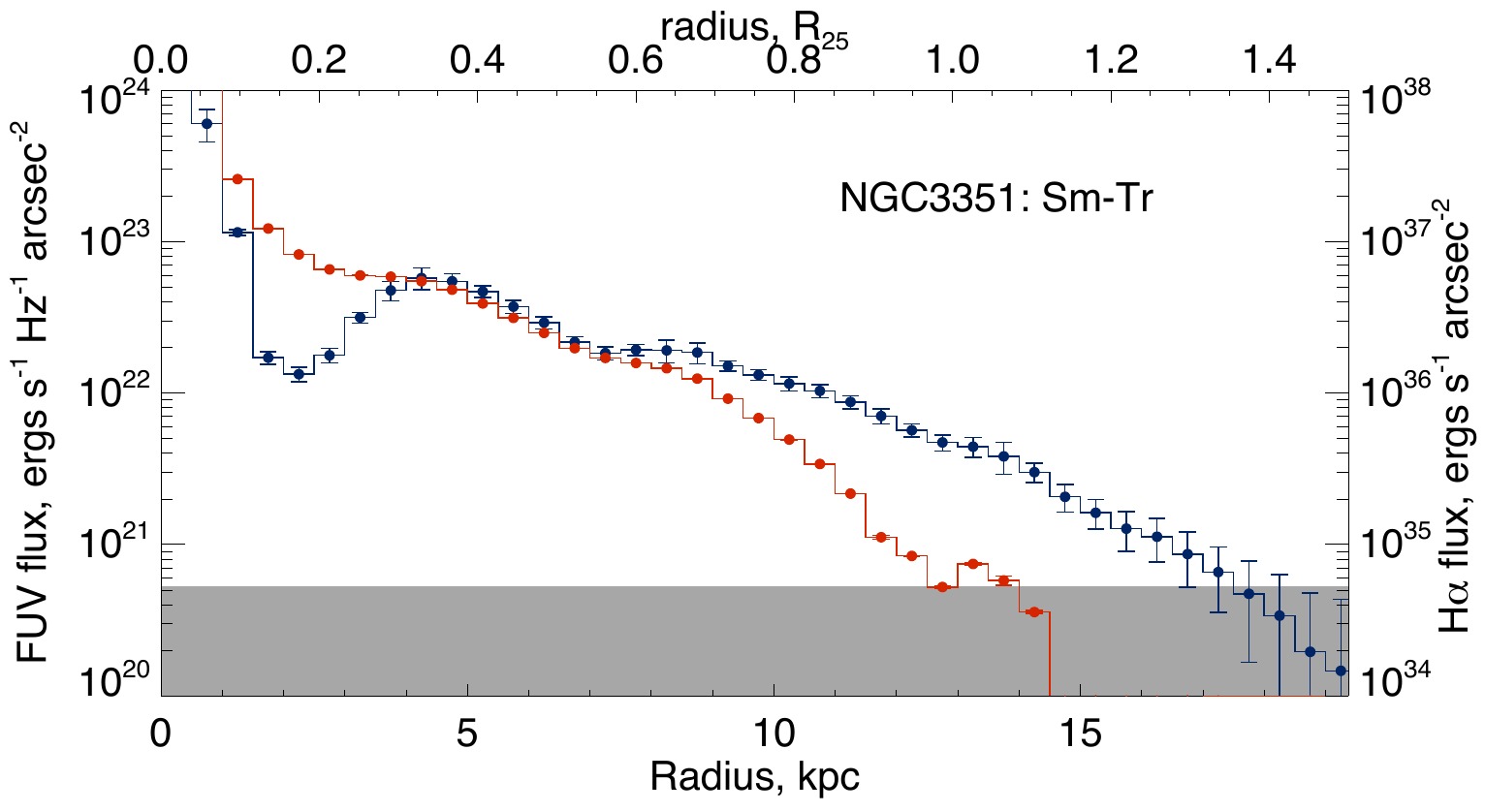}
	\includegraphics[width=84mm]{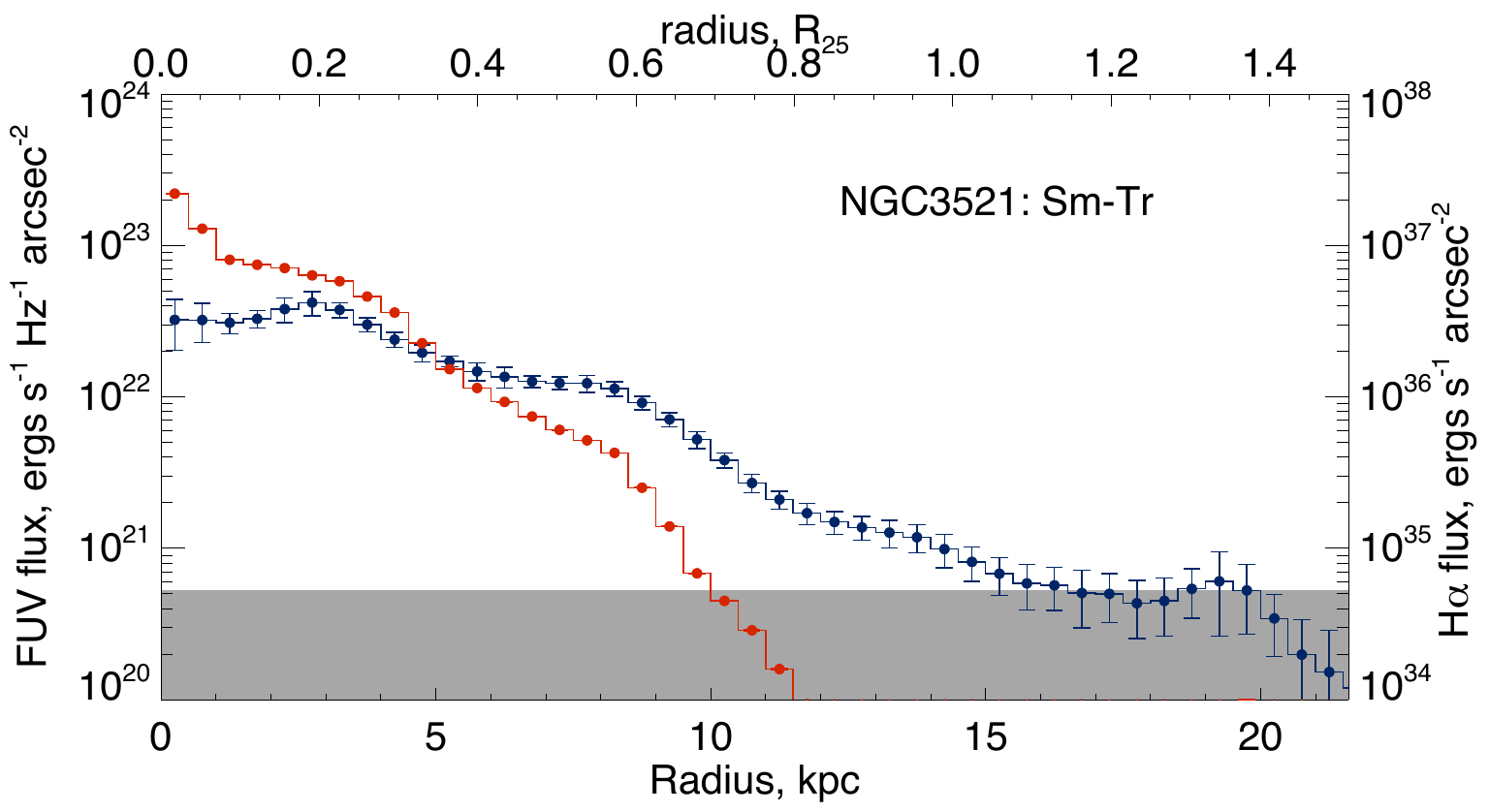}	

\end{figure}

\begin{figure}
	\includegraphics[width=84mm]{NGC3621_blurrnew_final.pdf}	
	\includegraphics[width=84mm]{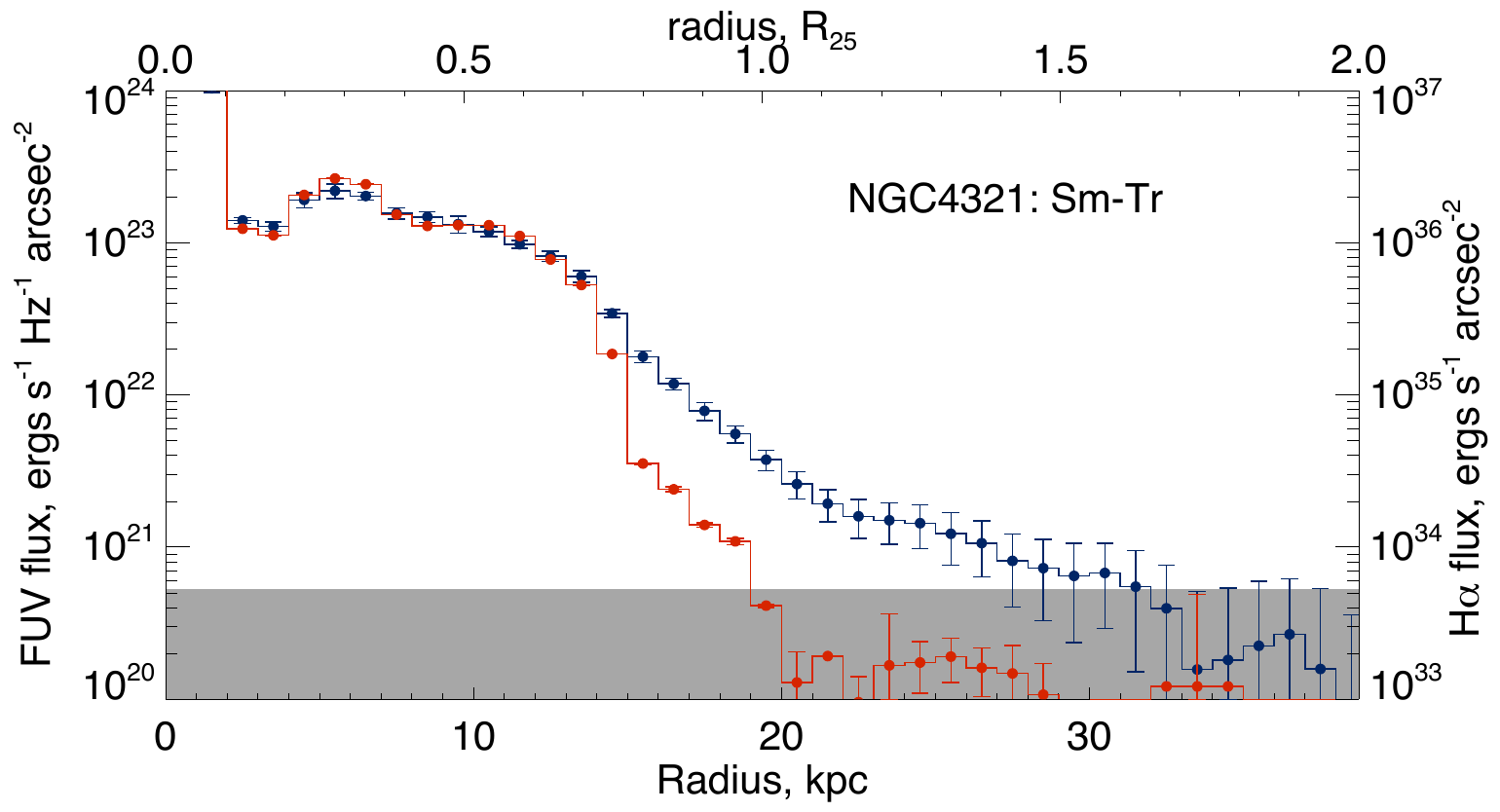}		
	\includegraphics[width=84mm]{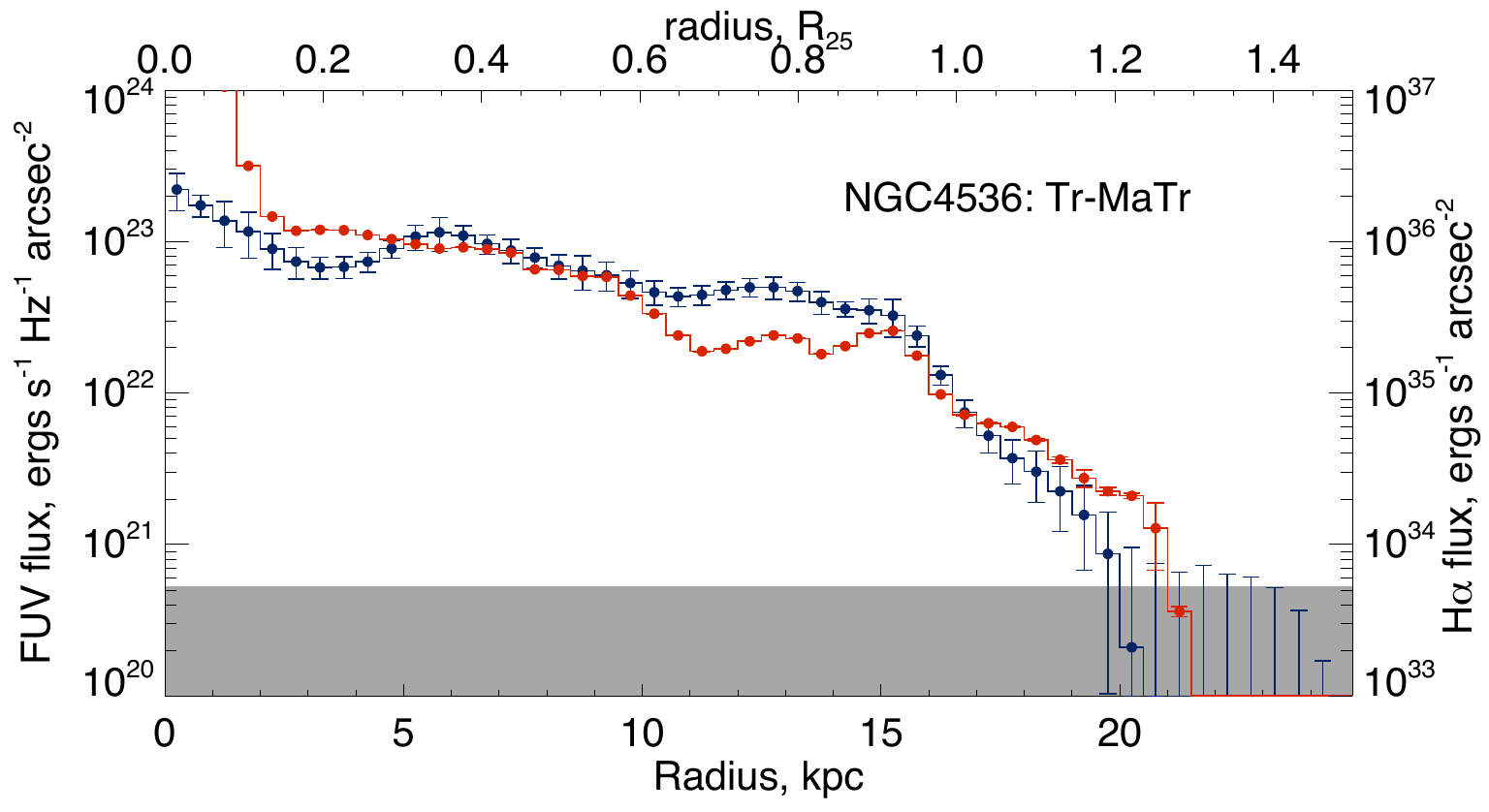}
	\includegraphics[width=84mm]{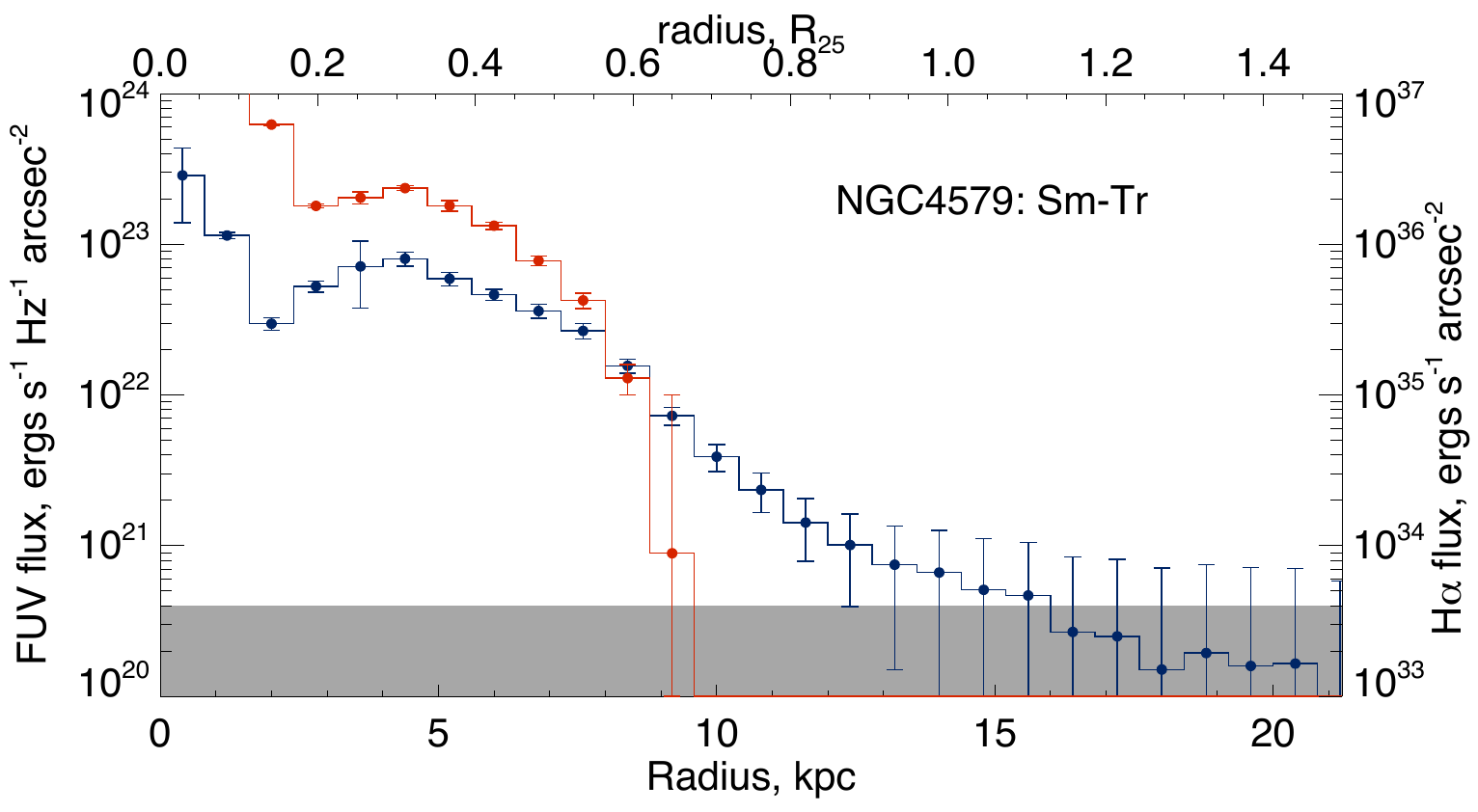}
	\includegraphics[width=84mm]{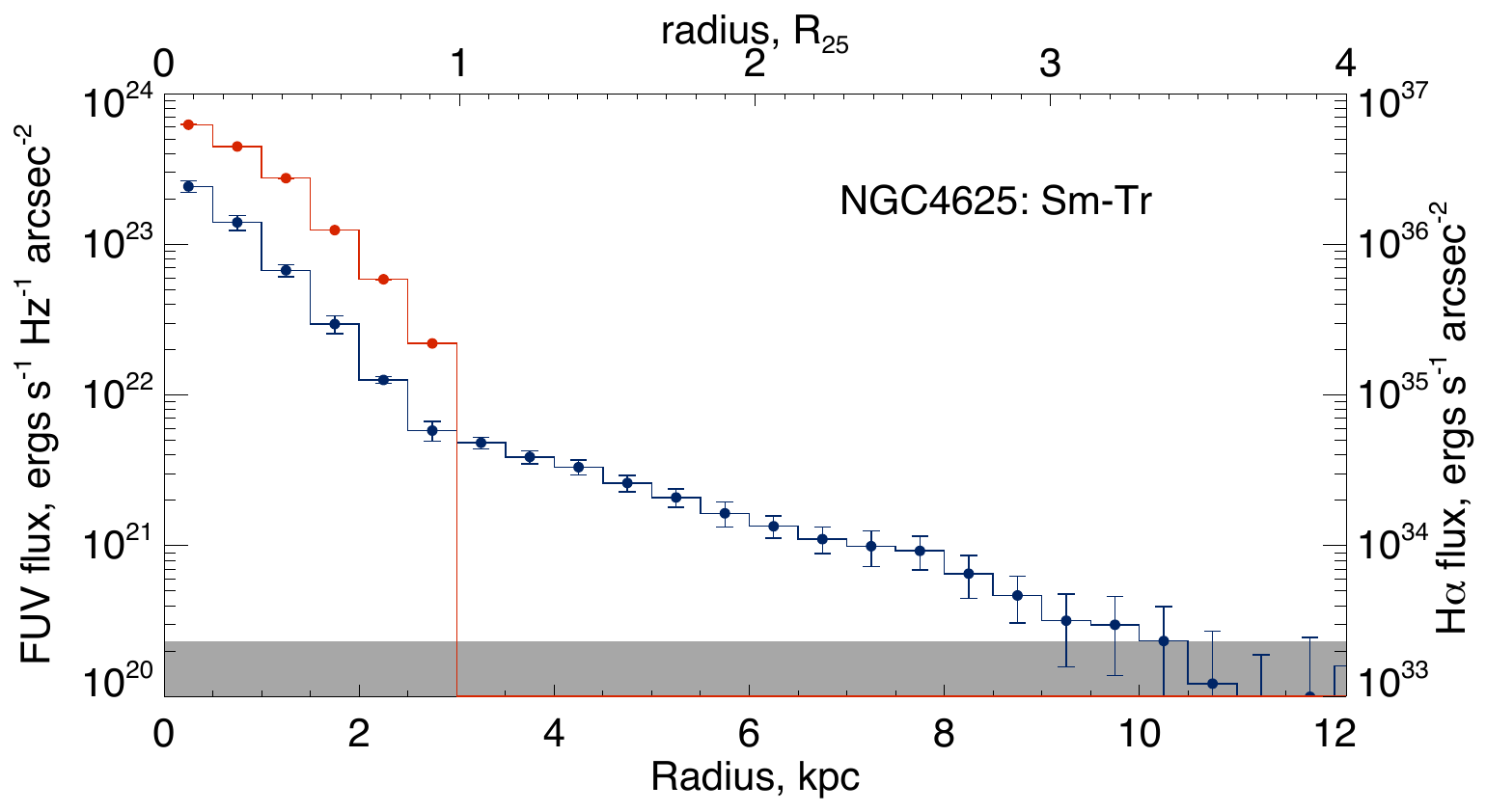}
\end{figure}

\begin{figure}
	\includegraphics[width=84mm]{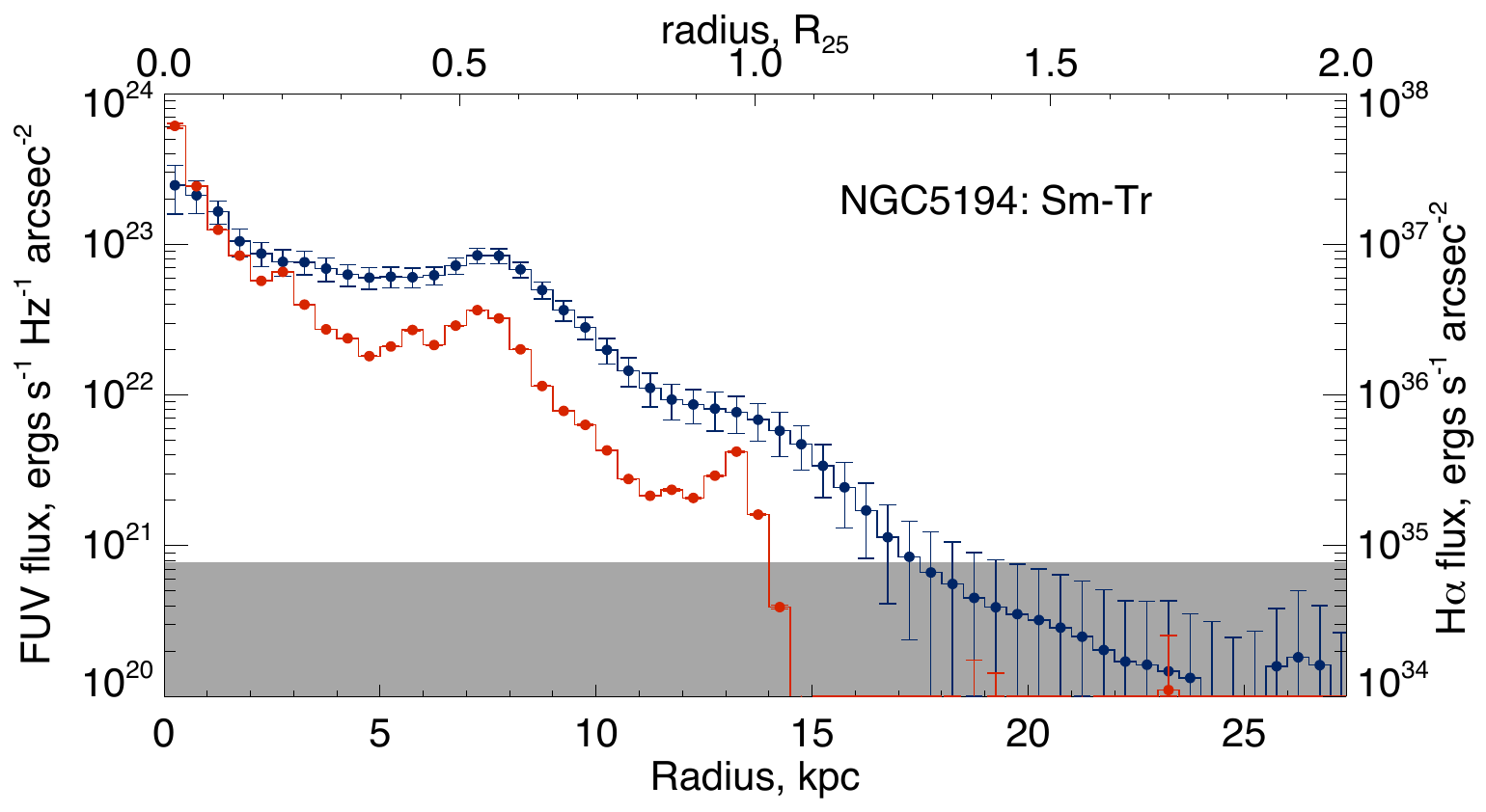}
	\includegraphics[width=84mm]{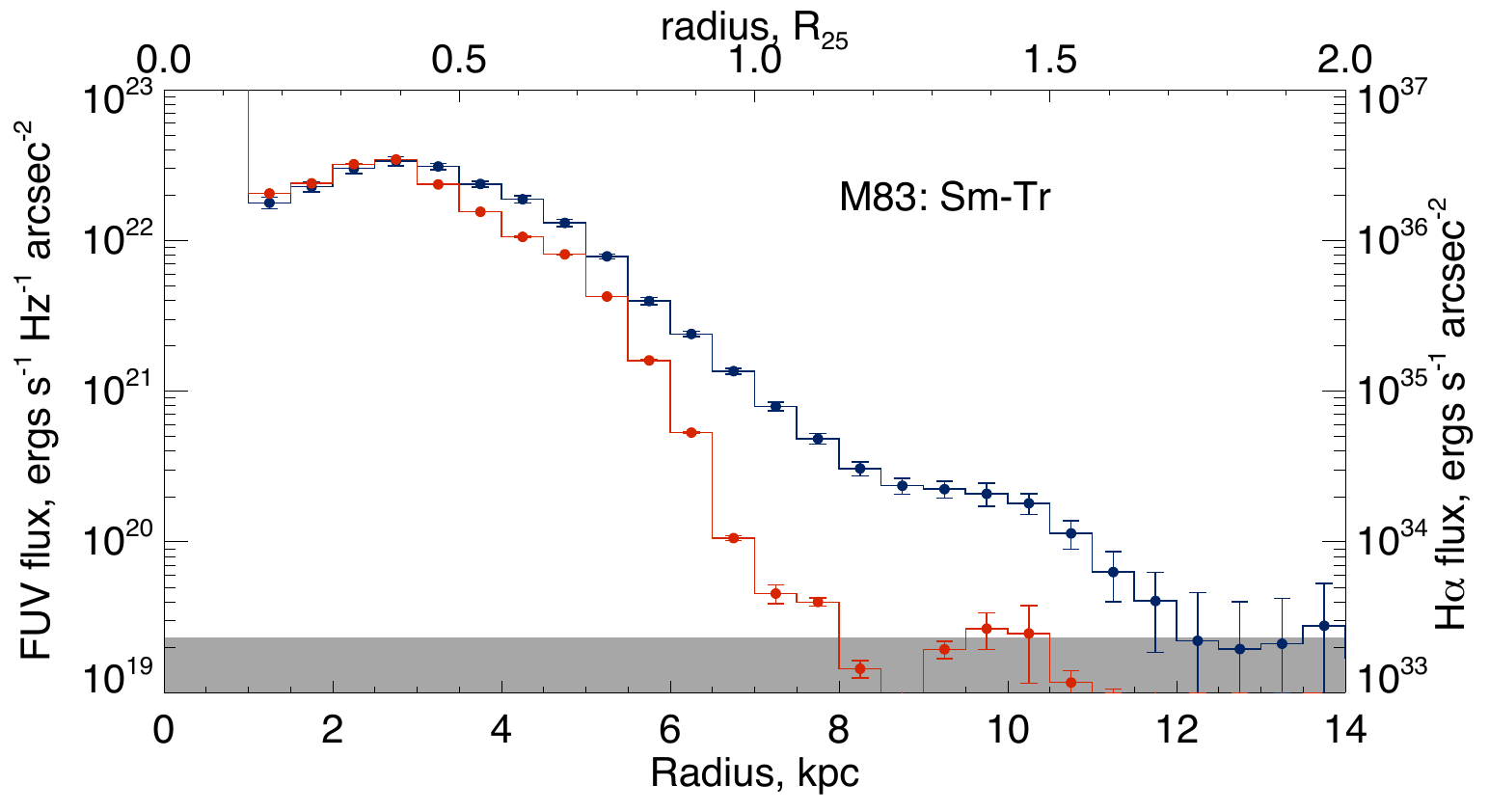}
	\includegraphics[width=84mm]{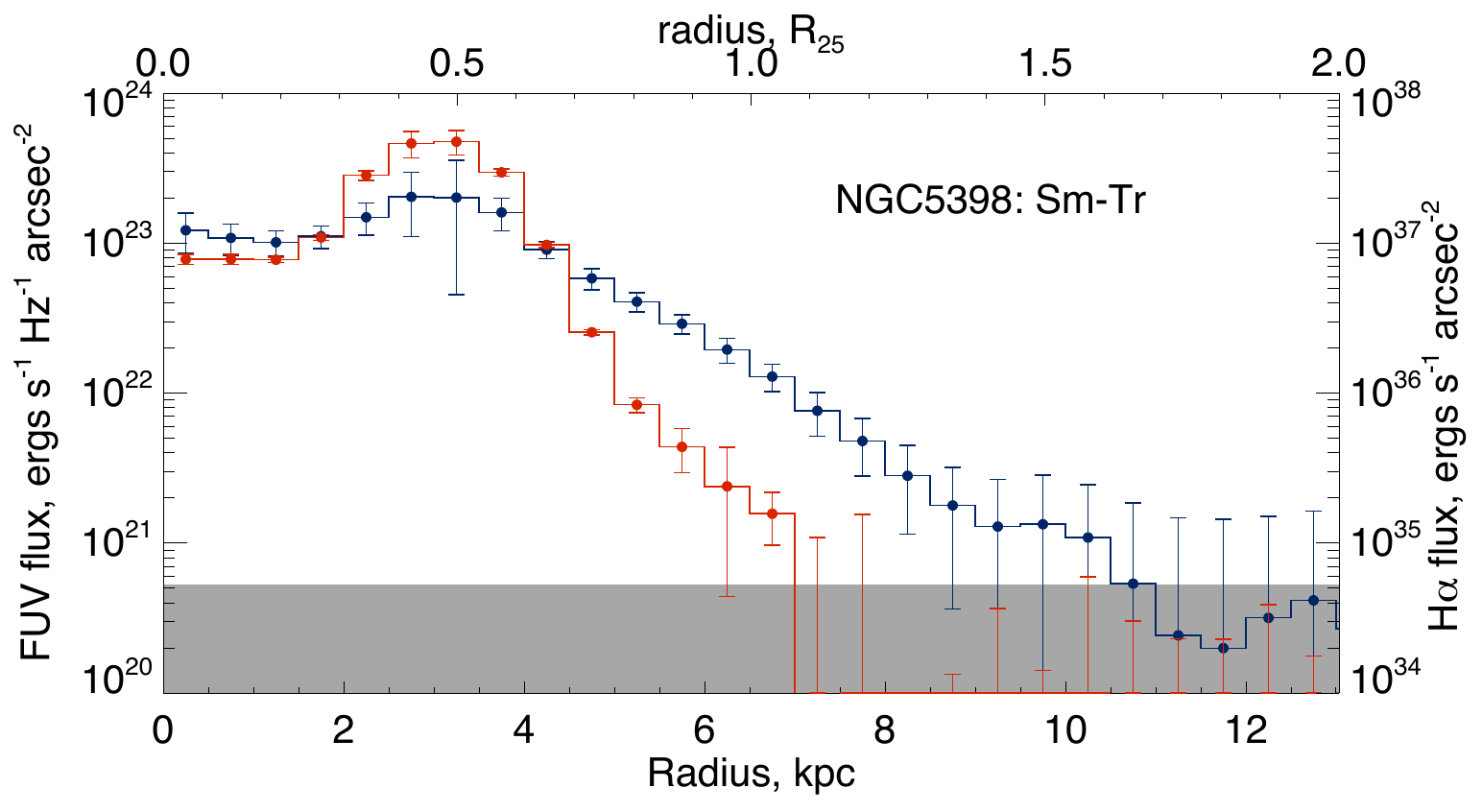}	
	\includegraphics[width=84mm]{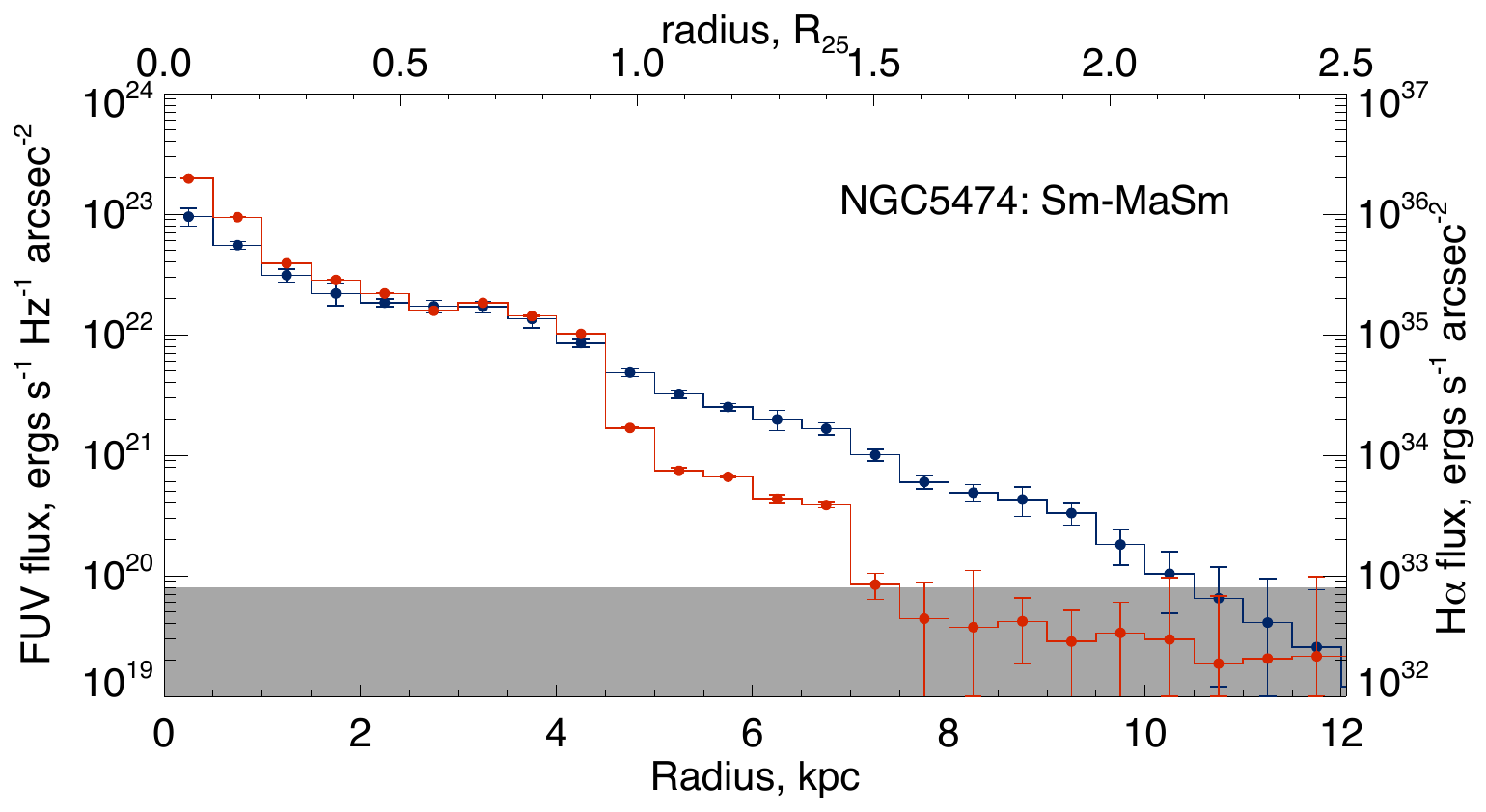}
	\includegraphics[width=84mm]{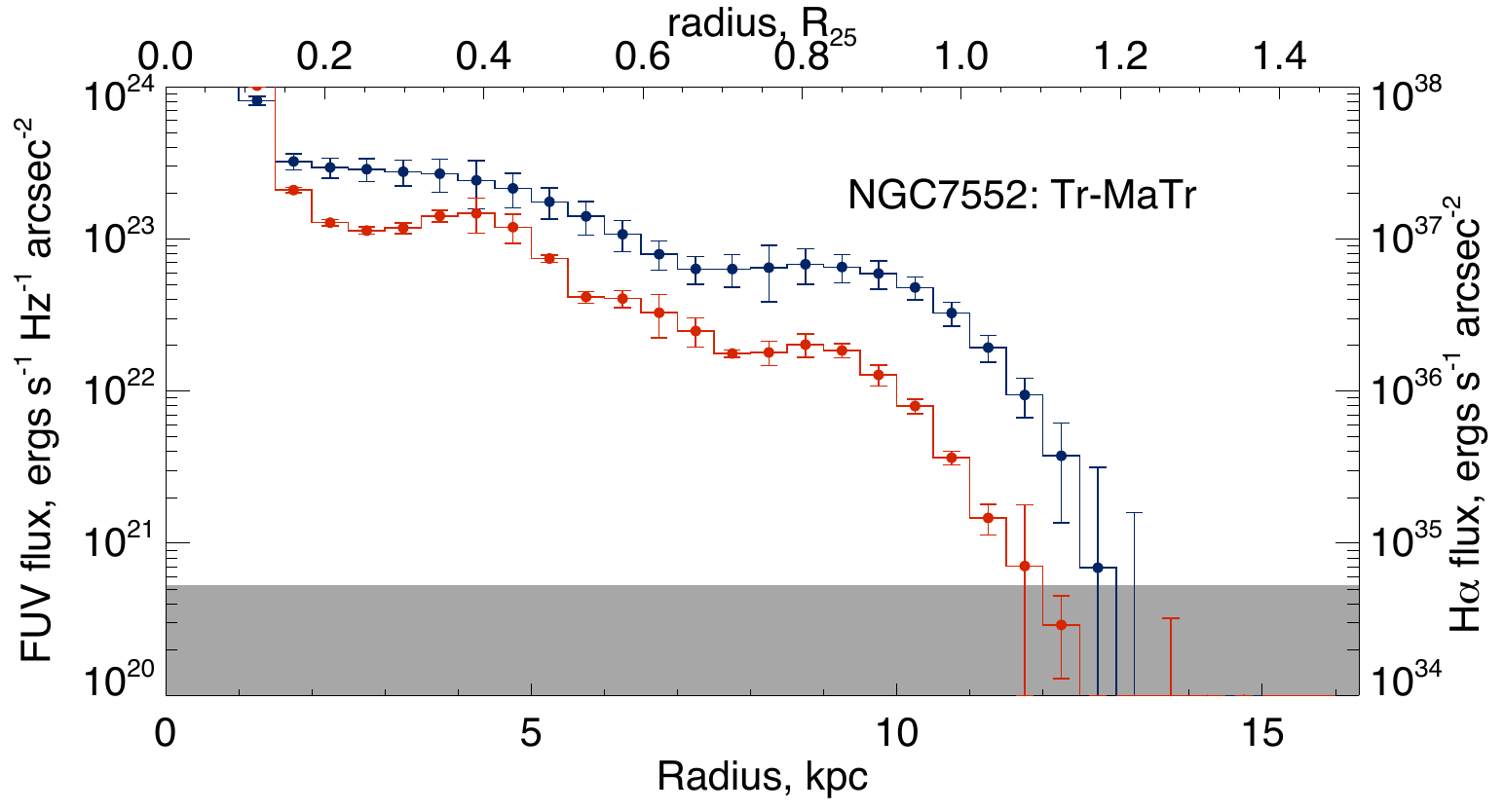}
\end{figure}
\begin{figure}
	\includegraphics[width=84mm]{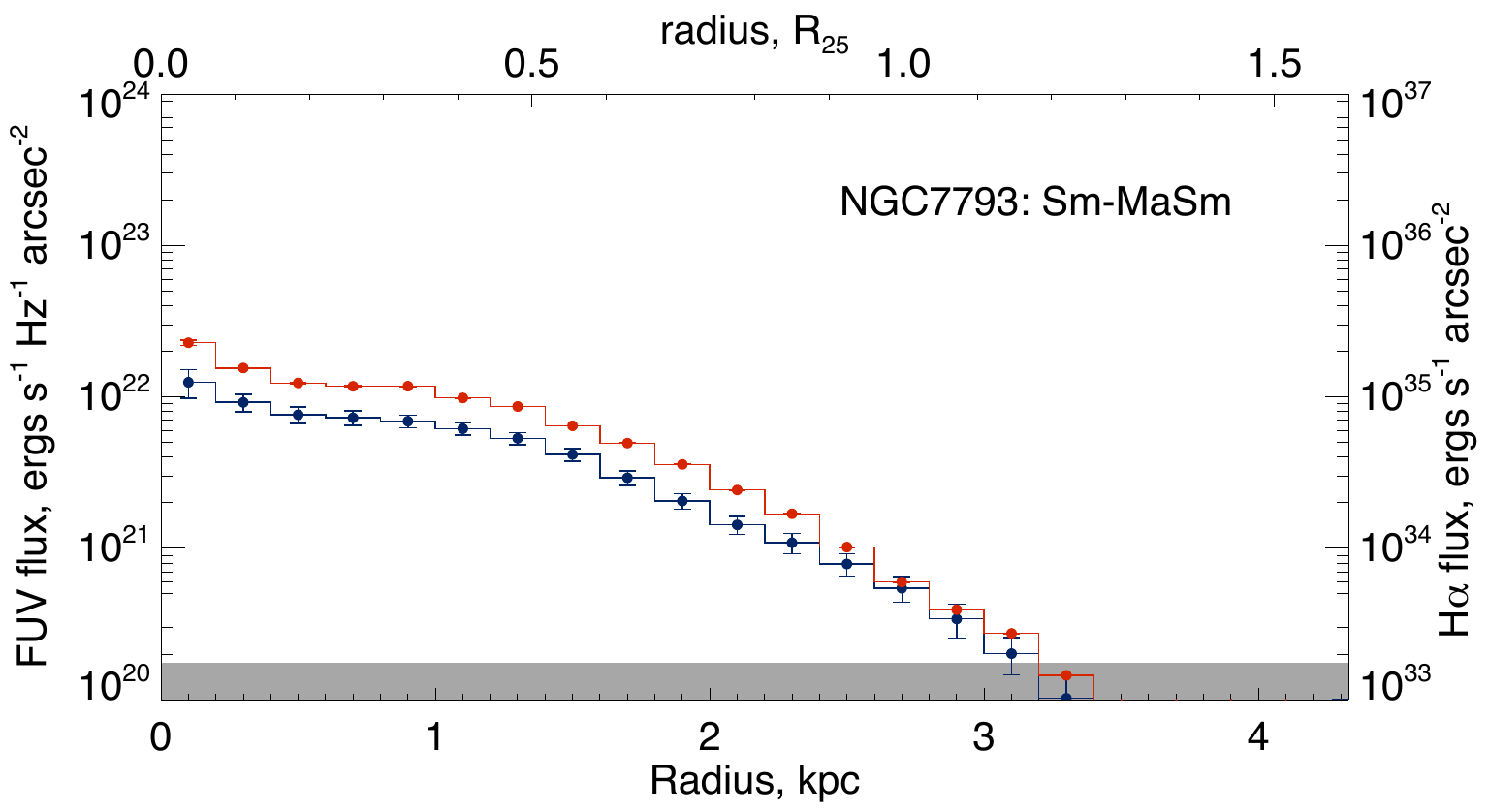}
\end{figure}


\label{lastpage}

\end{document}